\documentclass{aa}
\usepackage{txfonts}
\usepackage{natbib}
\bibpunct{(}{)}{;}{a}{}{,}
\usepackage{graphicx}
\newcommand{\teff}{$T_{\rm{eff}}$}
\newcommand{\logg}{$\log g$}
\newcommand{\lL}{\ifmmode \log \frac{L}{L_{\sun}} \else $\log \frac{L}{L_{\sun}}$\fi}
\newcommand{\mdot}{$\dot{M}$}
\newcommand{\myr}{M$_{\sun}$ yr$^{-1}$}
\newcommand{\vsini}{$V$~sin$i$}
\newcommand{\vinf}{$\varv_{\infty}$}

\newcommand{\vesc}{$\varv_{\rm esc}$}
\newcommand{\kms}{km~s$^{-1}$}
\newcommand{\msun}{M$_{\sun}$}
\newcommand{\zsun}{Z$_{\sun}$}
\newcommand{\mum}{$\mu$m}
\newcommand{\lya}{Ly$\alpha$}

\usepackage[normalem]{ulem}

\begin{document}

\title{Spectroscopic evolution of massive stars near the main sequence at low metallicity}
\author{F. Martins\inst{1}
\and A. Palacios\inst{1}
}
\institute{LUPM, Universit\'e de Montpellier, CNRS, Place Eug\`ene Bataillon, F-34095 Montpellier, France  \\
}

\offprints{Fabrice Martins\\ \email{fabrice.martins@umontpellier.fr}}

\date{Received / Accepted }

\abstract
{The evolution of massive stars is not fully understood. Several physical processes affect their life and death, with major consequences on the progenitors of core-collapse supernovae, long-soft gamma-ray bursts, and compact-object mergers leading to gravitational wave emission.}
{In this context, our aim is to make the prediction of stellar evolution easily comparable to observations.To this end, we developed an approach called "spectroscopic evolution" in which we predict the spectral appearance of massive stars through their evolution. The final goal is to constrain the physical processes governing the evolution of the most massive stars. In particular, we want to test the effects of metallicity.}
{Following our initial study, which focused on solar metallicity,  we investigated the low Z regime. We chose two representative metallicities: 1/5$^{\rm{}}$ and 1/30$^{\rm{}}$ \zsun. We computed single-star evolutionary tracks with the code STAREVOL for stars with initial masses between 15 and 150 \msun. We did not include rotation, and focused on the main sequence (MS) and the earliest post-MS evolution. We subsequently computed atmosphere models and synthetic spectra along those tracks. We assigned a spectral type and luminosity class to each synthetic spectrum as if it were an observed spectrum.}
{We predict that the most massive stars all start their evolution as O2 dwarfs at sub-solar metallicities contrary to solar metallicity calculations and observations. The fraction of lifetime spent in the O2V phase increases at lower metallicity.
  The distribution of dwarfs and giants we predict in the SMC accurately reproduces the observations. Supergiants appear at slightly higher effective temperatures than we predict. More massive stars enter the giant and supergiant phases closer to the zero-age main sequence, but not as close as for solar metallicity. This is due to the reduced stellar winds at lower metallicity. Our models with masses higher than $\sim$60 \msun\ should appear as O and B stars, whereas these objects are not observed, confirming a trend reported in the recent literature. At Z~=~1/30~\zsun, dwarfs cover a wider fraction of the MS and giants and supergiants appear at lower effective temperatures than at Z~=~1/5~\zsun. The UV spectra of these low-metallicity stars have only weak P-Cygni profiles. \ion{He}{ii}~1640 sometimes shows a net emission in the most massive models, with an equivalent width reaching $\sim$1.2 \AA. 
For both sets of metallicities, we provide synthetic spectroscopy in the wavelength range 4500-8000 \AA. This range will be covered by the instruments HARMONI and MOSAICS on the Extremely Large Telescope and will be relevant to identify hot massive stars in Local Group galaxies with low extinction. We suggest the use of the ratio of \ion{He}{i}~7065 to \ion{He}{ii}~5412 as a diagnostic for spectral type. Using archival spectroscopic data and our synthetic spectroscopy, we show that this ratio does not depend on metallicity. 
Finally, we discuss the ionizing fluxes of our models. The relation between the hydrogen ionizing flux per unit area versus effective temperature depends only weakly on metallicity. The ratios of \ion{He}{i} and \ion{He}{ii} to H ionizing fluxes both depend on metallicity, although in a slightly different way.}
{We make our synthetic spectra and spectral energy distributions available to the community.}


\authorrunning{Martins \& Palacios}
\titlerunning{Synthetic spectroscopy of massive stars at low Z}

\maketitle

\section{Introduction}
\label{s_intro}

Understanding the evolution and final fate of massive stars is of primordial importance now that observations of core-collapse supernovae, long-soft gamma-ray bursts (LGRBs), and compact-object mergers are becoming almost routine. However, many uncertainties still hamper unambiguous predictions from evolutionary models \citep[e.g.,][]{mp13}. Although mass loss \citep{cm86} and rotation \citep{mm00} have long been recognized as key drivers of stellar evolution, other processes significantly affect the way massive stars evolve. Magnetism, which is present at the surface of a minority of OB stars \citep{grunhut17}, may strongly impact the outcome of their evolution \citep{zsolt19}. An uncertain but potentially large fraction of massive stars have a companion that will modify the properties of the star compared to isolation \citep[e.g.,][]{moedistef13,kobul14,demink13,mahy20}.

Metallicity is another major driver of the evolution of massive stars. It modifies opacities and therefore the internal structure of stars. As a result,  massive metal-poor stars are usually hotter and more compact \citep{mm01}. The resulting steeper gradients are predicted to enhance the effects of rotation on stellar evolution \citep{mm00}, although direct observational confirmation is still lacking. At lower metallicity, radiatively driven winds are weaker \citep{vink01,mokiem07}, meaning that the effects of mass loss are reduced. The binary fraction at low metallicity is not well constrained: \citet{moedistef13} find no differences between the Magellanic Clouds and the Galaxy, while \citet{dw18} report a possible decrease of the binary fraction at lower metallicity among high-mass stars, in contrast to what is observed for low-mass stars \citep{rag10}. \citet{stanway20} studied how the uncertainties in binary parameters affect the global predictions of population-synthesis models. These latter authors concluded that varying the binary properties for high-mass stars leads to variations that do not exceed those caused by metallicity. The metallicity effects on rotation and mass loss also impact the occurrence of LGRBs. \citet{japelj18} and \citet{palmeiro19} show that low metallicity is favored for LGRBs, and there is a metallicity threshold above which they are seldom observed \citep{vergani15,perley16}. Metallicity therefore appears  to be a major ingredient of massive star evolution.

In the present paper, we discuss the role of metallicity in the spectroscopic appearance of massive stars on and close to the main sequence (MS). This extends the work we presented in \citet{mp17} in which we described our method to produce spectroscopic sequences along evolutionary tracks. This method consists in computing atmosphere models and synthetic spectra at dedicated points sampling an evolutionary track, and was pioneered by \citet{costar1} and recently revisited by us and \citet{groh13,groh14}. \citet{got17,got18} used a similar approach to investigate the ionizing properties of stars stripped of their envelope in binary systems. These latter authors found that such objects emit a large number of ionizing photons, equivalent to Wolf-Rayet stars. \citet{kub19} looked at the spectral appearance of stars undergoing quasi-chemically homogeneous evolution  \citep{maeder87,yoon06}, focusing on metal-poor objects (Z = 1/50 \zsun), for this type of evolution seems to be more easily achieved  at that metallicity \citep[e.g.,][]{brott11}. \citet{kub19} concluded that for most of their evolution, which proceeds directly leftward of the zero age main sequence (ZAMS), stars show only absorption lines in their synthetic spectra, therefore appearing as early-type O stars.
In the present work, similarly to \citet{mp17}, we focus on the MS and early post-MS evolution because these phases are the least affected by uncertainties \citep[see][]{mp13}. Our goal is to predict the spectral properties of stars at low metallicity, to compare them with observational data, and ultimately to provide constraints on stellar evolution. To this end, we selected two representative metallicities: 1/5 \zsun\ and 1/30 \zsun. The former is the classical value of the Small Magellanic Cloud (SMC), and the latter is on the low side of the distribution of metallicities in Local Group dwarf galaxies \citep{mcco05,ross15}. These two values of metallicity should therefore reasonably bracket the metal content of most stars that will be observed individually in the Local Group with next-generation telescopes such as the Extremely Large Telescope (ELT). In preparation for these future observations, we make predictions on the spectral appearance of hot massive stars in these metal-poor environments. We also provide classification criteria suitable for the ELT instruments.

In Sect.\ \ref{s_method} we describe our method. We present our spectroscopic sequences in Sect.\ \ref{s_specseq}, where we also define a new spectral type diagnostic. We present the ionizing properties of our models in Sect.\ \ref{s_ionHeII1640}. In this section we also discuss \ion{He}{ii}~1640 emission that is present in some of our models. Finally, we conclude in Sect.\ \ref{s_conc}. 

\section{Method}
\label{s_method}

\subsection{Evolutionary models and synthetic spectra}
\label{s_mod}

We computed evolutionary models for massive stars with the code \emph{STAREVOL} \citep{dmp09,amard16}. We assumed an Eddington grey atmosphere as outer boundary condition to the stellar structure equations. We used the \cite{asplund09} solar chemical composition as a reference, with Z$_\odot$~=~0.0134. A calibration of the solar model with the present input physics leads to an initial helium mass fraction $Y = 0.2689$ at solar metallicity. We used the corresponding constant slope $\Delta Y/\Delta Z=1.60$ (with the primordial abundance Y$_0$=0.2463 based on WMAP-SBBN by \citealt{Cocetal04}) to compute the initial helium mass fraction at Z = 2.69 $10^{-3}$ = 1/5 Z$_\odot$  and Z = 4.48~$10^{-4}$ = 1/30 Z$_\odot$, and to scale all the abundances accordingly. The OPAL opacities used for these models comply to this scaled distribution of nuclides. We did not include specific $\alpha$-element enhancement in our models. We described the convective instability using the mixing-length theory with $\alpha_{MLT} = 1.6304$, and we use the Schwarzschild instability criterion to define the boundaries of convective regions. We added a step overshoot at the convective core edge and adopt $\alpha_{ov} = 0.1 H_p $, with $H_p$ being the pressure scale height. We used the thermonuclear reaction rates from the NACRE II compilation \citep{Xu2013b} for mass number $A < 16$, and the ones from the NACRE compilation \citep{nacre} for more massive nuclei up to Ne. The proton captures on nuclei more massive than Ne are from \citet{longland} or \citet{iliadis}. The network was generated via the NetGen server \citep{Xu2013a}.\\
We used the mass-loss-rate prescriptions of \citet{vink01}, who account for the metallicity scaling of mass-loss rates \citep[see also][]{mokiem07}.
In order to account for the effect of clumping in the wind \citep{fullerton11}, the obtained mass-loss rates were divided by a factor of three \citep{cohen14}. This reduction is consistent with the revision of theoretical mass-loss rates proposed by \citet{lucy10}, \citet{kk17}, and \citet{bjorklund20}.

Along each evolutionary sequence, we selected points for which we computed an atmosphere model and the associated synthetic spectrum with the code \emph{CMFGEN} \citep{hm98}. \emph{CMFGEN} solves the radiative transfer and statistical equilibrium equations under non-LTE conditions using a super-level approach. The temperature structure is set from the constraint of radiative equilibrium. A spherical geometry is adopted to account for stellar winds. The input velocity structure is a combination of a quasi-static equilibrium solution below the sonic point and a $\beta$-velocity law above it (i.e., $\varv= \varv_{\infty} \times\ (1-R/r)^{\beta}$, where \vinf\ is the maximal velocity at the top of the atmosphere and $R$ is the stellar radius). We adopted \vinf\ = 3.0 $\times$ \vesc\ as in \citet{mp17}\footnote{We note that in the calculation of the mass-loss rates we use the recipe of \citet{vink01} which incorporates a ratio \vinf/\vesc\ of 2.6. This is slightly different from the value of 3.0 we use for the calculation of the synthetic spectra, but the difference is minimal: adopting 3.0 in the Vink formula would change the mass-loss rate by 0.08 dex, which is negligible.}.
This value is consistent with both observations \citep{garcia14} and theoretical predictions \citep{bjorklund20} in which \vinf/\vesc\ is in the ranges 1.0-6.0 and 2.5-5.5, respectively. We note that the observational study of \citet{garcia14} shows a correlation between terminal velocity and metallicity \citep[see also][]{leitherer92}, but no clear trend can be seen between the very scattered ratio \vinf/\vesc\ and metallicity. The velocity structure below the sonic point is iterated a few times during the atmosphere model calculation, taking the radiative force resulting from the radiation field and level populations into account. The density structure follows from the velocity structure and mass conservation equation. The models include the following elements: H, He, C, N, O, Ne, Mg, Si, S, Ar, Ca, Fe, and Ni. A total of about 7100 atomic levels\footnote{A super-level approach is used in \emph{CMFGEN} calculations. The $\sim$7100 levels are grouped in about 1800 super-levels.} and nearly 170000 atomic transitions are taken into account. Once the atmosphere model is converged, a formal solution of the radiative transfer equation is performed and leads to the synthetic spectrum in the wavelength range 10 \AA\ - 50 \mum. In that process, a depth-variable microturbulent velocity varying from 10 \kms\ at the bottom of the photosphere to 10\% of the terminal velocity at the top of the atmosphere is adopted.

\begin{figure}[t]
\centering
\includegraphics[width=0.49\textwidth]{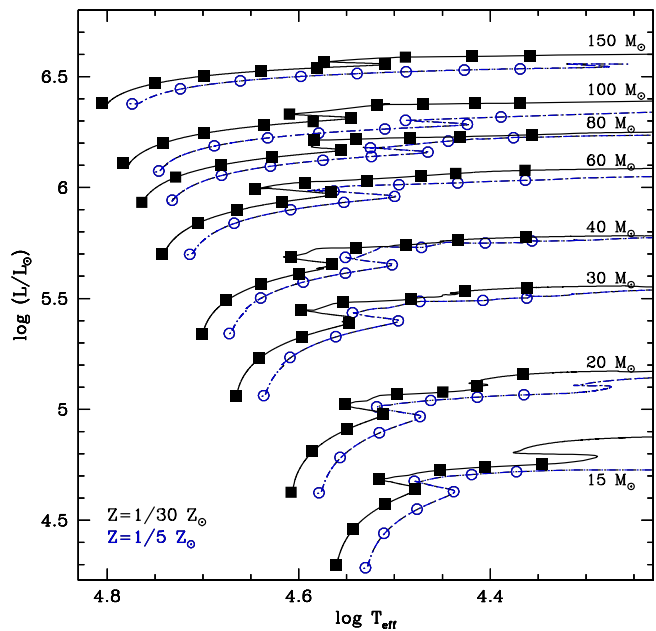}  
\caption{Hertzsprung-Russell diagram for the SMC (black lines and filled squares) and Z=1/30 \zsun\ (dot-dashed blue lines and open circles) cases. Lines are the \emph{STAREVOL} evolutionary tracks. Symbols are the points at which an atmosphere model and synthetic spectrum have been computed.}
\label{figHR}
\end{figure}

Figure\ \ref{figHR} shows the Hertzsprung-Russell diagram at the two selected metallicities. The optical spectra and spectral energy distributions (SEDs) are distributed through the \emph{POLLUX}\footnote{\url{http://pollux.oreme.org/}} database \citep{pollux10}. The parameters adopted for their computations are listed in Tables \ref{tab_smc} and \ref{tab_1on30}.

\subsection{Spectral classification}
\label{s_classif}

Once the synthetic spectra were calculated, we performed a spectral classification as if they were results of observations. We followed the method presented by \citet{mp17} with some slight adjustments. Our process can be summarized as follows:

\begin{itemize}

\item \textit{Spectral type}: The main classification criterion for O stars is the relative strength of \ion{He}{i}~4471 and \ion{He}{ii}~4542 as proposed by \citet{ca71} and quantified by \citet{mathys88}. For each spectrum, we therefore computed the equivalent width (EW) of both lines and calculated the logarithm of their ratio. A spectral type was assigned according to the Mathys scheme. For spectral types O9 to O9.7, we refined the classification using the criteria defined by \citet{sota11} and quantified by \citet{classif}, namely $\frac{EW(\ion{He}{i}~4144)}{EW(\ion{He}{ii}~4200)}$ and $\frac{EW(\ion{He}{i}~4388)}{EW(\ion{He}{ii}~4542})$. 
For B stars, we estimated the relative strength of \ion{Si}{iv}~4089 and \ion{Si}{iii}~4552. We used the atlas of \citet{wf90} to assign B-type sub-classes. Finally, for the earliest O stars (O2 to O3.5) we relied on the relative strength of \ion{N}{iii}~4640 and \ion{N}{iv}~4058 as defined by \citet{walborn02}.  

\item \textit{Luminosity class}: For O stars earlier than O8.5, the strength of \ion{He}{ii}~4686 was the main classification criterion. We used the quantitative scheme presented by \citet{classif} to assign luminosity classes. For stars with spectral type between O9 and O9.7, we used the ratio $\frac{EW(\ion{He}{ii}4686)}{EW(\ion{He}{i}4713)}$ defined by \citet{sota11} and quantified by \citet{classif}. 
For B stars, we relied mainly on the morphology of H$\gamma$ which is broad in dwarfs and gets narrower in giants and supergiants. 

\end{itemize}
For both spectral type and luminosity class assignment we discarded classification criteria based on the relative strengths of Si to He lines because they are metallicity dependent and this dependence is not quantified at metallicities different from solar.

\noindent For all stars, a final step in the classification process involved a direct comparison with standard stars. The spectra of these reference objects were retrieved from the GOSC catalog\footnote{\url{https://gosc.cab.inta-csic.es/}} for O stars and from the POLARBASE archive\footnote{\url{http://polarbase.irap.omp.eu/}} for B stars. The final spectral classes and luminosity classes are given in Tables \ref{tab_smc} and \ref{tab_1on30}.

\section{Spectroscopic sequences}
\label{s_specseq}

In this section we discuss the spectroscopic sequences along the evolutionary tracks that we obtained. We first describe general trends before examining the two selected metallicities.

\begin{figure*}[t]
\centering
\includegraphics[width=0.49\textwidth]{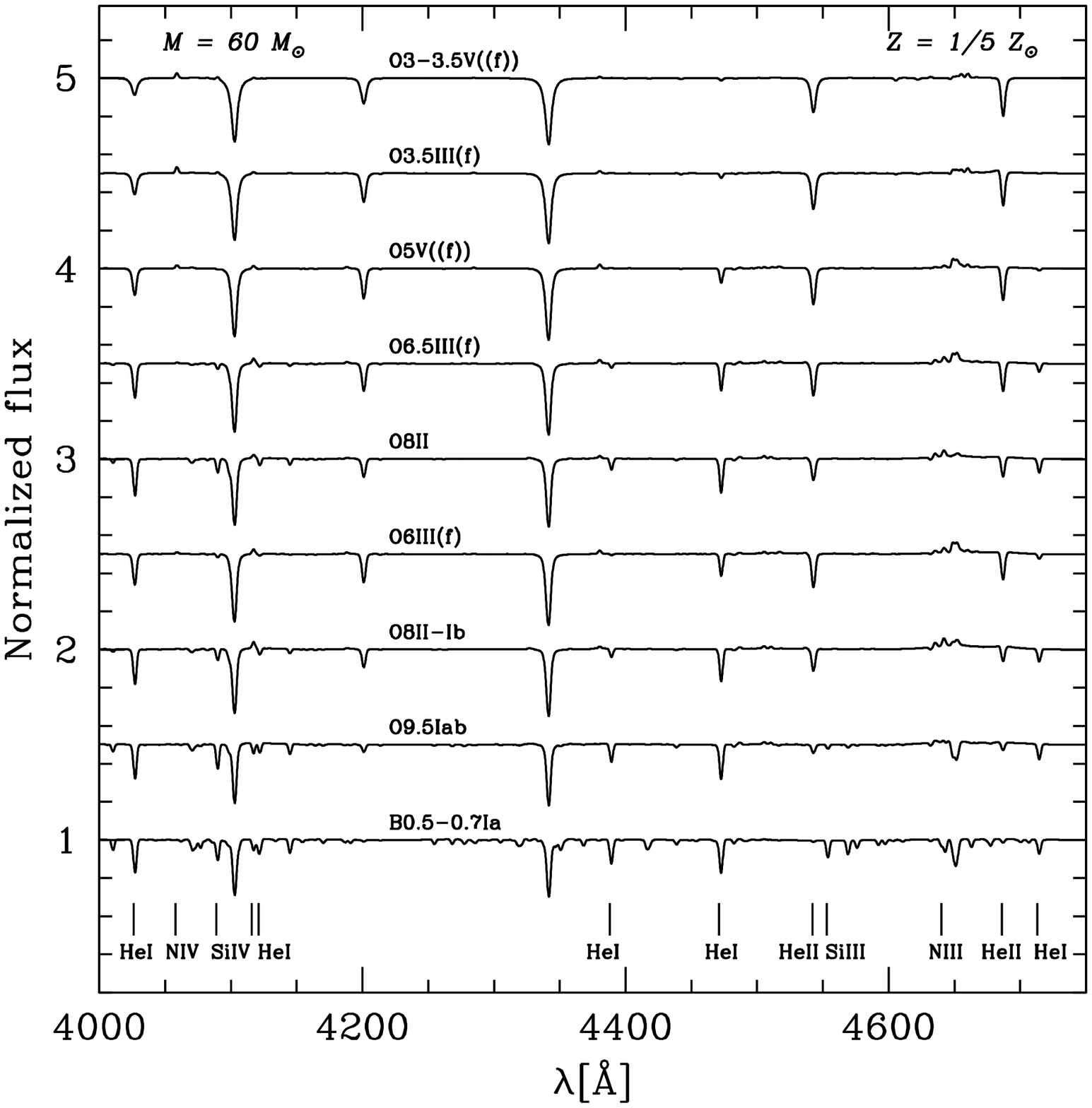}  
\includegraphics[width=0.49\textwidth]{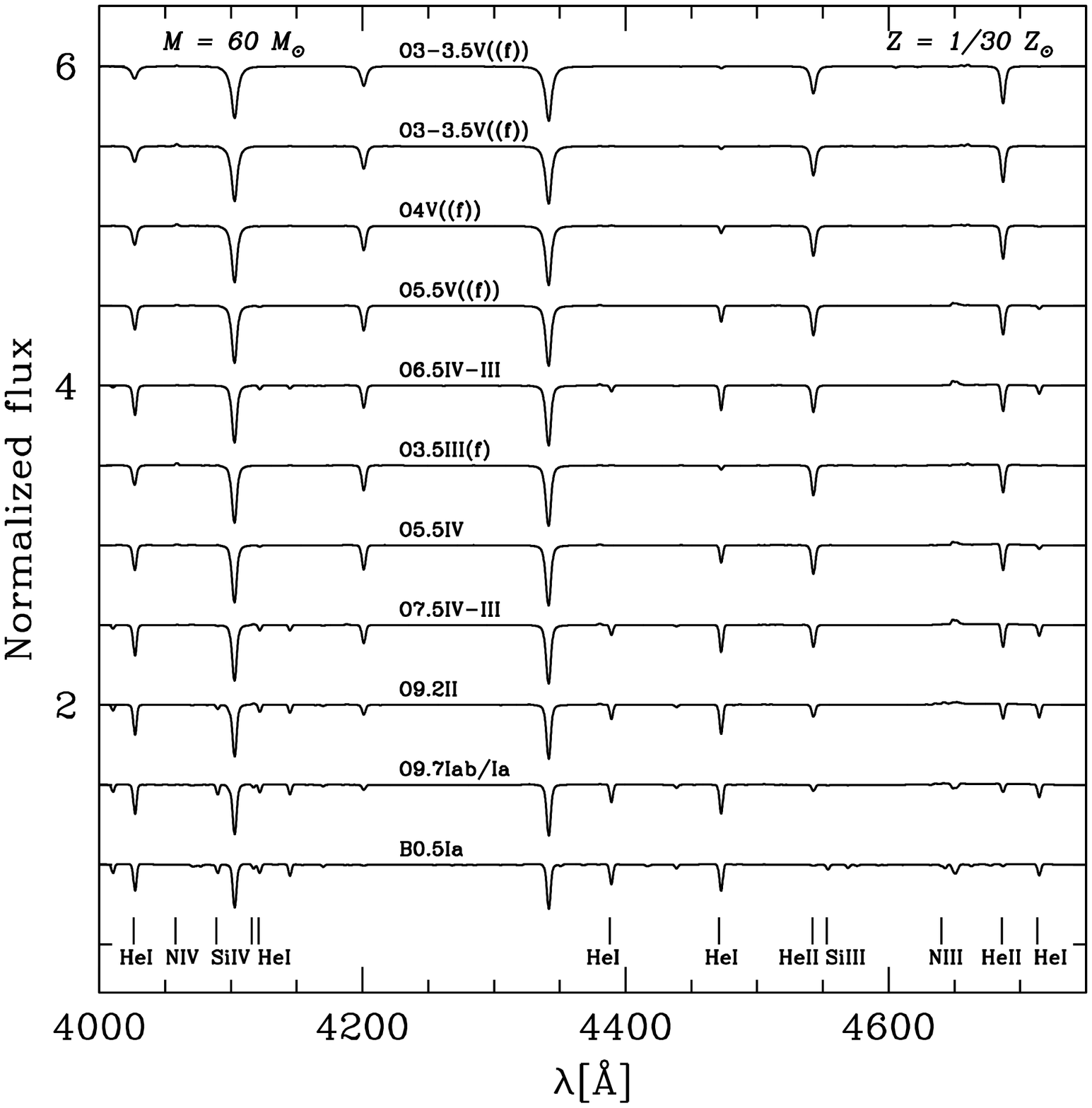}
\caption{Optical spectra of the sequence of models calculated along the 60 \msun\ track at SMC (left) and one-thirtieth (right) metallicity from the ZAMS at the top to the post-MS at the bottom. The main diagnostic lines are indicated at the bottom of the figure. The spectra have been degraded to a spectral resolution of $\sim$2500, similar to that of the GOSS survey.}
\label{fig_sv60opt}
\end{figure*}

\subsection{Example of spectroscopic sequences}
\label{s_example}

\begin{figure*}[t]
\centering
\includegraphics[width=0.49\textwidth]{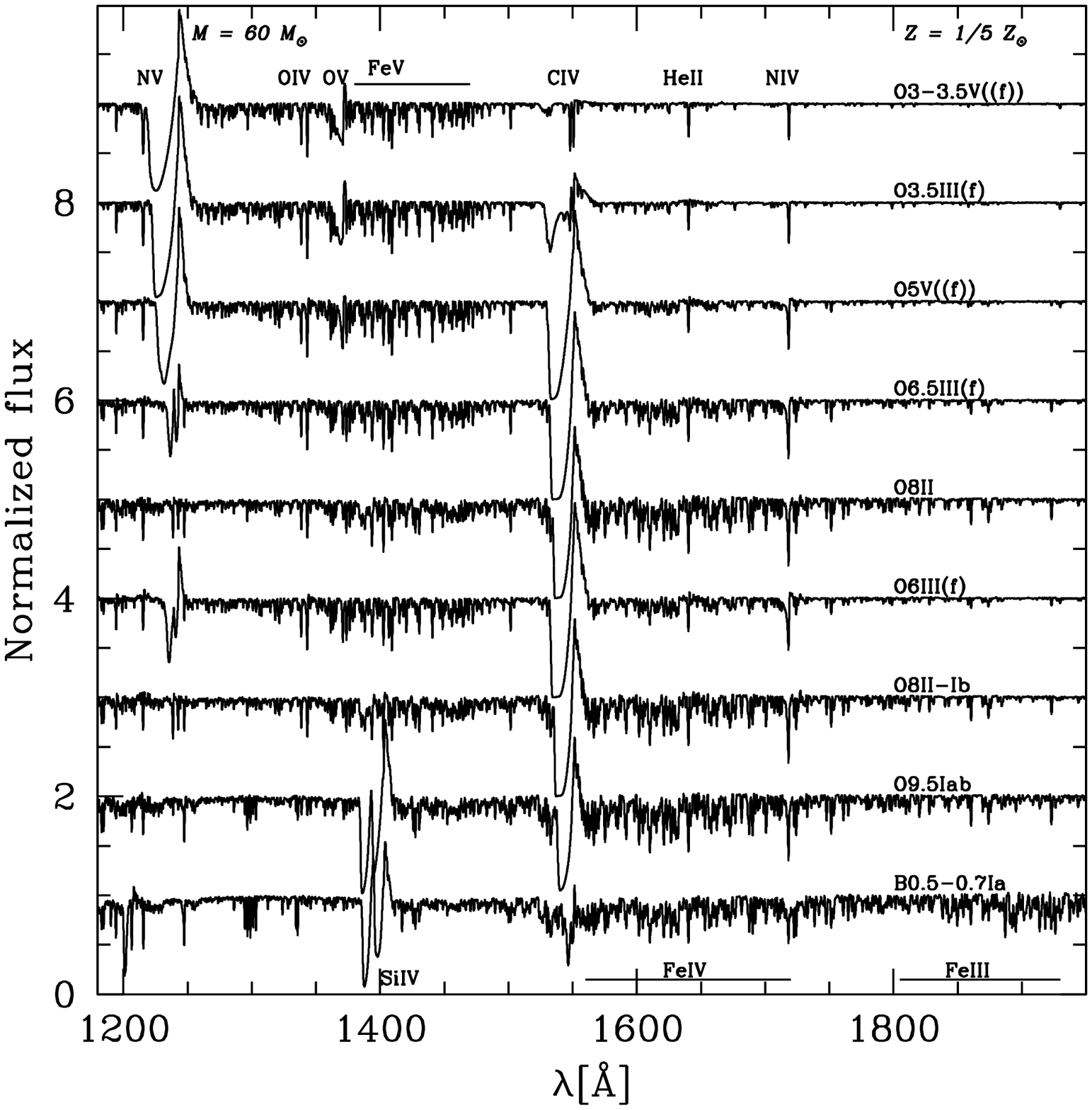}  
\includegraphics[width=0.49\textwidth]{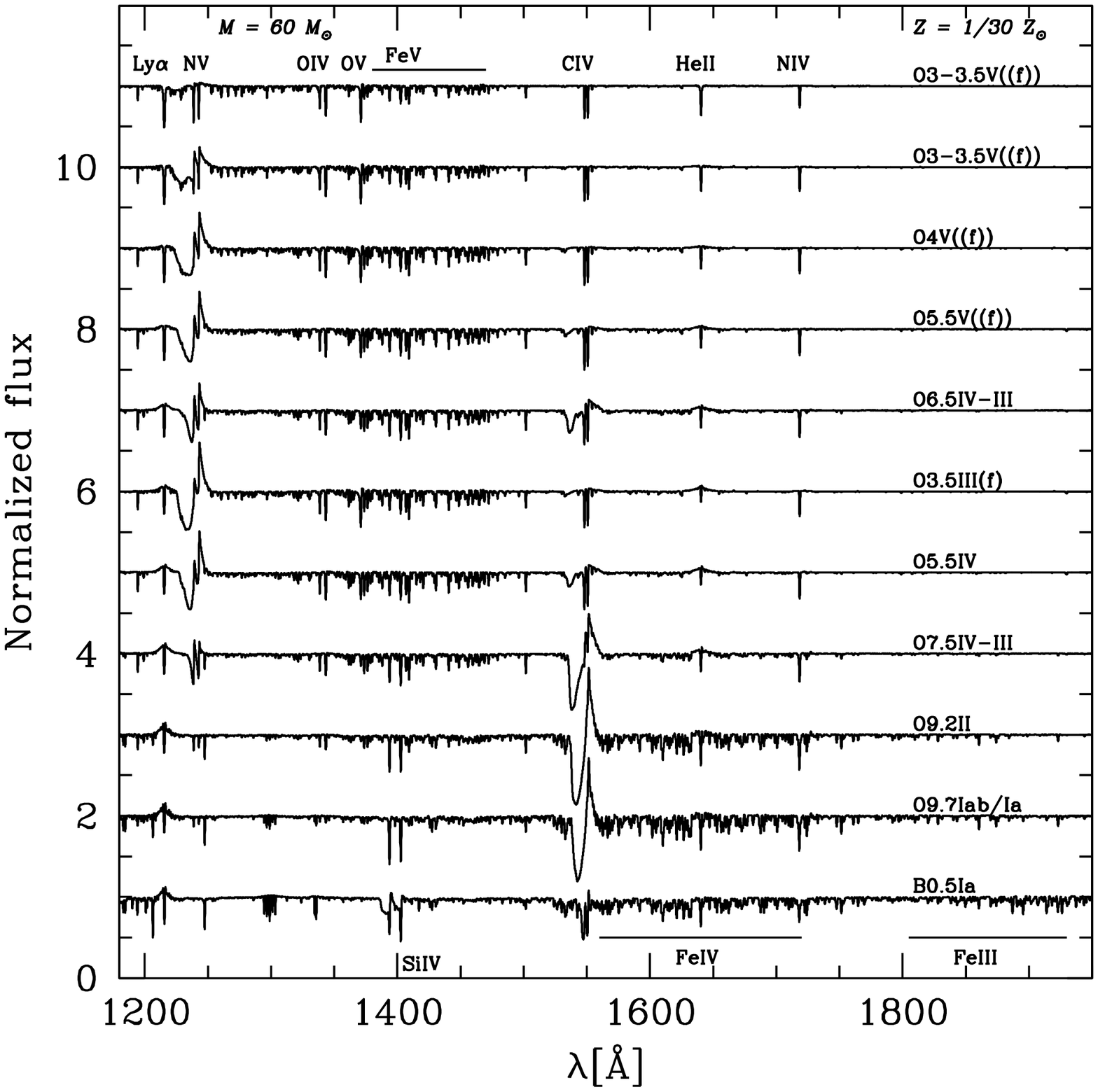}
\caption{Same as Fig.\ \ref{fig_sv60opt} but for the UV range. The spectra have been degraded to a spectral resolution of $\sim$16000 typical of HST/COS FUV observations.}
\label{fig_sv60uv}
\end{figure*}

We first describe full spectroscopic sequences for typical cases. In Fig.\ \ref{fig_sv60opt} we show the optical spectra computed along the 60 \msun\ tracks. According to our computations, the star appears as a O3-3.5 dwarf on the ZAMS and enters the post-MS evolution as a late-O/early-B supergiant. This is valid for both the SMC and one-thirtieth$^{\rm{}}$ solar metallicities. The evolution of the \ion{He}{i}~4471 to \ion{He}{ii}~4542 line ratio ---the main spectral type classification criterion \citep{ca71}--- is clearly seen. Figure~\ref{fig_sv60opt} highlights the reduction of the metal lines at lower metallicity: for Z~=~1/30~\zsun,\ silicon, nitrogen, and carbon lines are weaker than for a SMC metallicity. When comparing to Fig.\ 6 of \citet{mp17} which shows solar metallicity computations, the effect is even more striking. 
This effect is magnified in the ultraviolet range. Figure\ \ref{fig_sv60uv} shows the spectroscopic sequences for the same 60 \msun\ tracks, but between 1200 and 1900~\AA. First, the strong P-Cygni lines are severely reduced in the one-thirtieth$^{\rm{}}$ solar metallicity spectra. This is due to the reduction in both mass-loss rate and metal abundance. Second, the iron photospheric lines are weaker in the lower metallicity spectra. 
Figure\ \ref{fig_sv60uv} illustrates the change of iron ionization when \teff\ varies: at early spectral types, and therefore high \teff, \ion{Fe}{v} lines dominate the absorption spectrum around 1400 \AA; at late spectral types, it is \ion{Fe}{iv} lines and even \ion{Fe}{iii} lines in the coolest cases that are stronger.

Figures\ \ref{fig_sv20opt} and \ref{fig_sv20uv} in Appendix B show the optical and UV sequences followed by a 20 \msun\ star. Qualitatively, the trends are the same as for the 60~\msun\ star. Figure\ \ref{fig_ir} also in Appendix B displays the sequences of the 60 \msun\ stars in the K-band. At these wavelengths, the number of lines is reduced and there are very few metallic lines. The effects of metallicity are therefore difficult to identify. The \ion{C}{iv} lines around 2.06-2.08 \mum\ almost disappear at Z~=~1/30~\zsun. The \ion{N}{iii}/\ion{O}{iii} emission complex near 2.11 \mum\ is also reduced. These figures illustrate that the K-band is far from being an optimal tool with which to constrain stellar parameters and surface abundances, but importantly the figure also demonstrates that  the K-band cannot be used to reliably constrain metallicity effects in OB stars.

\subsection{Metallicity of the Small Magellanic Cloud}
\label{s_smc}

\smallskip
\begin{figure*}[t]
\centering
\includegraphics[width=0.49\textwidth]{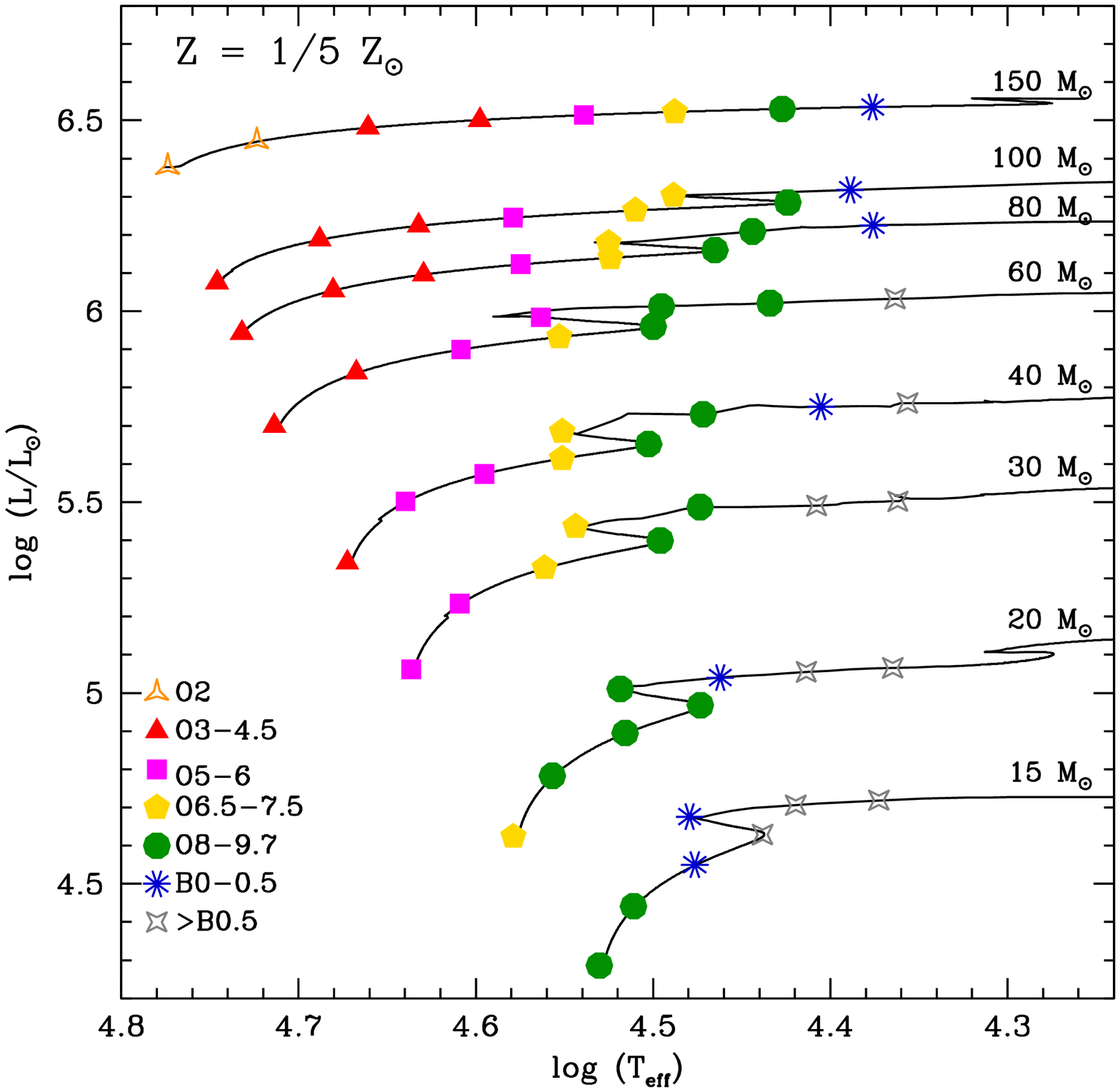}
\includegraphics[width=0.49\textwidth]{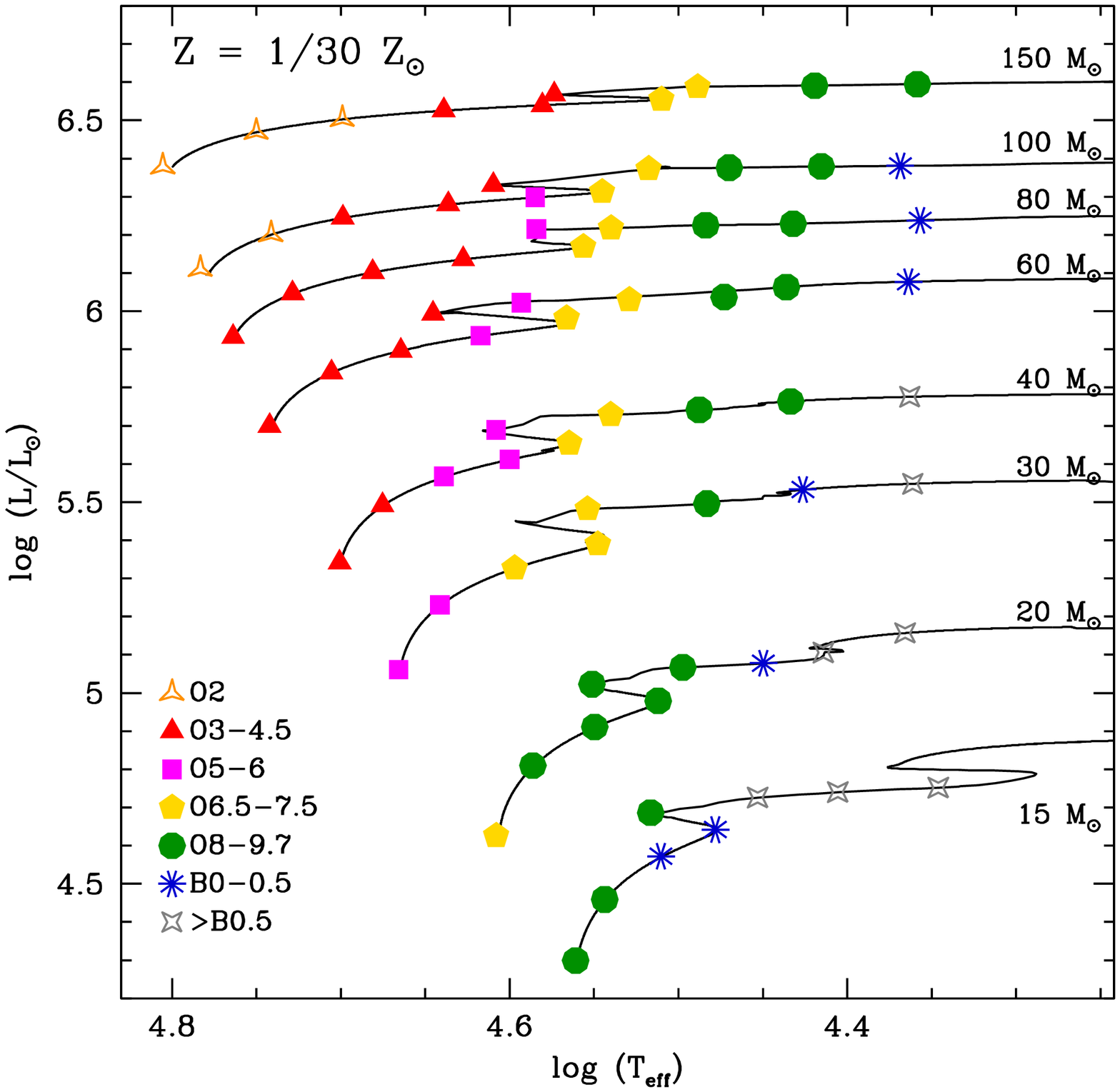}
\caption{Distribution of spectral types across the HR diagram at SMC (left panel) and one-thirtieth$^{\rm{}}$ solar (right panel) metallicity.}
\label{fig_hrST}
\end{figure*}

The left panel of Fig.~\ref{fig_hrST} shows the distribution of spectral types in the HR diagram at Z~=~1/5~\zsun. A given spectral type is encountered at slightly higher \teff\ for lower masses. This is caused by the higher surface gravity. For instance, the first model of the 20~\msun\ sequence is classified as O7.5. The same spectral type is attributed to the sixth model of the 150~\msun\ sequence. The surface gravity in these models is 4.38 and 2.98, respectively. At lower \logg,\ a lower \teff\ is required to reach the same ionization, and therefore the same spectral type \citep[see][]{martins02}. In the example given here, the \teff\ difference reaches 7000~K.

In Fig.~\ref{fig_hrST} the upper left part of the HR diagram is populated by stars earlier than O5. The number of such stars is higher than at solar metallicity \citep[see Fig.~7 of][]{mp17}. The reason for this is mainly the shift of the ZAMS and evolutionary tracks towards higher \teff\ at lower metallicity \citep{mm01}. Higher \teff, and therefore earlier spectral types, are therefore reached at lower metallicity.

\begin{figure*}[]
\centering
\includegraphics[width=0.49\textwidth]{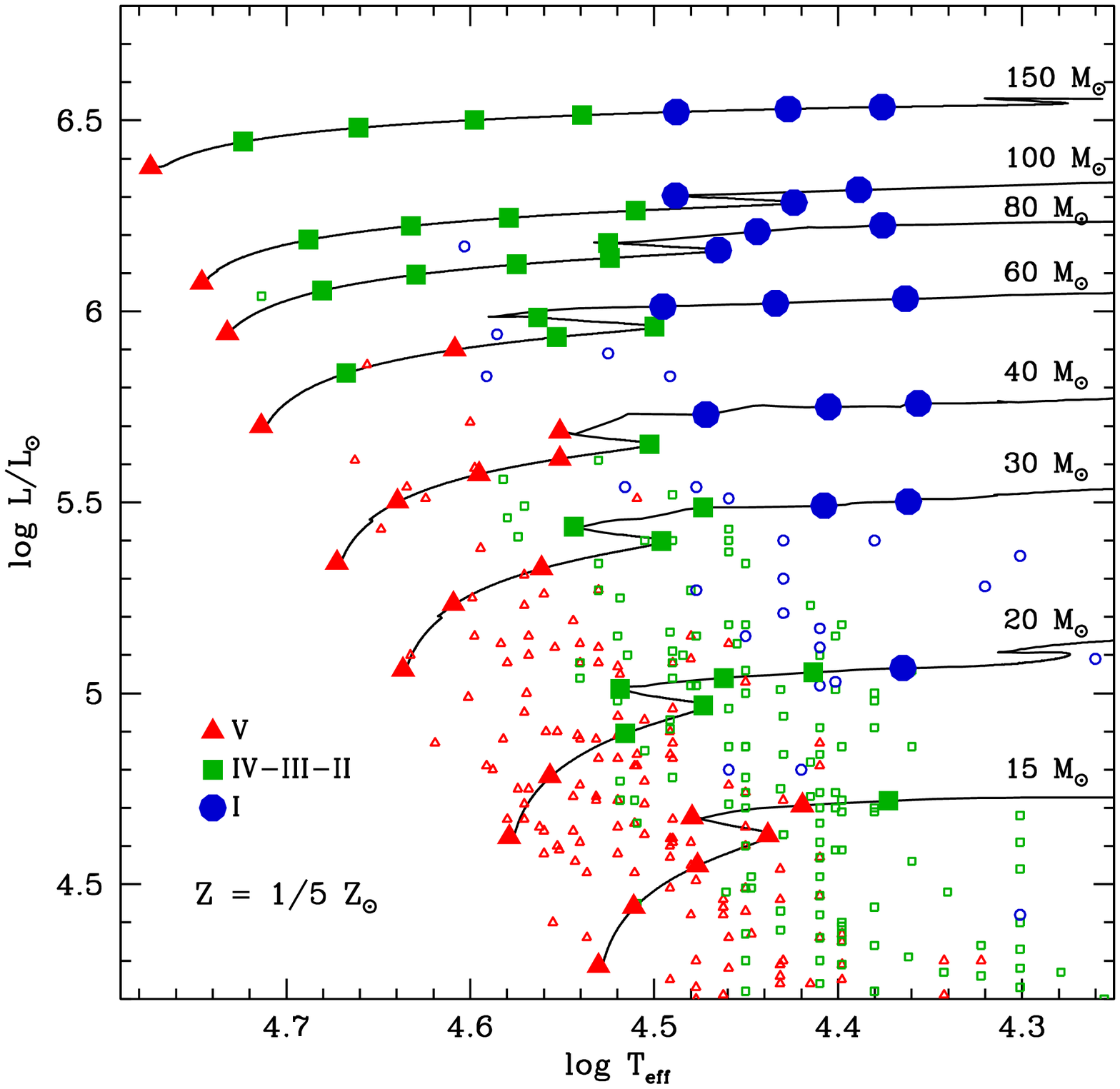}
\includegraphics[width=0.49\textwidth]{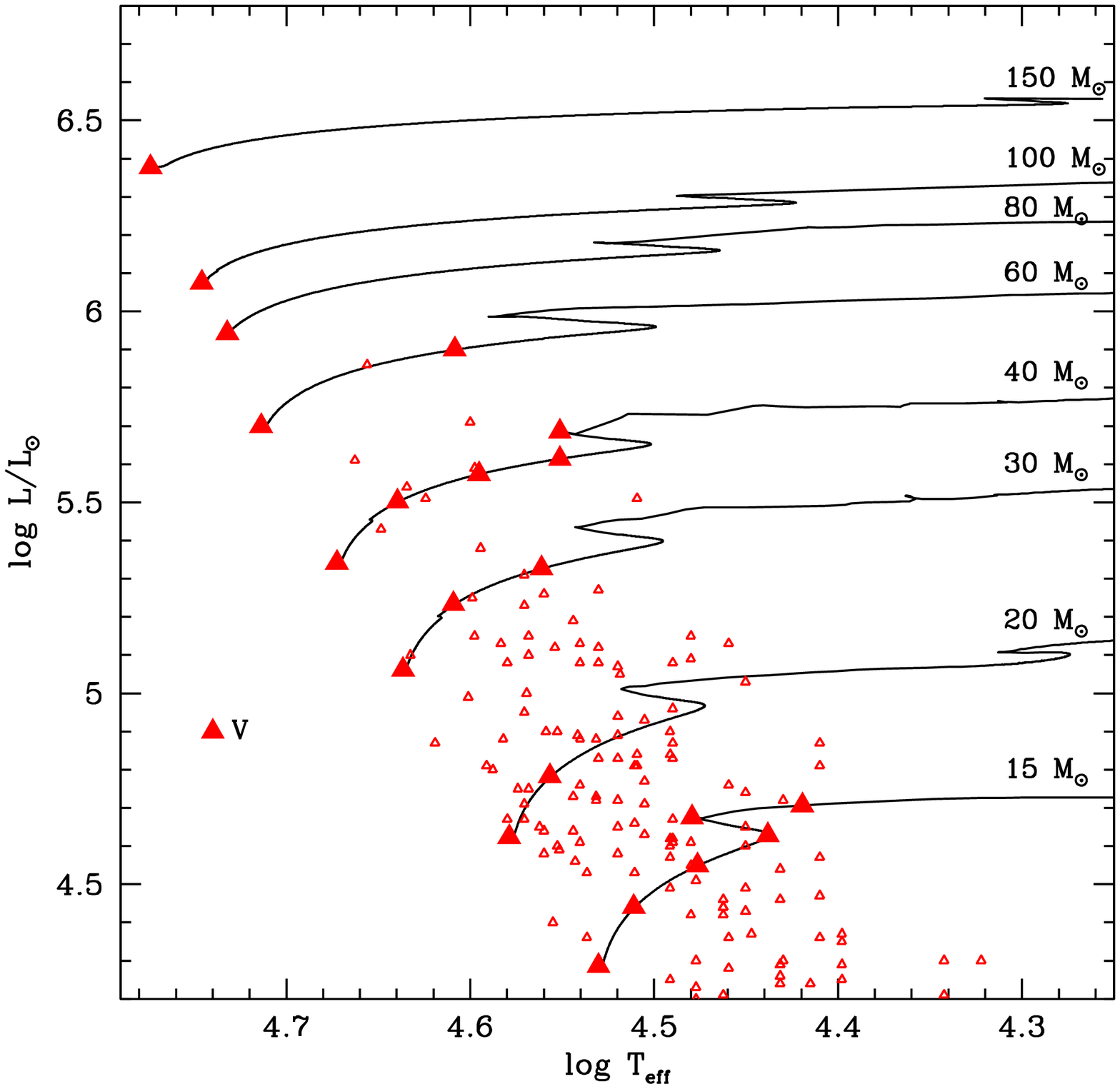}

\includegraphics[width=0.49\textwidth]{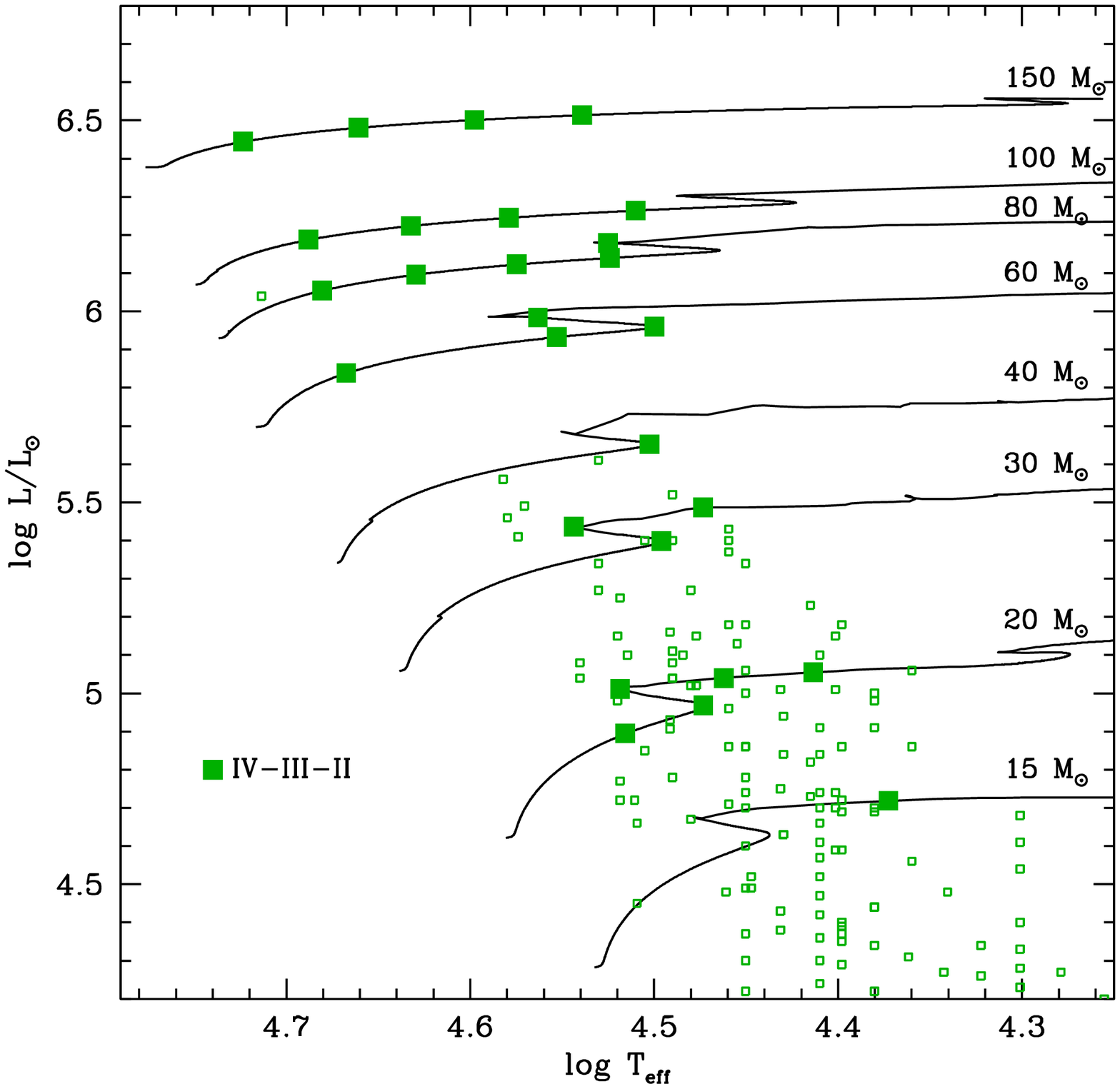}
\includegraphics[width=0.49\textwidth]{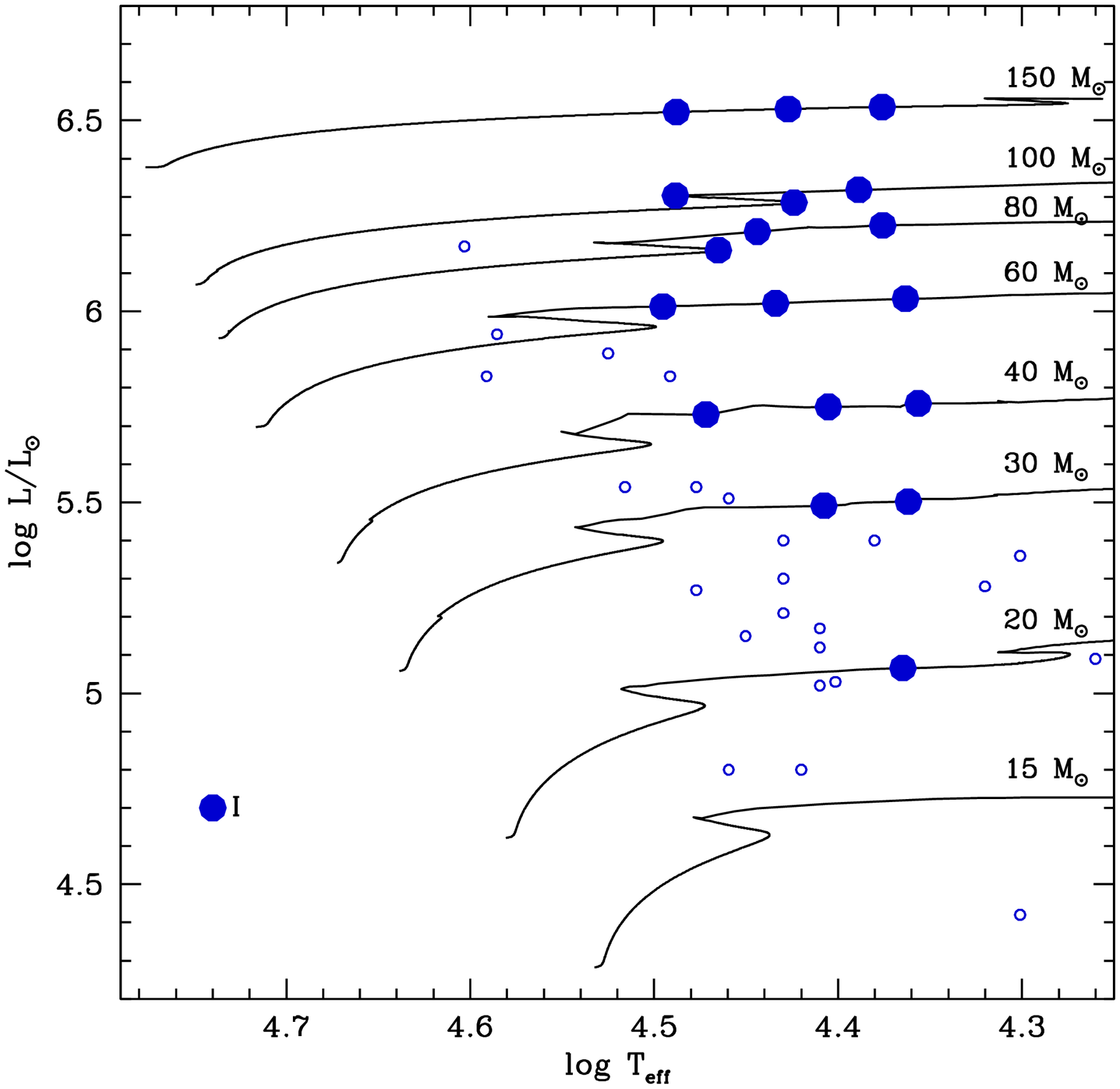}
\caption{Hertzsprung-Russell diagrams at Z = 1/5 Z$_\odot$ with the various luminosity classes indicated by symbols and colors. The top left panel shows all luminosity classes, while the top right (respectively bottom left and bottom right) panel focuses on dwarfs (respectively on giants and supergiants). Small open symbols are SMC stars analyzed by \citet{mokiem06}, \citet{jc13}, \citet{castro18}, \citet{rama19} and Bouret et al.\ (in prep.)}
\label{figHRLC}
\end{figure*}
 
In Fig.\ \ref{figHRLC} we show the distribution of luminosity classes in the HR diagram.
This distribution at the metallicity of the SMC is different from that obtained at solar metallicity \citep{mp17}. One of the key predictions of the solar case is that (super)giants may be found early on the MS. For instance, the 100 \msun\ track at solar metallicity is populated only by supergiants \citep[see][]{mp17}. For the SMC, giants and supergiants appear later in the evolution. This is simply understood as an effect of metallicity on stellar winds. As discussed by \citet{mp17}, most luminosity class diagnostics are sensitive to wind density. As mass-loss rates and terminal velocities are metallicity dependent \citep{leitherer92,vink01,mokiem07}, being weaker at lower Z, a supergiant classification is reached only for later evolutionary phases, where winds are stronger. In other words, two stars with the same effective temperature and luminosity but different metallicities will have the same position in the HR diagram but will have different luminosity classes. 
For similar reasons, \citet{mp17} showed that O2V stars were not encountered at solar metallicity, as confirmed by observations. For a star to have an O2 spectral type it needs to have a high effective temperature, above 45000 K. This is feasible for massive and luminous stars only. In the Galaxy, at high luminosities the winds are strong enough to impact the main luminosity class diagnostic line (\ion{He}{ii}~4686). Consequently, all O2 stars are either giants or supergiants. At the reduced metallicity of the SMC, \ion{He}{ii}~4686 is less filled with wind emission and a dwarf classification is possible. From Table \ref{tab_smc}, we see that O2V objects are found in the early MS of the 150 \msun\ track, and possibly also of the 80 and 100~\msun\ tracks (here the ZAMS models are classified O2-3V). The O2V classification is confined to the most massive stars but is not unexpected.  

In Fig.\ \ref{figHRLC} we also compare our predictions to the position of observed SMC stars.
According to our predictions, dwarfs cover most of the MS range for masses up to 40 \msun. Above that mass, giants appear soon after the ZAMS and are found over a large fraction of the MS. The observed distribution of dwarfs is relatively well accounted for by our models (see top right panel of Fig.\ \ref{figHRLC}). We note that there is a significant overlap between observed dwarfs and giants making a more quantitative comparison difficult. For instance, both luminosity classes are encountered near the terminal-age main sequence (TAMS) of the 20~\msun\ track. The three 20~\msun\ models immediately before, at, and immediately after the TAMS have luminosity classes IV, III-I, and IV, respectively (see Table~\ref{tab_smc}). This is globally consistent with observations.

We predict supergiants only at or after the TAMS, except for the 150~\msun\ track where they appear in the second part of the MS. Observations indicate that supergiants populate a hotter region of the HRD on average. This mismatch may be due to incorrect mass-loss rates in our computations that would produce weaker wind-sensitive lines (see below). If real, this phenomenon should also affect the position of giants (our predictions should be located to the right of the observed giants). Given the overlap between dwarfs and giants described above, we are not able to see if the effect is present. In our models, we introduce a mass-loss reduction by a factor of three due to clumping, which is a standard value for Galactic stars \citep{cohen14}. At the metallicity of the SMC, one may wonder whether this factor is the same. If it was smaller, wind-sensitive lines, which mostly scale with $\dot{M}/\sqrt{f}$ where $f$ is the clumping factor\footnote{We stress that theoretical predictions of mass-loss rates based on the calculation of radiative driving may not depend on the clumping factor ---and therefore on its potential metallicity dependence---  since clumping is usually small at the base of the atmosphere where most of the driving takes place \citep{sander20}.} would be slightly stronger than in our models. \citet{march07} concluded that there is no metallicity dependence of the clumping properties but their conclusion is based on a small sample of Wolf-Rayet stars. In addition these objects have winds that do not behave exactly as those of OB stars \citep[e.g.,][]{sander17}.
Finally, rotation, which is not included in our evolutionary models, could slightly strengthen winds and affect luminosity class determination. However, supergiants usually rotate slowly and this effect should be negligible.

As highlighted by \citet{rama19}, whose data are included in Fig.\ \ref{figHRLC}, there seems to be a quasi absence of observed stars above 40 \msun.
\citet{castro18} indicated that SMC stars were not observed above $\sim$40 \msun\ in the classical Hertzsprung-Russel diagram, but were found in the spectroscopic HR diagram \citep{lp14}, a modified diagram where the luminosity $L$ is replaced by $\frac{T_{eff}^4}{g}$ where $g$ is the surface gravity. \citet{castro18} attributed this difference to the so-called mass discrepancy problem, namely that masses determined from the HR diagram are different from those obtained from surface gravity \citep{herrero92,markova18}. \citet{dufton19} focused on NGC346 in the SMC and again found no stars more massive than 40 \msun\ in their HRD. The absence of the most massive OB stars in the SMC therefore seems to be confirmed by several independent studies relying on different atmosphere and evolutionary models. Since the distance to the SMC is well constrained, luminosities should be safely determined as well. \citet{rama19}  concluded that stellar evolution above 40 \msun\ in the SMC must be different from what is predicted at higher metallicity. These latter authors argued that quasi-chemically homogeneous evolution may be at work. This peculiar evolution is expected for fast-rotating stars \citep{maeder87,yoon06}: due to strong mixing, the opacity is reduced and the effective temperature increases along the evolution, instead of decreasing for normal MS stars. Consequently, stars evolve to the left part of the HRD, directly after the ZAMS. There is indeed evidence that at least some stars in the SMC follow this path \citep{martins09}. These objects are classified as early WNh stars; their effective temperatures are high and their chemical composition is closer to that of OB stars than to that of evolved Wolf-Rayet stars. These properties are consistent with quasi-chemically homogeneous evolution. \citet{jc03} and \citet{jc13} also suggested such evolution for the giant MPG~355. This giant is reported in the bottom left panel of Fig.~\ref{figHRLC}  as the open green square just right of the ZAMS at \lL\ $\sim$ 6.0. Its high nitrogen content and its peculiar position may be consistent with quasi-homogeneous evolution, although the measured \vsini\ remains modest \citep[120~\kms, see also][]{rg12}. Whether or not stellar evolution above $\sim$40~\msun\ follows a peculiar path in the SMC  is not established, but our study indicates that this possibility should be further investigated. A final, alternative possibility to explain the lack of stars more massive than about 60~\msun\ at SMC metallicity is a different star formation process or at least different star formation conditions compared to solar metallicity environments.

\begin{figure}[]
\centering
\includegraphics[width=0.49\textwidth]{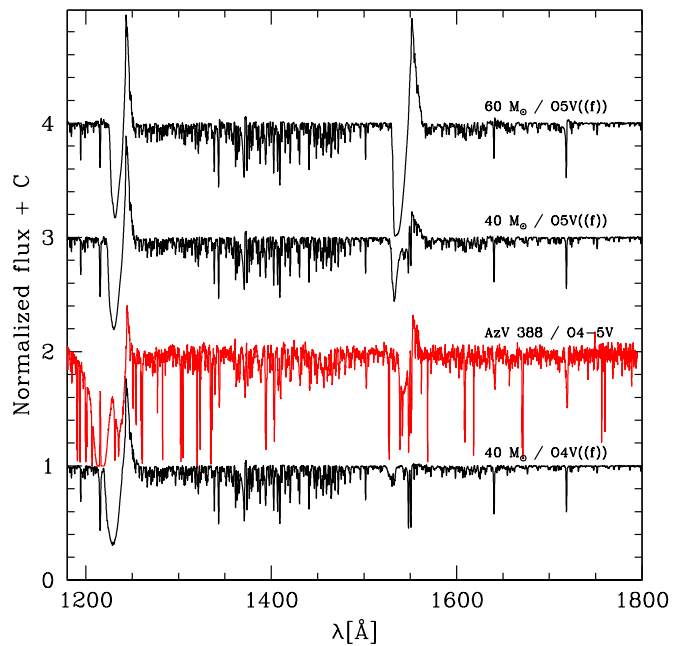}
\caption{Ultra-violet spectra of the models of the Z~=~1/5~\zsun\ series for which a spectral type O4V((f)) or O5V((f)) was attributed (see Table~\ref{tab_smc}). The HST/COS spectrum of the O4-5V star AzV~388 in the SMC, from \citet{jc13}, is inserted in red. The main lines are indicated. The initial masses of the models are also given.}
\label{fig_O4V}
\end{figure}

We conclude this section by commenting in general on the behavior of optical and UV spectra. In Fig.\ \ref{fig_O4V} we illustrate that stars displaying similar helium lines in the optical, and therefore of similar spectral type, can have different UV spectra. Here, we focus on the models of the Z~=~1/5~\zsun\ grid that have been classified as O4V((f)) or O5V((f)). These correspond to stars with initial masses ranging from 40 to 60 \msun. We see that despite having similar spectral types, the strength of the wind features increases with initial mass. More massive stars are also more luminous and, as mass-loss rates are sensitive to luminosity \citep[e.g.,][]{bjorklund20}, this translates into stronger P-Cygni features. However the winds are not strong enough to cause \ion{He}{ii}~4686 to enter the regime of giants or supergiants and the models remain classified as dwarfs. For a given mass, the luminosity effect is also observed as the star evolves off the ZAMS: the \ion{C}{iv}~1550 line is stronger in the 40~\msun\ model classified as O5V((f)), which is also more evolved and more luminous than the 40~\msun\ model classified as O4V((f)) (see Table \ref{tab_smc}).\\
For comparison, and as a sanity check, we added the spectrum of the SMC star AzV~388 (O4-5V) in Fig.\ \ref{fig_O4V}. The goal is not to provide a fit of the observed spectrum but to assess whether or not our models are broadly consistent with typical features observed in the SMC. The two strongest lines of AzV~388 (\ion{N}{v}~1240 and \ion{C}{iv}~1550) have intensities comparable to our 40~\msun\ model classified O5V((f)). \citet{jc13} determined \teff\ = 43100~K and \lL\ = 5.54 for AzV~388. These properties are very similar to those of our O5V((f)) model (\teff\ = 43614~K and \lL\ = 5.50, see Table~\ref{tab_smc}). The morphology of UV spectra we predict is therefore broadly consistent with what is observed in the SMC. The larger \vinf\ in our model (3496 \kms\ versus 2100 \kms\ for AzV~388  according to \citealt{jc13}) explains the larger blueward extension of the P-Cygni profiles in our model.

\subsection{One-thirtieth$^{\rm{}}$ solar metallicity}
\label{s_1on30}

We now turn to the Z = 1/30 \zsun\ grid. Before discussing the predicted spectroscopic sequences, we first look at how our evolutionary tracks compare with other computations in this recently explored metallicity regime.

\subsubsection{Comparison of evolutionary models}
\label{s_1on30_mod}

\begin{figure}[]
\centering
\includegraphics[width=0.49\textwidth]{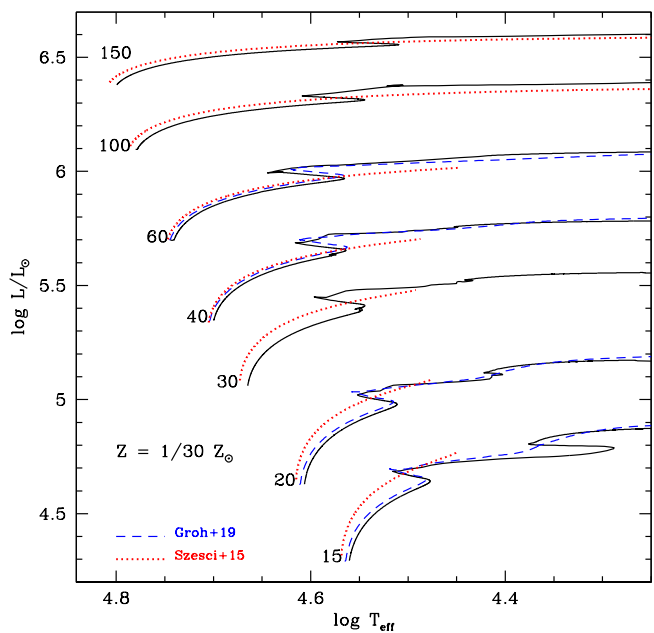}
\caption{Comparison of evolutionary models without rotation at Z~=~1/30~\zsun\ (our calculations in black solid lines), Z~=~1/35 \zsun\ from \citet{groh19}  (blue dashed lines), and Z~=~1/50~\zsun\ \citet{szecsi15}  (red dotted lines). Initial masses are indicated at the beginning of each evolutionary sequence. For the \citet{szecsi15} 40 and 60~\msun\ tracks the actual values of the initial masses are respectively 39 and 59~\msun.}
\label{comptracks1on30}
\end{figure}

In this section we consider the tracks of \citet{szecsi15} and \citet{groh19} which assume Z = 1/50 \zsun\ and Z = 1/35 \zsun\ respectively. The comparisons for tracks with similar masses are shown in Fig.\ \ref{comptracks1on30}. In general, we find good agreement between all predictions. The Groh et al. and Sz\'esci et al. tracks start at slightly higher \teff\ than our models. This is easily explained by the metallicity differences, with stars with less metals having higher \teff. The tracks by \citet{groh19} are on average 0.02-0.03 dex more luminous. Additionally, they have very similar shapes to our tracks, especially near the TAMS. The tracks by \citet{szecsi15} are 0.03 to 0.12 dex more luminous than ours. The difference is larger (smaller) for lower (higher) initial masses and is mainly attributed to the smaller metallicity.
Due to the very large core overshooting they adopt in their models (more than three times as large as the one adopted in this work and in \citealt{groh19}) as commented in their paper, the non-rotating models by \cite{szecsi15} reach the TAMS at much lower effective temperature than our models. This can be seen on their \citep{szecsi15} low mass tracks, which are interrupted before they reach the short contracting phase translated into a hook at the TAMS of classical models. For the 100~\msun\ and 150~\msun\ models, the Szecsi et al. models become underluminous compared to ours near $\log$ $T_{\rm{eff}}$=4.55, because they are still undergoing core H burning while our models have switched to core He burning and have undergone thermal readjustment at the end of core H burning.\\
Let us finally note the hooks in our 15 and 20 \msun\ tracks below $\log T_{\rm{eff}} \approx 4.4$. These correspond to the onset of core helium burning and are a known feature of models with very moderate overshooting and a core convection defined by the Schwarzschild criterion \citep{soh1959,iben1966,kww2012}.

\subsubsection{Spectroscopic sequences at  Z = 1/30 \zsun}
\label{s_1on30_spec}

The right panel of Fig.\ \ref{fig_hrST} as well as Table \ref{tab_1on30} reveal that, above 40~\msun, stars at Z = 1/30 \zsun\ spend almost the entire MS as O2 to O6 stars, with a significant fraction of the MS spent in the earliest spectral types (i.e., <O4.5). We predict that 100 and 150~\msun\ stars spend a non-negligible part of their evolution as O2 stars. We therefore expect a large fraction of early-type O stars in young massive clusters in this metallicity range. For comparison, NGC~3603, one of the youngest and most massive cluster in the Galaxy, has fifteen O3-O4 stars but no O2 star \citep{melena08}. The reason for this is the higher effective temperature of lower metallicity stars \citep[e.g.,][]{mokiem07a}, and their corresponding earlier spectral types.
In our spectroscopic sequences at Z~=~1/30~\zsun\ most of the O2 stars are dwarfs.

\citet{kub19} calculated theoretical spectra of metal-poor stars (Z = 1/50 \zsun) following quasi chemically homogeneous evolution. This type of evolution is different from the one followed in our computations, because it requires that rotational mixing be taken into account. However, the ZAMS models of \citet{kub19} can be compared to our results. Kub\'atov\'a et al.\ assign a spectral type O8.5-O9.5V to their ZAMS 20 \msun\ model, for which \teff\ = 38018 K, \lL\ = 4.68, and \logg~=~4.35.
These parameters are close to our 20 \msun\ ZAMS model (see Table\ \ref{tab_1on30}) which we classify as O7.5V((f)). The slightly larger \teff\ and \logg\ in our model easily explain the small difference in spectral type. The 60 \msun\ ZAMS model of Kub\'atov\'a et al.\ has  \teff\ = 54954 K, \lL\ = 5.75, and \logg\ = 4.39, again very similar to our corresponding 60~\msun\ model. We assign a O3-3.5V((f)) classification to our model, while \citet{kub19} prefer $<$O4III. We therefore agree on the spectral type but find a different luminosity class. The latter is based in the strength of \ion{He}{ii}~4686. As we use a mass-loss rate that is about 0.5 dex smaller than  that used by Kub\'atov\'a et al., we naturally predict a weaker \ion{He}{ii}~4686 emission, which explains the different luminosity classes. The global spectral classification between both sets of models is therefore relatively consistent, considering that different metallicities are used (1/30 \zsun\ for us, 1/50 \zsun\ for Kub\'atov\'a et al.).  

\smallskip

\begin{figure}[]
\centering
\includegraphics[width=0.49\textwidth]{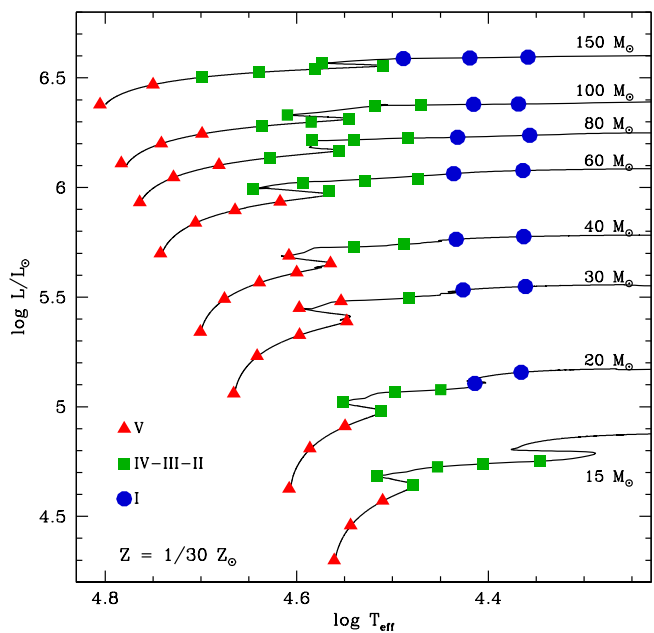}
\caption{Hertzsprung-Russell diagram at Z = 1/30 \zsun\ with the various luminosity classes indicated by symbols and colors.}
\label{figHRLC1on30}
\end{figure}

The distribution of luminosity classes in our predicted spectra is shown in Fig.\ \ref{figHRLC1on30}. Compared to the Galactic case \citep[see][]{mp17}, the match between MS and luminosity class V is almost perfect up to M$\sim$60~\msun. Giants populate an increasingly large fraction of the MS at higher masses. At a metallicity of 1/30 \zsun, and below 60 \msun, a dwarf luminosity class is therefore quasi-equivalent to a MS evolutionary status. For M=15~\msun\ we do not predict supergiants even in the early phases of the post-MS evolution that we cover (they may appear later on, at lower \teff). In our computations, supergiants are seen only in the post-MS phase of stars more massive than 20~\msun.

There is so far only one O star detected in a Z = 1/30 \zsun\ galaxy \citep[Leo~P,][]{evans19}. There are a few hot massive stars detected in Local Group galaxies with metallicities between that of the SMC and 1/10 \zsun\ \citep{bresolin07,evans07,garcia13,hosek14,tramper14,camacho16,garcia18,garcia19}. An emblematic galaxy in the low-metallicity range is I~Zw~18 \citep[Z$\sim$1/30-1/50 \zsun,][]{izotov99} in which \citet{izotov97} reported the detection of Wolf-Rayet stars \citep[see also][]{brown02}. No OB star has yet been observed in I~Zw~18 in spite of strong nebular \ion{He}{ii}~4686 emission \citep{kehrig15} which is difficult to reproduce with standard stellar sources \citep[e.g.,][]{schaerer19}. Comparison of the distribution of spectral types and luminosity classes at Z = 1/30 \zsun\ is therefore not feasible at present. \citet{garcia17} showed in their Fig.~2 a HR diagram for stars in Local Group galaxies with Z$\sim$1/5-1/10 \zsun. The most massive objects are O stars with masses $\sim$60 \msun. The absence of more massive stars that, according to our predictions, should appear as early-O type stars, may be an observational bias. Alternatively, this absence may also extend the results obtained in the SMC: the most massive OB stars may be absent in these low-metallicity environments, for a reason that remains unknown.

The right panels of Figs.\ \ref{fig_sv60opt} and \ref{fig_sv60uv} show the spectroscopic sequences of the 60 \msun\ models at Z = 1/30 \zsun\ (see Figs.~\ref{fig_sv20opt} and \ref{fig_sv20uv} for the 20~\msun\ models). As previously noted, the line strengths in the UV range are strongly reduced compared to the SMC case. Most lines usually showing a P-Cygni profile in OB stars are almost entirely in absorption. For M=60~\msun\ only \ion{N}{v}~1240 and \ion{C}{iv}~1550 show a P-Cygni profile in early/mid O dwarfs and late-O/early-B supergiants, respectively. A similar behavior is observed for the most massive star we study (M=150~\msun, see Fig.\ \ref{fig_sv150uv}). For M=20~\msun, \ion{C}{iv}~1550 is the only line developing into a weak P-Cygni profile. According to the scaling of mass-loss rates with metallicity \citep[$\dot{M} \propto\ Z^{0.7-0.8}$, see][]{vink01,mokiem07}, these rates should be approximately three to four times lower at Z = 1/30 \zsun\ than at SMC metallicity (1/5 \zsun) and about 15 times lower than in the Galaxy. \citet{jc15} and \citet{garcia17} show HST UV spectra of O stars in IC~1613, WLM, and Sextans A, three Local Group galaxies with metallicities between 1/5 and 1/10 \zsun. In IC~1613 and WLM (Z = 1/5 \zsun), the P-Cygni profiles are weak but still observable; their strength is comparable to that of SMC stars \citep[see Fig. 4 of][]{garcia17}. In the spectrum of the Sextans A O7.5III((f)) star presented by \citet{garcia17}, most wind-sensitive lines are in absorption. Other O stars in Sextans A show the same behavior (M. Garcia, private communication). In view of the lower metallicity of Sextans A (Z = 1/10 \zsun), this is consistent with the expectation of the reduction of mass-loss rates at lower metallicity. 

\subsection{Optical wavelength range of the ELT }
\label{s_elt}

\begin{figure*}[ht]
\centering
\includegraphics[width=0.99\textwidth]{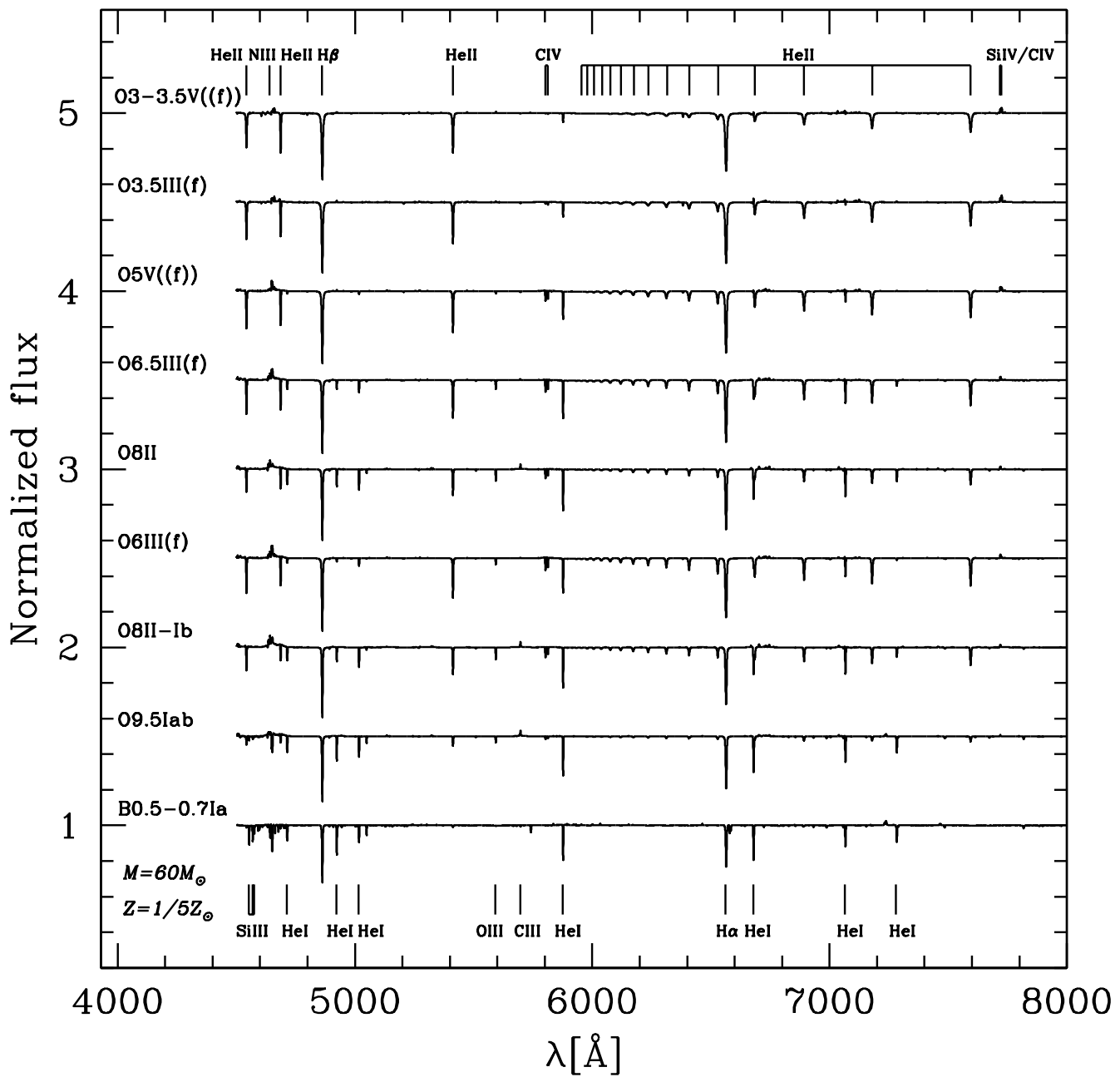}
\caption{Spectra of the sequence of models calculated along the 60 \msun\ track at SMC metallicity in the blue spectral range of ELT/HARMONI and ELT/MOSAIC. The main diagnostic lines are indicated. The spectra have been degraded to a spectral resolution of  approximately 5000, which is typical of the ELT instruments. A rotational velocity of 100 \kms\ has been considered for all spectra.}
\label{fig_sv60_mosaic_smc}
\end{figure*}

\begin{figure*}[ht]
\centering
\includegraphics[width=0.99\textwidth]{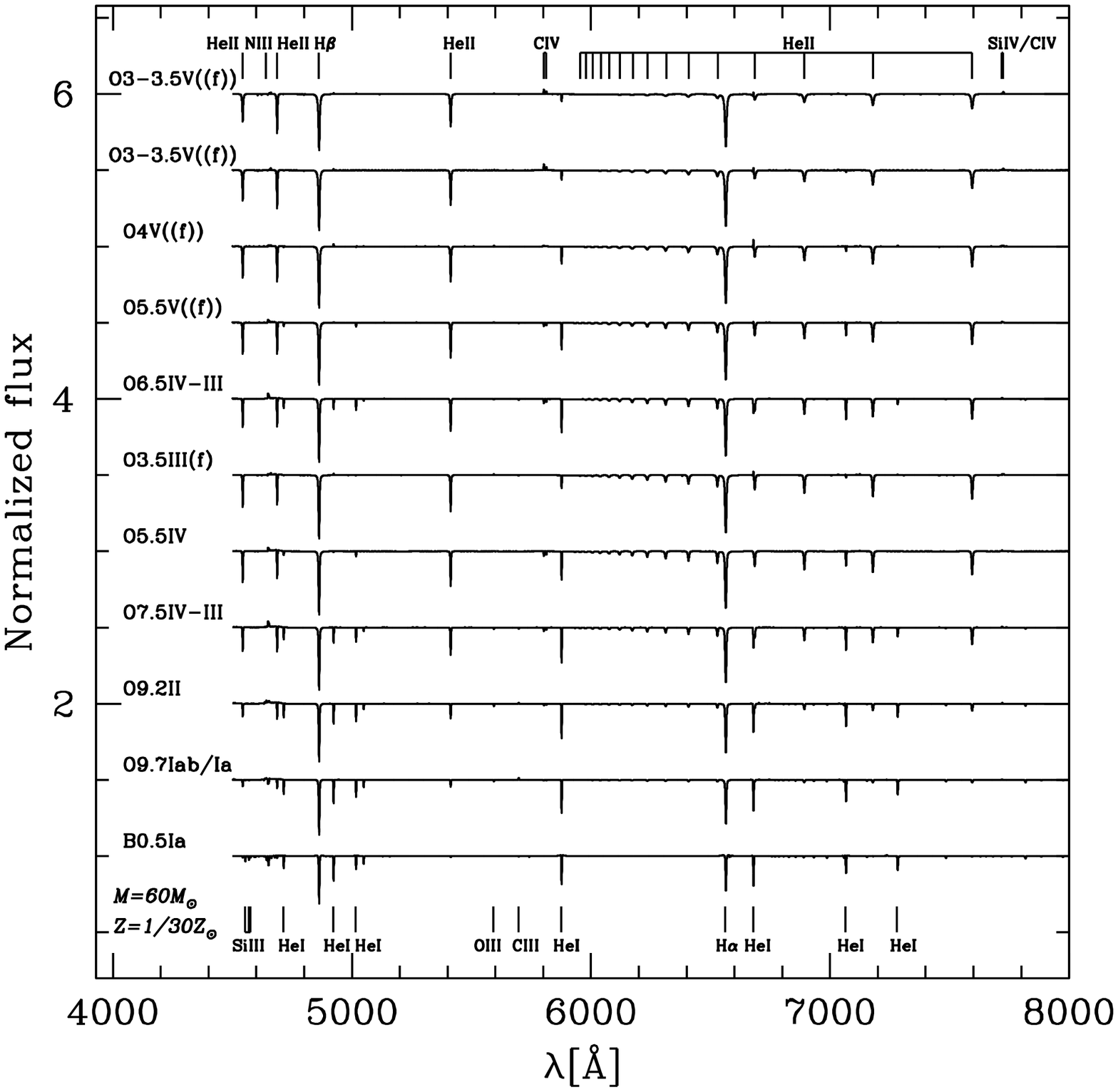}
\caption{Same as Fig.\ \ref{fig_sv60_mosaic_smc} but for the Z=1/30 Z$_{\odot}$ case.}
\label{fig_sv60_mosaic_z1on30}
\end{figure*}

Local Group dwarf galaxies are prime targets to hunt for OB stars beyond the Magellanic Clouds \citep{camacho16,garcia13,evans19}; most of them have low metallicity \citep{mcco05}. Current facilities barely collect low-spectral-resolution and low-signal-to-noise-ratio data for a few of their OB stars. The advent of the new generation of ground-based ELTs assisted with sophisticated adaptive-optics systems will likely lead to a breakthrough in the detection of low-metallicity massive stars. In particular, two instruments planned for the European ELT will have integral-field units or multi-objects spectroscopic capabilities: HARMONI and MOSAIC\footnote{\url{http://www.eso.org/sci/facilities/eelt/instrumentation/index.html}}. These instruments will have resolving power of at least a few thousand and will have a wavelength coverage from $\sim$4500 \AA\ to the K-band. They will therefore not  entirely cover the classical optical range  from which most of the spectroscopic diagnostic lines have been defined \citep{ca71,walborn72,mathys88,sota14,classif}.

In Figs.\ \ref{fig_sv60_mosaic_smc} and \ref{fig_sv60_mosaic_z1on30} we show our predicted spectra for 60 \msun\ stars at SMC and one-thirtieth$^{\rm{}}$ solar metallicities. We focus on the wavelength range 4500-8000 \AA\ which will be probed by HARMONI and MOSAIC. We selected this range for the following reasons: it contains a fair number of lines from different elements; at these wavelengths, OB stars emit more flux than in the near-infrared; Local Group dwarf galaxies have relatively low extinction \citep{tramper14,garcia19}. We therefore anticipate that it will be more efficient to detect and characterize new OB stars in this wavelength range.

Figures\ \ref{fig_sv60_mosaic_smc} and \ref{fig_sv60_mosaic_z1on30} show that several \ion{He}{I} and \ion{He}{II} lines are present in the selected wavelength range. In particular,  many \ion{He}{ii} lines from the n=5 (ground-state principal quantum number equal to 5) series are visible. The change in ionization when moving from the hottest O stars to B stars is clearly seen. For instance, the \ion{He}{II} lines at 5412 \AA\ and 7595 \AA\ weaken when the \ion{He}{I} lines at 5876 \AA\ and 7065 \AA\ strengthen. Effective temperature determinations based on spectral features observed with ELT instruments should therefore be relatively straightforward provided nebular lines do not produce too much contamination. H$\beta$, a classical indicator of surface gravity \citep{martins11,sergio20}, is also available. At slightly longer wavelengths, between 8000 and 9000 \AA\ (a range that will be covered by HARMONI and MOSAIC but not shown here), the Paschen series offers numerous hydrogen lines that are also sensitive to \logg\ \citep{negue10}. Surface gravity will therefore be easily determined from ELT observations. 

The wavelength considered in Figs.~\ref{fig_sv60_mosaic_smc} and \ref{fig_sv60_mosaic_z1on30} contains a few lines from carbon, nitrogen, and oxygen, but less than the bluer part (3800-4500 \AA). The strongest lines are \ion{C}{iv}~5805-5812, \ion{N}{iii}~4640, and \ion{O}{iii}~5592. At longer wavelengths, there are even fewer CNO lines (see Fig.\ref{fig_ir} for the K-band). The determination of CNO abundances of OB stars will therefore be more difficult than in the more classical optical and UV spectra where tens of lines are available \citep[e.g.,][]{mimesO}. \ion{Si}{iv}~7718, which is found next to \ion{C}{iv}~7726, is a relatively strong line in the earliest O stars that may turn useful for metallicity estimates.

H$\alpha$ will be observed by ELT instruments. It is a classical mass-loss-rate indicator because the photospheric component is filled with wind emission \citep{repolust04}. However, below about $10^{-7}$ \myr\ the wind contribution vanishes. Other hydrogen lines from the Paschen and Brackett series are present in the JHK bands, but they are weaker than H$\alpha$. Since mass-loss rate scales with metallicity \citep{vink01,mokiem07} we anticipate that only upper limits on this parameter will be obtained for all but the most luminous and evolved OB stars in low-metallicity environments, unless complementary UV data are acquired.

\begin{figure*}[t]
\centering
\includegraphics[width=0.49\textwidth]{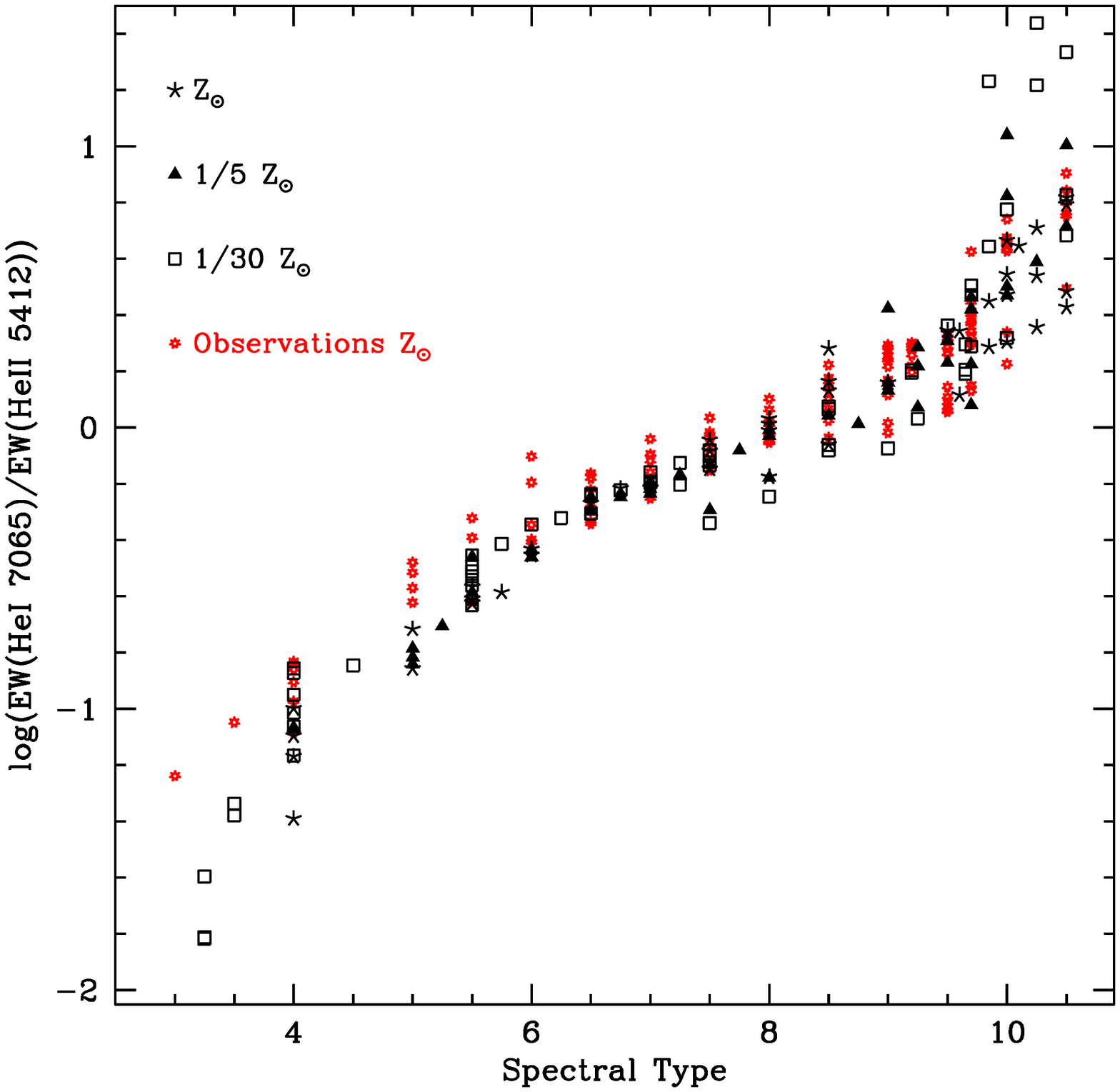}
\includegraphics[width=0.49\textwidth]{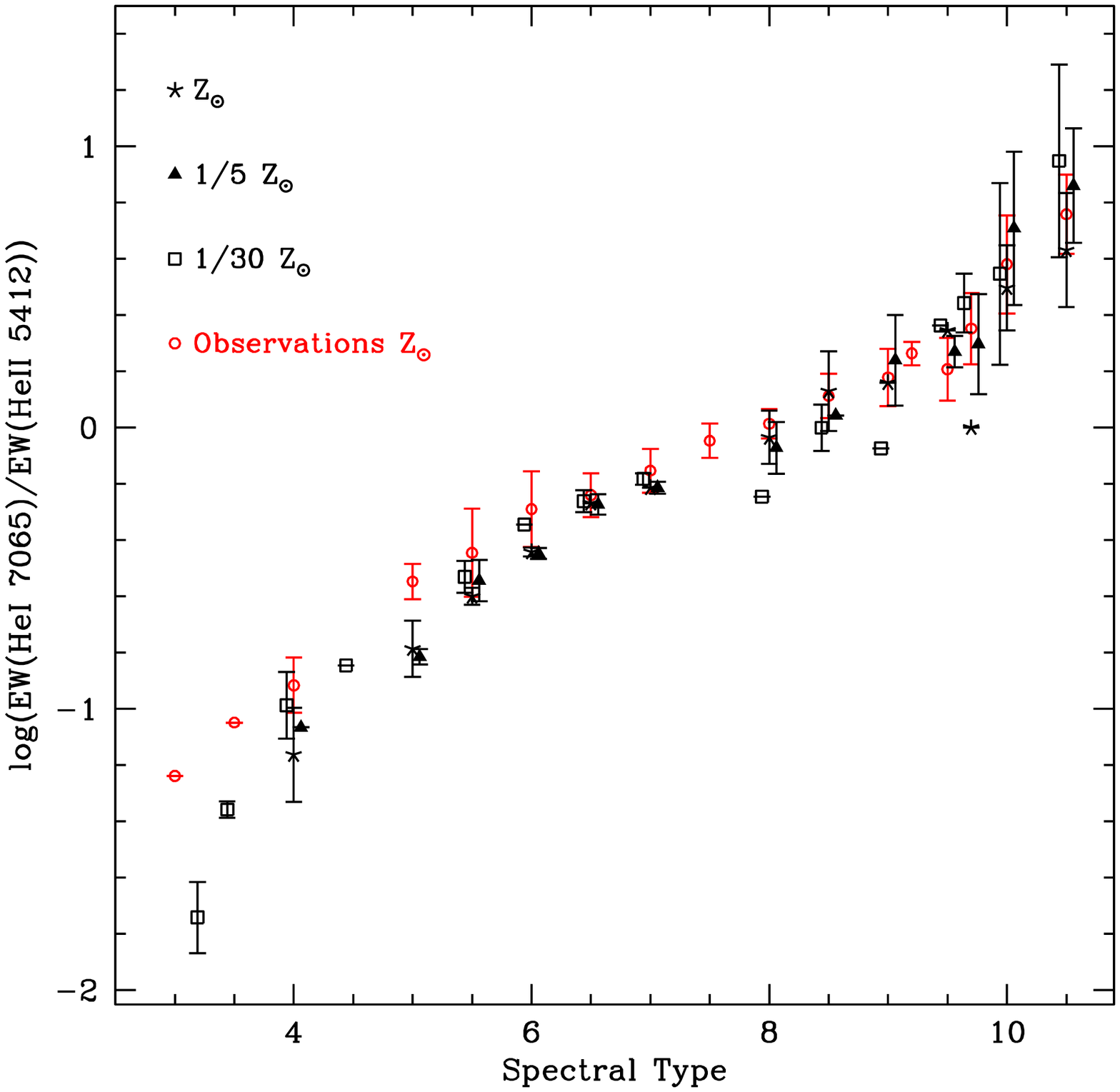}
\caption{Ratio of the EW of \ion{He}{i}~7065 to \ion{He}{ii}~5412 as a function of spectral type for the low-metallicity models calculated in the present study, the solar metallicity models of \citet{mp17} and observations of Galactic stars collected from archives (open red circles - see text for details). \textit{Left panel}: All data points shown. When no unique spectral type was assigned to a model (e.g., O6-6.5) the average was used (i.e., O6.25). Numbers above ten correspond to B stars (with ten being B0, and 10.5 being B0.5). \textit{Right panel}: Same as left panel but showing only the average value of the EW ratio for each spectral type. In that panel the spectral types of the 1/5 \zsun\ (1/30 \zsun) models have been shifted by +0.03 (-0.03) for clarity. We also considered only ``official'' spectral types, that is, we excluded for example 6.25 when a spectral type O6-6.5 was assigned to a model.}
\label{ew_st}
\end{figure*}

Based on the evolution of spectral lines seen in Figs.~\ref{fig_sv60_mosaic_smc} and \ref{fig_sv60_mosaic_z1on30} we have identified a potential criterion for spectral classification in the wavelength range 4500-8000 \AA\ that will be covered by both HARMONI and MOSAIC. Helium lines are the prime diagnostics of spectral types among O stars \citep{ca71,mathys88}. We measured the EW of various \ion{He}{i} and \ion{He}{ii} lines, computed their ratios, and plotted them against the estimated spectral types. We did this for the two sets of models (SMC and one-thirtieth$^{\rm{}}$ solar metallicity). More specifically, we considered \ion{He}{i}~4713, \ion{He}{i}~4920, \ion{He}{i}~5876, \ion{He}{i}~7065, \ion{He}{ii}~4542, and \ion{He}{ii}~5412. We find that the ratio EW(\ion{He}{i}~7065)/EW(\ion{He}{ii}~5412) shows a monotonic and relatively steep evolution through spectral types. In addition, the two lines are not particularly close to the blue part of the spectral range considered, where detectors may be less efficient. We show the trends we obtained in Fig.~\ref{ew_st}. There is no difference among the two metallicities: at a given spectral type, the EW ratios of the two metallicities overlap (see right panel of Fig.~\ref{ew_st}). To further investigate the potential of this indicator, we added our solar metallicity models \citep[from][]{mp17}. Again, the EW ratios are similar to the lower metallicity models. A final check was made by incorporating measurements from Galactic stars: these are the red points in Fig.~\ref{ew_st}. We relied on archival data from CFHT/ESPaDOnS, TBL/NARVAL, and ESO/FEROS. The details of the data are given in Appendix \ref{ap_archive}. The reduced observed spectra were normalized and EWs were measured in the same way as for the model spectra. We see that from spectral types O5.5 to B0.5 the agreement between the observed EW ratios and the model ratios is excellent. We note a small offset at earlier spectral types (O3 to O5). This may be caused by several factors: (1) the small number of observed spectra in that spectral type range; (2) the use of additional criteria ---namely nitrogen lines--- to refine spectral classification, particularly at O3, O3.5 and O4; and (3) the increasing weakness of \ion{He}{i}~7065 in that range and consequently the stronger impact of neighboring \ion{Si}{iv} and \ion{C}{iv} lines, the modeling of which needs to be tested. We also stress that at spectral type O5 a similar offset was observed in the classical EW(\ion{He}{i}~4471)/EW(\ion{He}{ii}~4542) ratio shown in Fig.~1 of \citet{classif}.
In view of these results, we advocate the ratio EW(\ion{He}{i}~7065)/EW(\ion{He}{ii}~5412) as a reliable spectral type criterion in the wavelength range 4500-8000 \AA, especially for spectral types between O5.5 and B0.5. It can be used for classification of O and early-B stars in Local Group galaxies observed with the ELT.

\section{Ionizing properties and \ion{He}{ii}~1640 emission}
\label{s_ionHeII1640}

In this section we describe the ionizing properties of our models and study their dependence on metallicity. We also describe the morphology of \ion{He}{ii}~1640 in our models, a feature that depends on the ionizing power of stars in star-forming galaxies.

\subsection{Ionizing fluxes}
\label{s_ionflux}

Here we first discuss the hydrogen ionizing flux before turning to the helium ionizing fluxes. All ionizing fluxes of our models are given in Tables~\ref{tab_smc} and \ref{tab_1on30}.

\subsubsection{Hydrogen ionizing flux}
\label{s_qH}

\begin{figure}[]
\centering
\includegraphics[width=0.4\textwidth]{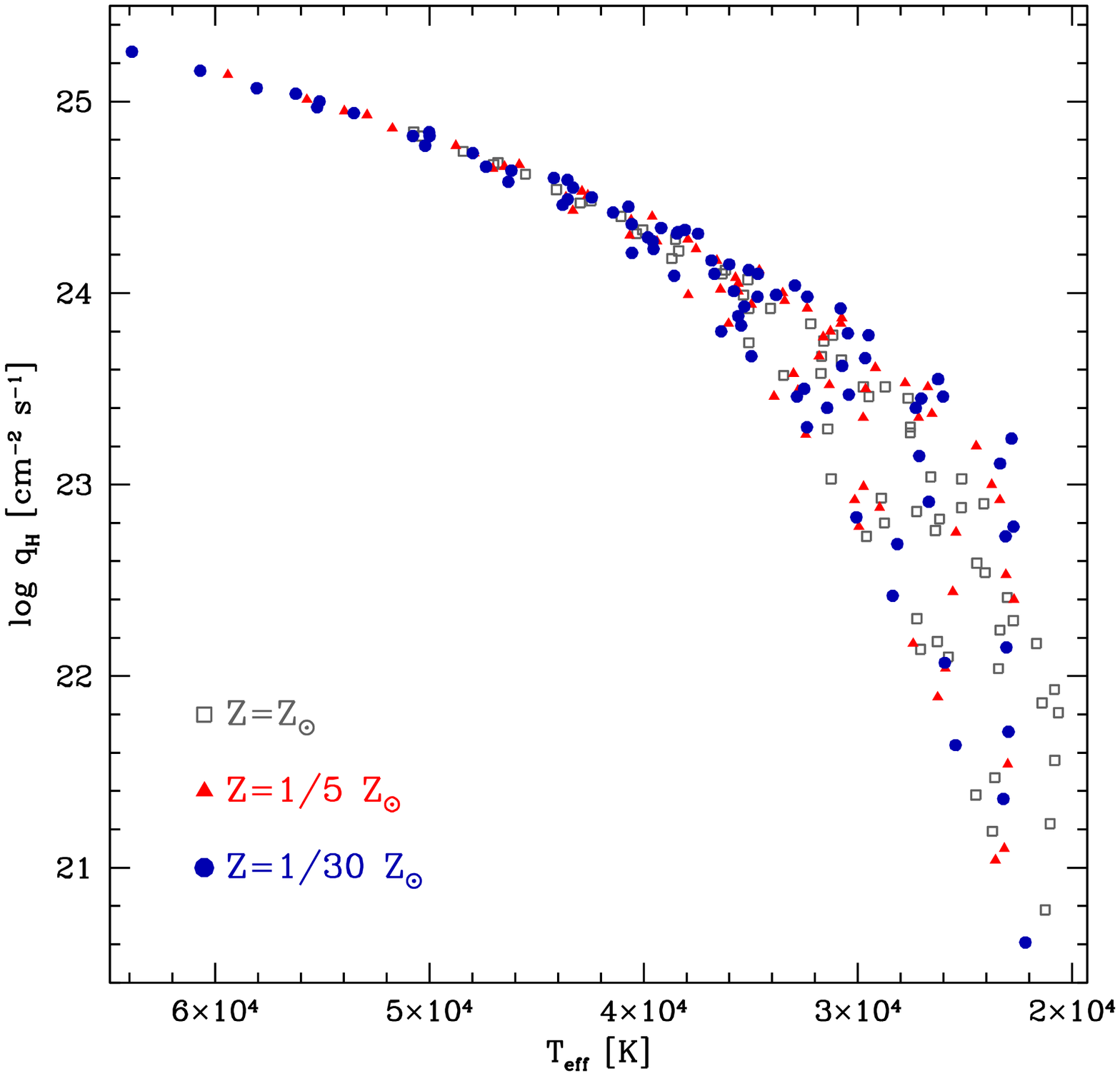}

\includegraphics[width=0.4\textwidth]{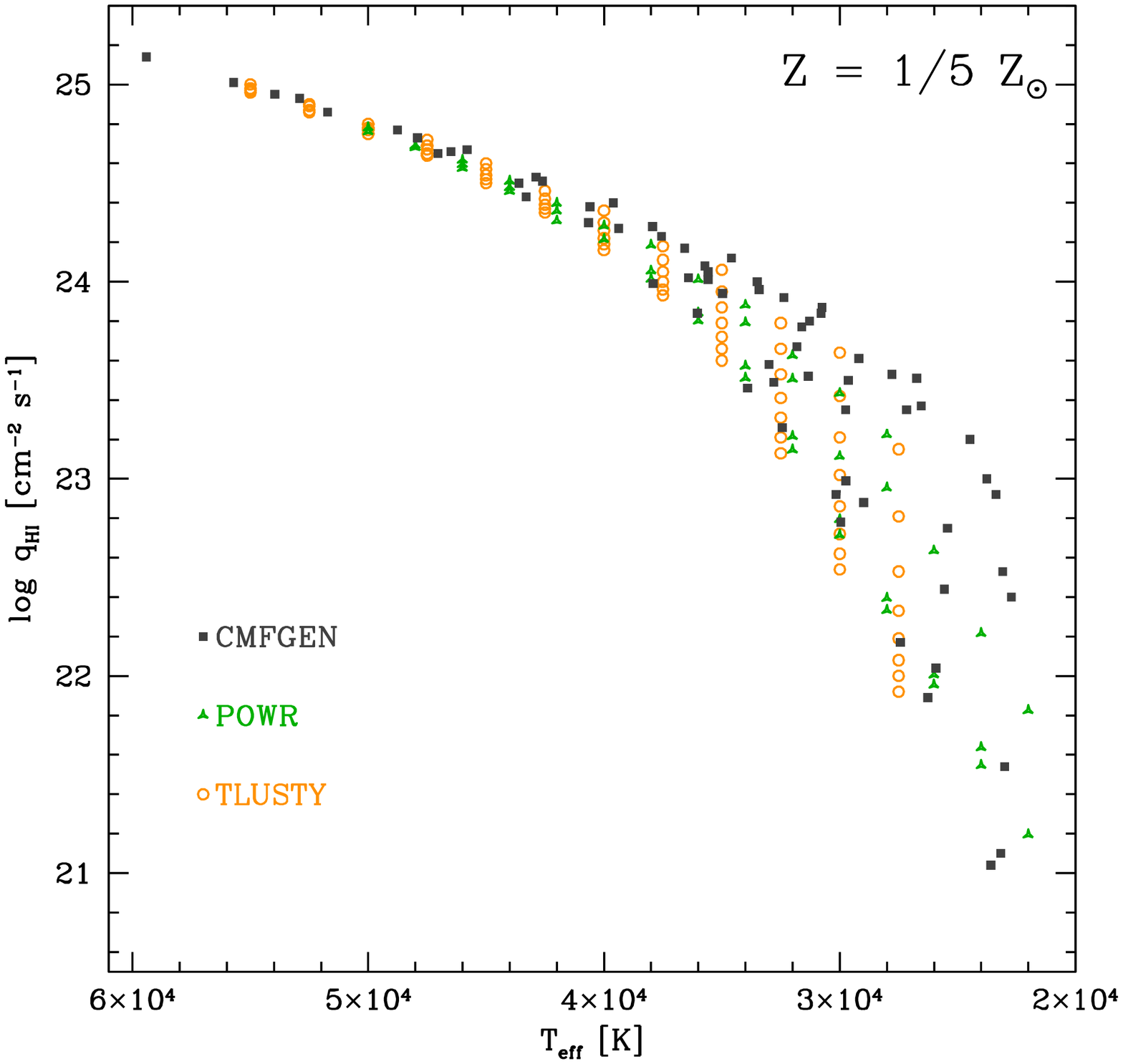}

\includegraphics[width=0.4\textwidth]{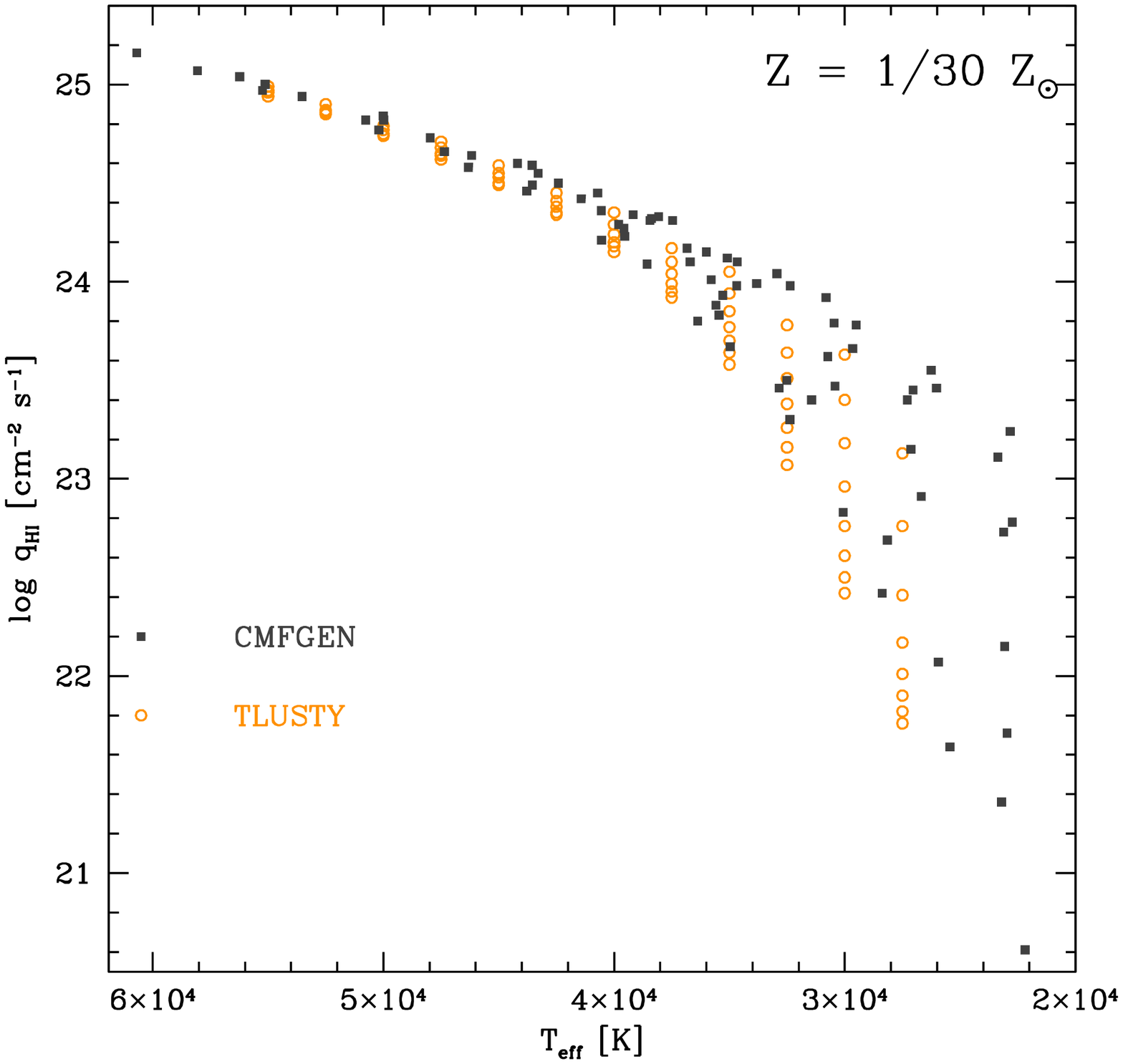}
\caption{\ion{H}{i} ionizing fluxes per unit surface area as a function of effective temperature. The upper panel shows the ionizing fluxes for the two metallicities considered in this work and our solar metallicity models \citep{mp17}. The middle and bottom panel show the 1/5 \zsun\  and 1/30 \zsun\ models, respectively, which are compared to TLUSTY \citep{lh03} and PoWR \citep{hainich19} models.}
\label{fig_qH}
\end{figure}

In Fig.\ \ref{fig_qH} we compare the ionizing fluxes per unit surface area -- $q(\mathrm{H})$\footnote{Where $q(\mathrm{H})=\frac{Q(\mathrm{H})}{4 \pi\ R^2}$with $R$ the stellar radius.} -- for three metallicities: solar, one-fifth$^{\rm{}}$ solar, and one-thirtieth$^{\rm{}}$ solar (see top panel). At the highest \teff\ the relation between $\log q(\mathrm{H})$ and \teff\ is very narrow. When \teff\ decreases, a dispersion in $\log q(\mathrm{H})$ for a given \teff\ appears. This is explained by the effect of surface gravity on SEDs \citep[see detailed physics in][]{ah85} and the wider range of surface gravities covered by cooler models. Indeed, a look at Fig.~\ref{figHR} and Tables~\ref{tab_smc} and \ref{tab_1on30} indicates that the hottest models correspond to MS stars with high surface gravities, while lower \teff\ models can be either MS or post-MS models, with a wide range of \logg. Figure~\ref{fig_qH} does not reveal any strong metallicity dependence of the relation between hydrogen ionizing fluxes (per unit surface area) and effective temperature. At high \teff\ the (small) dispersion of $q(\mathrm{H})$ for a given \teff\ is larger than any variation of $q(\mathrm{H})$ with Z that may exist. At the lowest effective temperatures, the lower limit of the $q(\mathrm{H})$ values is the same for all metallicities. The upper boundary of $q(\mathrm{H})$ is located slightly higher at low Z. We stress that because luminosities are higher at lower Z for a given \teff\ (see Fig.~\ref{figHR}), radii are also larger and consequently $Q(\mathrm{H})$ are higher (for a given \teff). 

\begin{figure}[]
\centering
\includegraphics[width=0.49\textwidth]{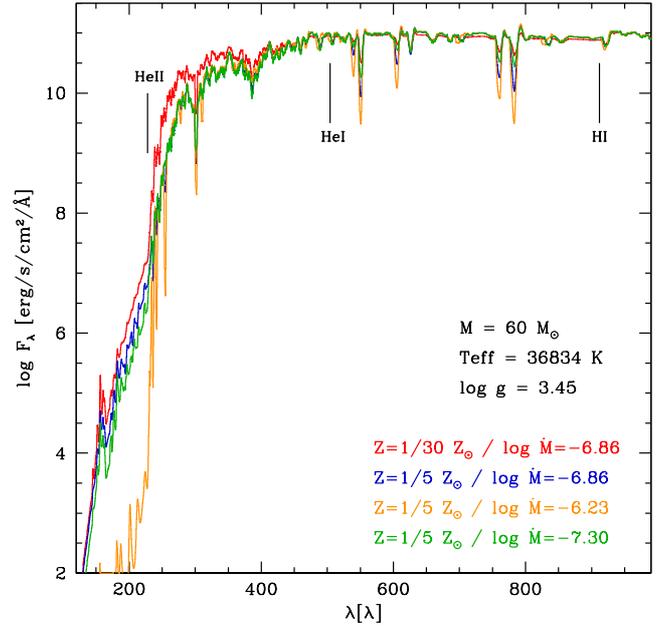}
\caption{Effect of metallicity on the SED. The initial model (red line) is the fifth model of the 60 \msun\ series at Z~=~1/30~\zsun. The blue line shows the same model for which the metallicity has been changed to 1/5~\zsun, all other parameters being kept constant. In the model shown by the orange line, in addition to metallicity, the mass-loss rate has been increased by a factor 4.2 according to \mdot\ $\propto$ $Z^{0.8}$. Finally, in the model shown in green, the mass-loss rate has been reduced down to $\log \dot{M} = -7.30$.
The \ion{H}{i}, \ion{He}{i,} and \ion{He}{ii} ionizing edges are indicated by vertical black lines.}
\label{fig_qHeffectZ}
\end{figure}

Figure~\ref{fig_qHeffectZ} illustrates how the SED changes when the metal content and mass-loss rate are modified, all other parameters being kept constant. As discussed at length by \citet{costar3} the variations in opacity and wind properties affect the SED. An increase of the metal content from 1/30$^{\rm{}}$ \zsun$^{\rm{}}$ to 1/5$^{\rm{}}$ \zsun\ strengthens the absorption due to lines. The consequence is a reduction of the flux where the line density is the highest.  This is particularly visible in Fig.~\ref{fig_qHeffectZ} between 250 and 400 \AA. A stronger opacity also affects the continua, especially the \ion{He}{ii} continuum below 228 \AA. However, in the case illustrated in Fig.~\ref{fig_qHeffectZ}, we also note that the redistribution of the flux from short to long wavelengths (due to increased opacities and to ensure luminosity conservation) takes place mainly below the hydrogen ionizing edge: the flux in the lowest metallicity model is higher (smaller) than the flux in the Z~=~1/5~\zsun\ model below (above) $\sim$550 \AA. But above 912 \AA,\ both models have the same flux level. Consequently, $\log q(\mathrm{H})$ is almost unchanged (24.15 vs. 24.17). Figure~\ref{fig_qHeffectZ} also reveals that variations in mass-loss rate for the model investigated here have little effect on the hydrogen ionizing flux, whereas the \ion{He}{ii} ionizing flux is affected (see following section).

In the middle and bottom panels of Fig.~\ref{fig_qH} we compare our hydrogen ionizing fluxes to the results of \citet{lh03} obtained with the code TLUSTY and \citet{hainich19} obtained with the code PoWR \citep{sander15}. For the latter we used the "moderate" mass-loss grid\footnote{Data have been collected at this address \url{http://www.astro.physik.uni-potsdam.de/PoWR/}} and we checked that the choice of mass-loss rates does not impact the conclusions. At high \teff\ the values of $q(\mathrm{H})$ of the three sets of models are all consistent within the dispersion. At lower \teff\ our predictions have the same lower envelope as \citet{hainich19}, while the plane-parallel models of \citet{lh03} have slightly lower fluxes. Our ionizing fluxes reach higher values than the two other sets of models for a given \teff. These differences are readily explained by the wider range of \logg\ covered by our models. Taking \teff\ $\sim$27000 K as a representative case, the grids of \citet{lh03} and \citet{hainich19} do not include models with \logg\ $<$ 3.0 while we have a few models with \logg\ $\sim$2.7. The models of \citet{lh03} also reach higher \logg\ (up to 4.75) which explains the small difference in the minimum fluxes. The same conclusions are reached at Z~=~1/30~\zsun. 
Different sets of models therefore agree well as far as the hydrogen ionizing fluxes per unit area are concerned. 

\subsubsection{Helium ionizing fluxes}
\label{s_qHe}

In this section we now focus on the ratios of helium to hydrogen ionizing fluxes because they are a common way of quantifying the hardness of a stellar spectrum. It is also a convenient way of investigating the effects of metallicity on stellar SEDs.

\begin{figure*}[]
\centering
\includegraphics[width=0.49\textwidth]{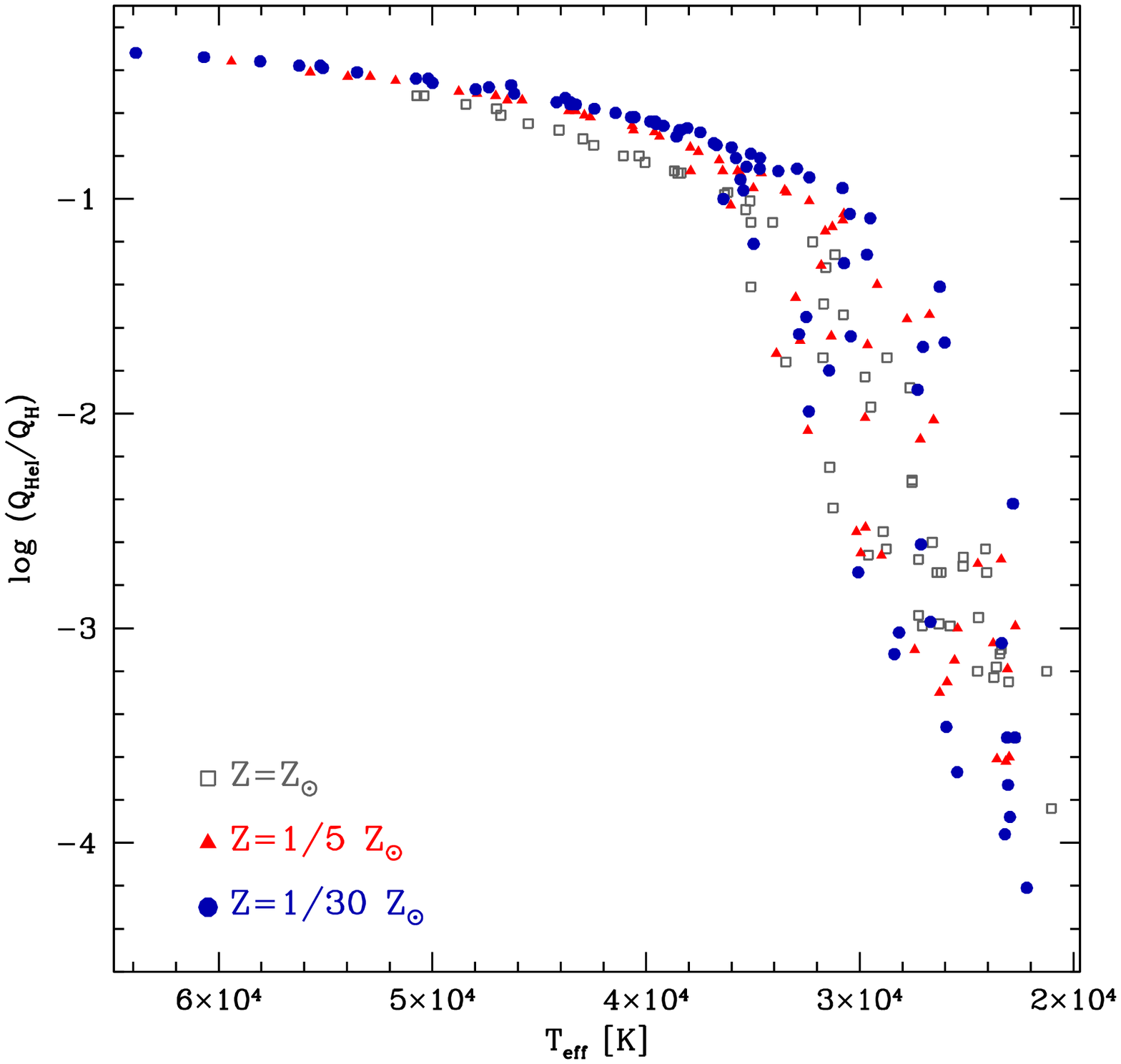}
\includegraphics[width=0.49\textwidth]{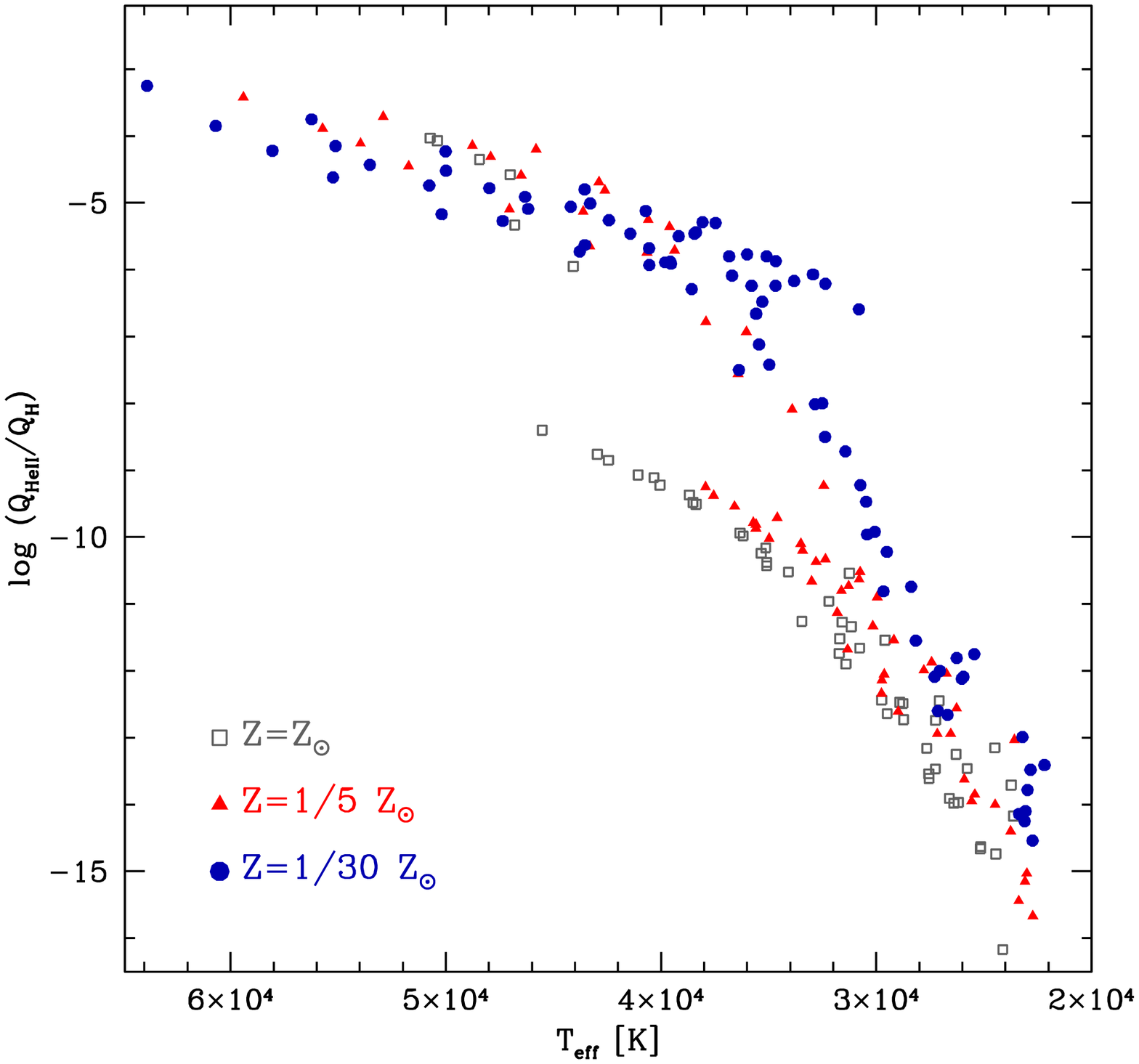}
\caption{Ratio of \ion{He}{i} (left) or \ion{He}{ii} (right) to \ion{H}{i} ionizing fluxes as a function of effective temperature for the two metallicities considered in this work. We have also added the models of \citet{mp17} at solar metallicity.}
\label{fig_ionflux}
\end{figure*}

Figure\ \ref{fig_ionflux} shows the ratios of \ion{He}{i} and \ion{He}{ii} to \ion{H}{I} ionizing fluxes as a function of \teff\ for the two metallicities considered in the present study. We have also added our results for the solar metallicity calculations of \citet{mp17}. The $\frac{Q(HeI)}{Q(\mathrm{H})}$ ratio displays a very well-defined sequence down to $\sim$35000~K  for each metallicity. At lower temperatures, the ratios drop significantly and the dispersion increases mainly because of the strong reduction of the $\ion{He}{i}$ ionizing flux. The general trend of the $\frac{Q(HeII)}{Q(\mathrm{H})}$ is similar: a shallow reduction as \teff\ decreases down to a temperature that depends on the metallicity (see below) followed by a sharp drop. The dispersion at high \teff\ is larger than that of the $\frac{Q(HeI)}{Q(\mathrm{H})}$ ratio. 

This latter ratio shows a weak but clear metallicity dependence at \teff~$>$ 35000~K in the sense that lower metallicity stars have higher ratios. The difference between solar and one-thirtieth solar metallicity reaches  $\sim$0.2 dex at most. For \teff\ $<$ 35000~K, the larger dispersion blurs any metallicity dependence that may exist, although lower metallicity models reach on average higher ratios (the upper envelope of the distribution of Z~=~1/30~\zsun\ points is located above that of Z~=~1/5~\zsun\ and \zsun\ ones). The higher $\frac{Q(HeI)}{Q(\mathrm{H})}$ ratio at lower metallicity is mainly explained by the smaller effects of line blanketing when the metal content is smaller. With reduced line opacities, and since in OB stars most lines are found in the (extreme-)UV part of the spectrum, there is less redistribution of flux from short to long wavelength \citep[e.g.,][]{martins02}. This effect is seen in Fig.~\ref{fig_qHeffectZ} between 250 and 400 \AA\ as explained before.\\

The metallicity dependence of the $\frac{Q(HeII)}{Q(\mathrm{H})}$ ratio is of a different nature. At high effective temperatures, the three sets of models have about the same ratios for a given \teff, given the rather large dispersion. At low temperatures, more metal-poor models produce higher $\frac{Q(HeII)}{Q(\mathrm{H})}$ ratios. At intermediate temperature, the difference between the three metallicities considered is best explained by a displacement of the \teff\ at which the $\frac{Q(HeII)}{Q(\mathrm{H})}$ ratio drops significantly. This "threshold \teff" as we refer to it in the following is located at about 45000~K for solar metallicity models, $\sim$35000~K for Z~=~1/5~\zsun\ , and $\sim$31000~K at Z~=~1/30~\zsun. We return to an explanation of this behavior below.

\begin{figure*}[]
\centering
\includegraphics[width=0.49\textwidth]{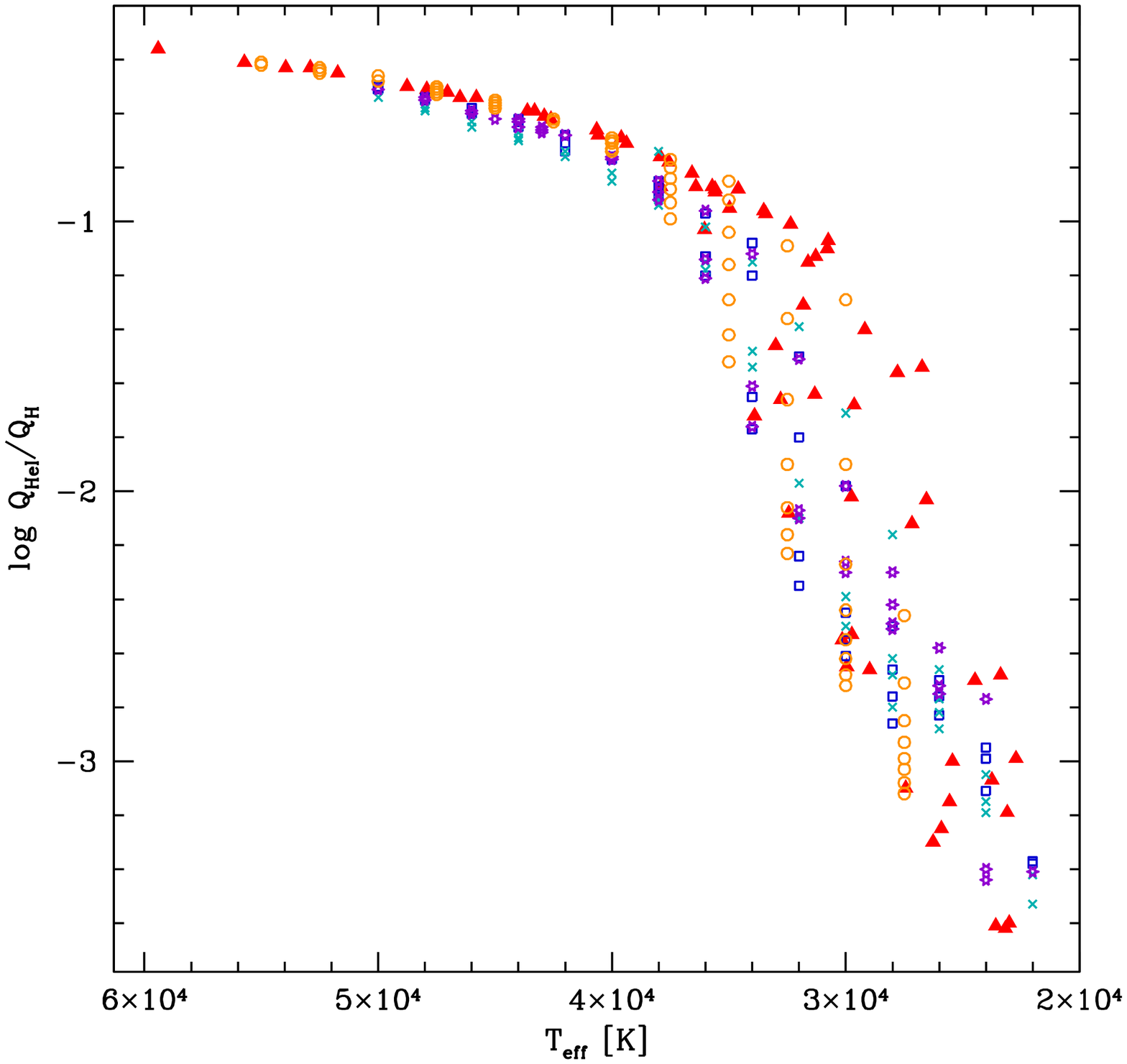}
\includegraphics[width=0.49\textwidth]{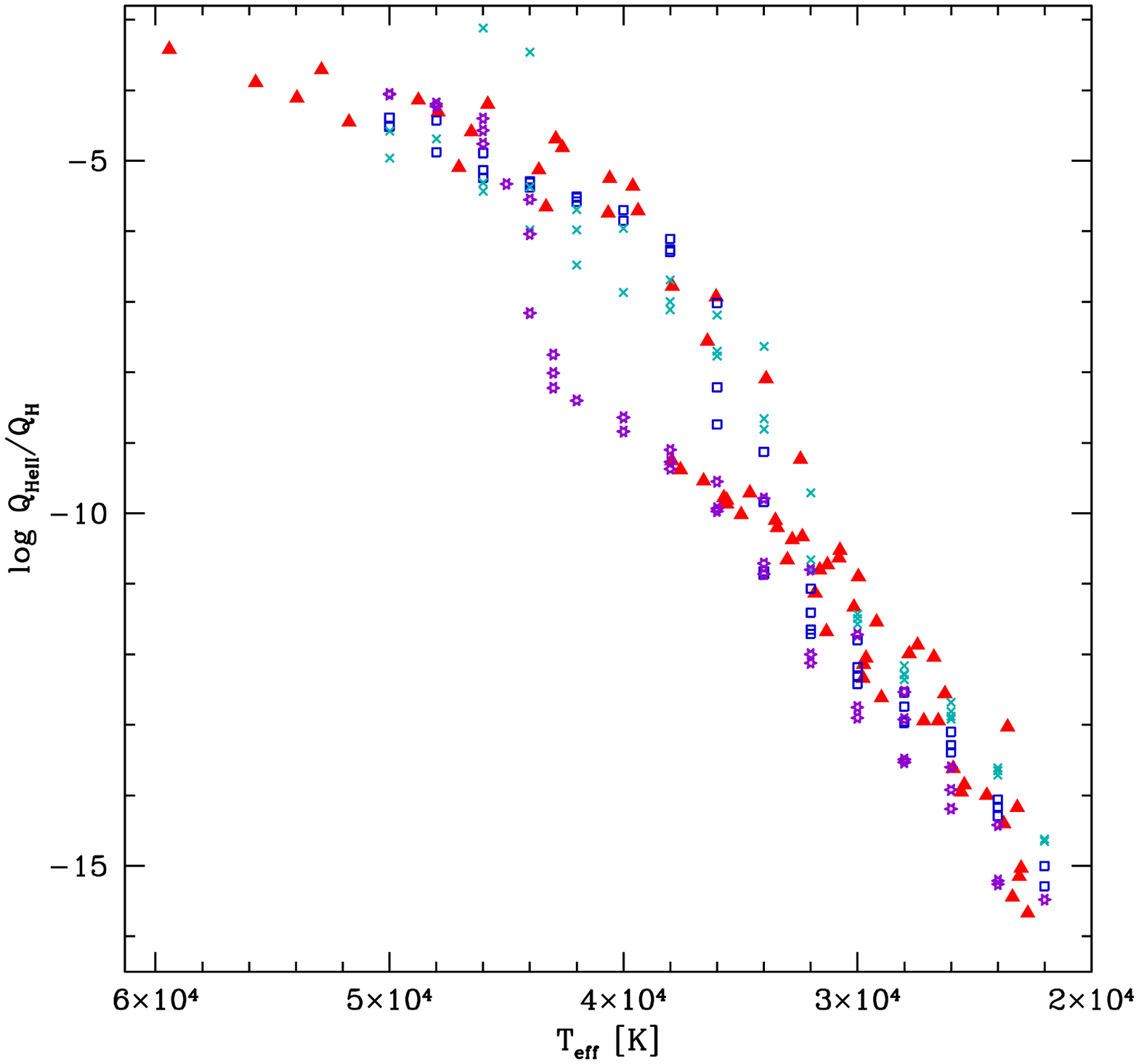}
\caption{Same as Fig.\ \ref{fig_ionflux} but for our SMC models (red filled triangles) and comparison models of \citet{lh03} (open orange circles) and \citet{hainich19}. For the latter, cyan crosses, blue squares, and purple stars correspond to low, intermediate, and high mass-loss rates, respectively.}
\label{fig_powr}
\end{figure*}

Beforehand we compare in Fig.\ \ref{fig_powr} our ionizing flux ratios to the predictions of \citet{hainich19} for Z~=~1/5~\zsun. The computations of these latter authors assume three sets of mass-loss rates (low, moderate, and high according to their nomenclature). We show them all in Fig.\ \ref{fig_powr}. We also add the results of \citet{lh03}. The general shape of the $\frac{Q(HeI)}{Q(\mathrm{H})}$--\teff\ relation is the same in the three sets of computations: the main drop happens at about the same \teff. For the highest temperatures, the ratios are the same in our study and that of \citet{lh03}. Between $\sim$35000 and 50000~K, the models of \citet{hainich19} are smaller by $\sim$0.1 dex. For the $\frac{Q(HeII)}{Q(\mathrm{H})}$ ratio\footnote{The \ion{He}{ii} ionizing fluxes are not available for the models of \citet{lh03} because these are plane-parallel models and wind effects are important.}, the high mass-loss-rate models of \citet{hainich19} experience a drop at higher \teff\ than all other computations (ours, and those of Hainich et al.  with lower mass-loss rates). 

This behavior is similar to what we observe in the right panel of Fig.\ \ref{fig_ionflux}:  different threshold \teff\ at different metallicities. The physical reason for this is an effect of mass-loss rates. \citet{gabler89,gabler92} and \citet{costar3} studied the effects of stellar winds on the \ion{He}{ii} ionizing continuum. We refer to these works for details on the physical processes. In short, because of the velocity fields in accelerating winds, lines (in particular resonance lines) are Doppler-shifted throughout the atmosphere. They therefore absorb additional, shorter wavelength photons compared to the static case, a process known as desaturation. As a consequence, the lower level population is pumped into the higher level. The ground level opacity is reduced, leading to stronger continuum emission \citep{gabler89}. \citet{costar3} showed that this effect works as long as the recombination of doubly ionized helium into \ion{He}{ii} is moderate. On the other hand, if recombinations are sufficiently numerous, the \ion{He}{ii} ground-state population becomes overpopulated and the opacity increases, causing a strong reduction of the \ion{He}{ii} ionizing flux. Recombinations depend directly on the wind density and are therefore more numerous for high mass-loss rates. \\

The effects described immediately above are clearly seen in Fig.~\ref{fig_qHeffectZ}. Let us now focus on the models at Z~=~1/5~\zsun.
Starting with the model with the smallest mass-loss rate ($\log \dot{M} = -7.30$), an increase up to $\log \dot{M} = -6.86$ translates into more flux below 228~\AA. This is the regime of desaturation. A subsequent increase by another factor 4 (up to $\log \dot{M} = -6.23$) leads to a drastic reduction in the flux shortward of 228~\AA. With such a high mass-loss rate, and therefore density, recombinations dominate the physics of the \ion{He}{ii} ionizing flux.

The right panel of Fig.~\ref{fig_powr} indicates that the PoWR models with the highest mass-loss rates have the smallest $\frac{Q(HeII)}{Q(\mathrm{H})}$ ratios, at least below 45000~K. This is fully consistent with the recombination effects. For the highest \teff\ the wind ionization is so high that even for strong mass-loss rates the \ion{He}{ii} ground-level population remains small. We verified that the same behavior is observed in our models. To this end, we ran new calculations for our solar metallicity grid, reducing the mass-loss rates. For selected models with \teff\ between 35000 and 43000~K, we find that the $\frac{Q(HeII)}{Q(\mathrm{H})}$ is increased up to the level of the low-metallicity models when mass-loss rates are reduced by a factor between 4 to 40. A stronger reduction of mass-loss rate is required for lower \teff. This is expected because at lower \teff\ the ionization is lower and a stronger reduction of recombinations is required to have a small ground-state opacity. As a sanity check we verified that in the initial models, with low $\frac{Q(HeII)}{Q(\mathrm{H})}$ ratios, \ion{He}{ii} is the dominant ion in the outer wind where the \ion{He}{ii} continuum is formed \citep[see also][]{schmutzhamann86}. In the models with lower mass-loss rates that have higher $\frac{Q(HeII)}{Q(\mathrm{H})}$ ratios, doubly ionized helium is the dominant ion in that same region, confirming the smaller recombination rates when mass-loss rates are reduced. 

We conclude that our computations do show a significant metallicity dependence of the $\frac{Q(HeII)}{Q(\mathrm{H})}$ ratio. This dependence is best described by the position of the threshold \teff\ at which the sudden drop between high and low $\frac{Q(HeII)}{Q(\mathrm{H})}$ ratios occurs. The position of this threshold temperature is physically related to mass-loss rates, as first demonstrated by \citet{schmutzhamann86}. As mass-loss rates depend on Z \citep{vink01,mokiem07}, the $\frac{Q(HeII)}{Q(HI}$ ratio also depends on metallicity. \ion{He}{ii} ionizing fluxes are therefore sensitive to prescriptions of mass-loss rates used in evolutionary and atmosphere models.

\subsection{\ion{He}{ii}~1640 emission}
\label{s_heii}

An interesting feature of our UV spectroscopic sequences is the presence of \lya\ and \ion{He}{ii}~1640 emission in some of the models with the highest masses (see last column of Table~\ref{tab_smc} and \ref{tab_1on30}). Figure\ \ref{fig_sv150uv} displays the most illustrative cases.

\ion{He}{ii}~1640 emission is a peculiar feature of some young massive clusters and star-forming galaxies both locally and at high redshift. It can be relatively narrow, and therefore considered of nebular nature, or broader and produced by stars \citep[e.g.,][]{cassata13}. So far, the only stars known to produce significant \ion{He}{ii}~1640 emission are Wolf-Rayet stars \citep{bri08,gv15,paul19}. Nebular \ion{He}{ii} emission requires ionizing photons with wavelengths shorter than 228 \AA. Possible sources for such hard radiation are (in addition to Wolf-Rayet stars themselves) population III stars \citep{schaerer03}, massive stars undergoing quasi-chemically homogeneous evolution \citep{kub19}, stripped binary stars \citep{got17}, X-ray binaries \citep{schaerer19}, and radiative shocks \citep{allen08}.

\citet{saxena19} report EW values of $\sim$1-4 \AA\ in a sample of \ion{He}{ii}~1640-emitting galaxies at redshift 2.5-5.0 \citep[see also][]{steidel16,patricio16}. Slightly larger values (5 to 30 \AA) are given by \citet{nanaya19} at redshifts from 2 to 4, while values lower than 1 \AA\ are also reported by \citet{senchyna17} in nearby galaxies. All these measurements include both stellar and nebular contributions. The integrated, mainly stellar \ion{He}{ii}~1640 emission of R136 in the LMC is 4.5\AA\ \citep{paul16,paul19}. This value is similar to other (super) star clusters in the Local Universe \citep{chandar04,leitherer18}. For comparison, the EW of our models with a net emission\footnote{The corresponding Ly$\alpha$ emission is $\lesssim$2.5 \AA.} reaches a maximum of $\sim$1.2 \AA.

\begin{figure*}[t]
\centering
\includegraphics[width=0.49\textwidth]{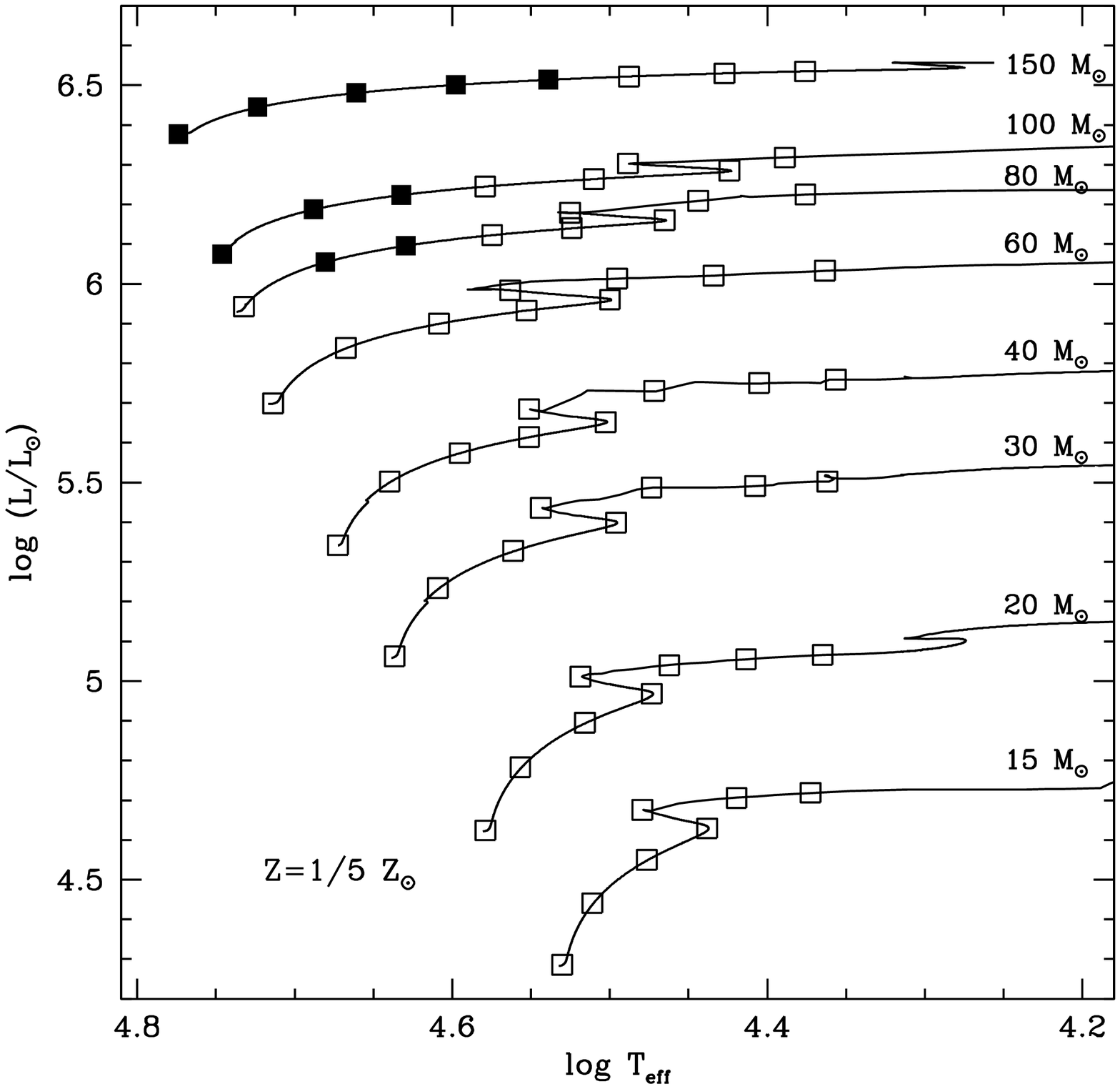}
\includegraphics[width=0.49\textwidth]{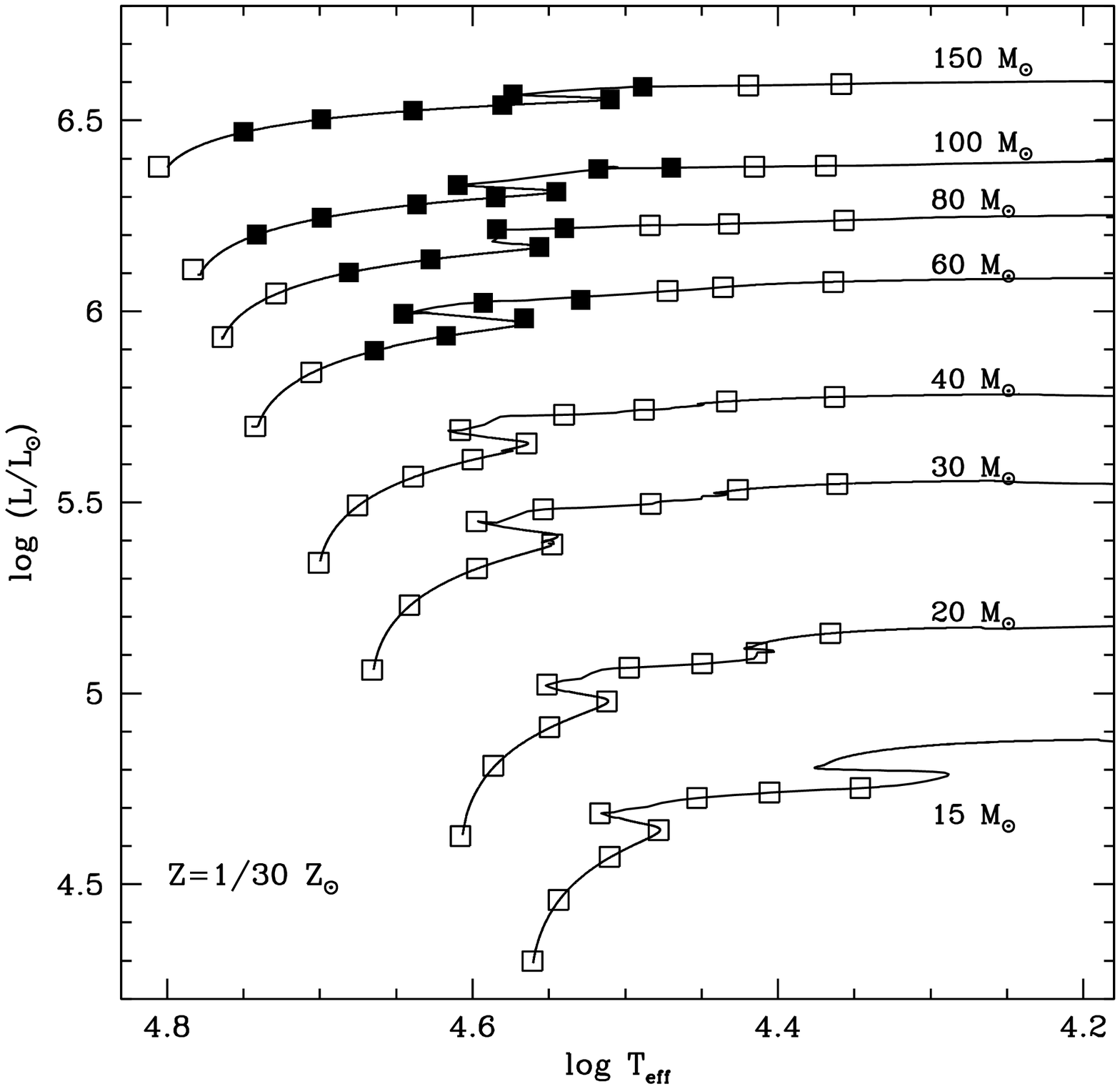}
\caption{HR diagram (Z = 1/5 \zsun, left panel; Z = 1/30 \zsun, right panel) showing the position of the models with \ion{He}{ii}~1640 emission (absorption) in black filled (open) squares.}
\label{figHeII}
\end{figure*}

\begin{figure}[t]
\centering
\includegraphics[width=0.49\textwidth]{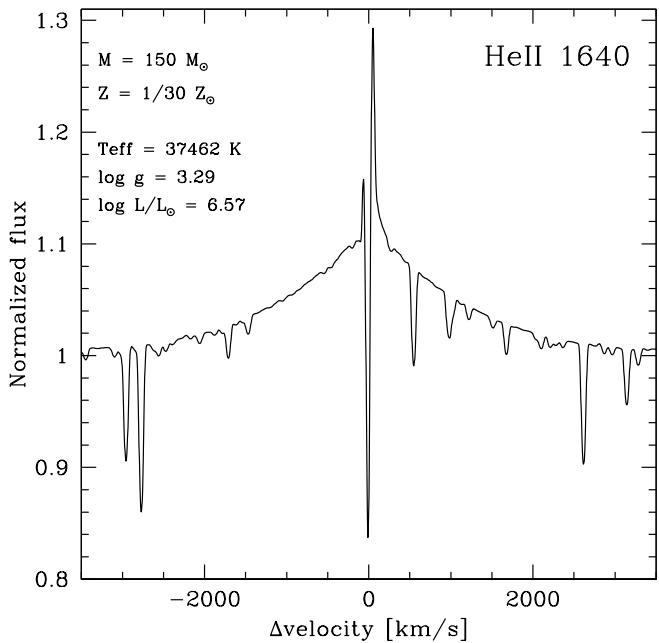}
\caption{\ion{He}{ii}~1640 profile of the model with the strongest emission.}
\label{figHeIIprof}
\end{figure}

\citet{gv15} studied very massive Wolf-Rayet stars with metallicities down to 0.01 \zsun. These latter authors showed that such objects have significant \ion{He}{ii}~1640 emission that could explain observations in some super-star clusters \citep{cassata13,wofford14}.
Figure~\ref{figHeII} shows the location of our models with a net  \ion{He}{ii}~1640. 
At the metallicity of the SMC, these are found above 80 \msun\ and in the first part of the MS.
At Z~=~1/30~\zsun\ stellar \ion{He}{ii}~1640 emission is produced in stars more massive than 60~\msun\ and  these stars are found mainly close to the TAMS, although their location extends to earlier phases at higher masses. \ion{He}{ii}~1640 emission appears at ages between 0 and $\sim$2.5~Myr (Z~=~1/5~\zsun) and between $\sim$1.5 and $\sim$4 Myr (Z~=~1/30~\zsun). Compared to \citet{gv15}, we therefore predict emission in lower mass stars, which are likely more numerous in young star clusters. These may therefore contribute to the integrated light of young stellar populations.
Nonetheless, we stress that our models always have \ion{He}{ii}~4686 in absorption. Consequently, if low-metallicity stars appear as we predict, they cannot account for the emission in that line observed in a number of star-forming galaxies \citep[e.g.,][]{kehrig15,kehrig18}.
The different location of \ion{He}{ii}~1640 emission stars in the HRD at the two metallicities considered is explained as follows. For Z~=~1/5~\zsun\ winds are stronger and therefore very hot stars are more likely to show emission. Conversely, at higher metallicity there are more metallic lines on top of the \ion{He}{ii}~1640 profile (see Fig.\ \ref{figHeIIprof}). At the temperatures typical of the TAMS, these lines are more numerous than at the ZAMS. At Z~=~1/5~\zsun\ they are strong enough to produce an absorption that counter-balances the underlying \ion{He}{ii}~1640 emission. Because of the effect of these metallic lines, EWs are on average larger at lower metallicity (see Tables~\ref{tab_smc} and \ref{tab_1on30}). Additionally, at lower Z, winds are weaker and \ion{He}{ii}~1640 does not develop an emission profile close to the ZAMS, where wind densities are too small. \ion{He}{ii}~1640 emission is therefore observed closer to the ZAMS (TAMS) at higher (lower) metallicity.

Figure\ \ref{figHeIIprof} shows a zoom on the \ion{He}{ii}~1640 line of the model with the strongest emission along the 150 \msun\ sequence (Z~=~1/30~\zsun). The profile has relatively broad wings extending up to $\pm$2000 \kms. The central part is composed of a narrow component ($\sim$250 \kms\ wide) with two emission peaks separated by a narrow absorption component. This narrow component is likely affected by nebular emission when present in integrated observations of stars and their surrounding nebula.

\section{Conclusion}
\label{s_conc}

We  present calculations of synthetic spectroscopy along evolutionary tracks computed at one-fifth and one-thirtieth solar metallicity. The models cover the MS and the early post-MS phases. Stellar-evolution computations were performed with the code \emph{STAREVOL}, while atmosphere models and synthetic spectra were calculated with the code \emph{CMFGEN}. Our models cover the mass range 15-150 \msun. For each mass, we provide spectroscopic evolutionary sequences. This study extend our work at solar metallicity presented in \citet{mp17}.

Our spectroscopic sequences all start as O dwarfs (early, intermediate, or late depending on initial mass) and end (in the early post-MS) as B giants or supergiants. The most massive stars are predicted to begin their evolution as O2V stars, contrary to solar metallicity computations for which such stars are not expected and not observed. The fraction of O2V stars increases when metallicity decreases. 

At the metallicity of the SMC (Z = 1/5 \zsun) and below 60~\msun\ stars spend a large fraction of the MS as dwarfs (luminosity class V) although the region near the TAMS is populated by giants (luminosity class IV, III, and II). Above 60 \msun, models enter the giant phase early on the MS. Our predictions reproduce the observed distribution of dwarfs and giants in the SMC  relatively well. For supergiants, the distribution we predict is located at lower \teff\ than observed. We confirm results presented by \citet{castro18} and \citet{rama19}, which show that,  from the HR diagram, there seems to be a lack of stars more massive than $\sim$60 \msun\ in the SMC. We predict that stars with masses higher than 60 \msun\ should be observed as O and B stars with luminosities higher than $10^6$ L$_{\odot}$, but almost no such star is reported in the literature. Whether this is an observational bias or an indication of either a peculiar evolution or a quenching of the formation of the most massive stars in the SMC is not clear.

At Z = 1/30 \zsun, a larger fraction of the MS is spent in the luminosity class V, even for the most massive models. Below 60~\msun, the MS is populated only by luminosity class V objects. The appearance of giants and supergiants is pushed to lower \teff\ at low Z. This is caused by the reduced wind strength \citep[see][]{mp17}. 
This reduction in the strength of wind-sensitive lines with metallicity is striking in the UV spectra. At one-thirtieth solar metallicity, only weak P-Cygni profiles in \ion{N}{v}~1240 and \ion{C}{iv}~1550 are sometimes observed.

We also present spectroscopic sequences in the wavelength range 4500-8000 \AA\ that will be covered by the instruments HARMONI and MOSAICS on the ELT. Hot massive stars will be best observed at these wavelengths in Local Group galaxies with low extinction. We advocate the use of the ratio of \ion{He}{I}~7065 to \ion{He}{ii}~5412 as a new spectral class diagnostics. Using archival high-resolution spectra and our synthetic spectra, we show that this ratio is a robust criterion for spectral typing, and is independent of metallicity. 

We provide the ionizing fluxes of our models. The relation between hydrogen-ionizing fluxes per unit area and \teff\ does not depend on metallicity. On the contrary, we show that the relations $\frac{Q(HeI)}{Q(\mathrm{H})}$ versus \teff\ and $\frac{Q(HeII)}{Q(\mathrm{H})}$ versus \teff\ both depend on metallicity, although in a different way. Both relations show a shallow decrease when \teff\ diminishes until a sharp drop at a characteristic \teff. Below this latter point of characteristic \teff,\ the ratios of  ionizing fluxes decrease faster. For $\frac{Q(HeI)}{Q(\mathrm{H})}$, at a given \teff, low-metallicity stars have higher ratios above the drop encountered at $\sim$35000~K. For $\frac{Q(HeII)}{Q(\mathrm{H})}$, it is the position of the drop that is affected, being located at higher \teff\ for stars with higher metallicity. This behavior is rooted in the metallicity dependence of mass-loss rates.

Finally, we highlight that in some models for the most massive stars, we predict a net emission in \ion{He}{ii}~1640, a feature observed in some star-forming galaxies but difficult to reproduce in population synthesis models. The emission we predict is stronger at lower metallicity, reaching a maximum EW of the order of 1.2~\AA. The line profile is composed of broad wings and a narrow core and is present in a region of the HRD near the ZAMS (TAMS) at Z~=~1/5 \zsun\ (Z~=~1/30 \zsun). 

Our SEDs and synthetic spectra are made available to the community through the POLLUX database.

\section*{Acknowledgments}

We thank Andreas Sander for a prompt referee report. We warmly thank John Hillier for making the code CMFGEN available to the community and for constant help with it. We thank Daniel Schaerer for discussion on the HeII~1640 emission in star-forming galaxies.
This work made use of the Polarbase database (developed and maintained by CNRS/INSU, Observatoire Midi-Pyr\'en\'ees and Universit\'e Toulouse III) and of the services of the ESO Science Archive Facility.

\begin{appendix}
\label{ap_tab}

\section{Stellar parameters, spectral classification, and ionizing fluxes}
\label{ap_param}

In Tables \ref{tab_smc} and \ref{tab_1on30} we gather the parameters adopted for the computation of the synthetic spectra along the evolutionary sequences. We also give the resulting spectral types and luminosity classes, as well as the \ion{H}{i} , \ion{He}{i,} and \ion{He}{ii} ionizing fluxes. Finally we provide the EW of \ion{He}{ii}~1640.

\begin{table*}[ht]
\begin{center}
\caption{Atmosphere model parameters (\teff, luminosity, surface gravity, mass-loss rate, wind terminal velocity), associated spectral types, ionizing fluxes, and EW of \ion{He}{ii}~1640 for the SMC metallicity grid} \label{tab_smc}
\begin{tabular}{lccccclcccc}
\hline
$M$        &  $T_{\rm eff}$  &  $\log(L/L_{\odot})$ & $\log g$ & $\log \dot{M}$ & $\varv_{\infty}$ & Spectral & $\log Q(\mathrm{H})$ & $\log Q(\ion{He}{i})$ & $\log Q(\ion{He}{ii})$ & EW(\ion{He}{ii}~1640)\\    
$M_{\odot}$  &  K          &                   &          &               & \kms       & Type  & [s$^{-1}$] & [s$^{-1}$] & [s$^{-1}$] & [\AA] \\
\hline
150& 59406 & 6.38 & 4.28 & -5.97 & 5272 & O2V((f))          &   50.24 & 49.88 & 46.82 & -0.23 \\
   & 52909 & 6.45 & 4.01 & -5.70 & 4408 & O2III(f)          &   50.30 & 49.87 & 46.59 & -0.85 \\
   & 45803 & 6.48 & 3.72 & -5.55 & 3666 & O2-3III-I         &   50.32 & 49.78 & 46.12 & -0.86 \\
   & 39604 & 6.50 & 3.44 & -5.52 & 3071 & O3-3.5III-I       &   50.30 & 49.61 & 44.94 & -0.51 \\
   & 34600 & 6.51 & 3.19 & -5.59 & 2637 & O5-5.5III         &   50.26 & 49.38 & 40.55 & -0.09 \\
   & 30752 & 6.52 & 2.98 & -5.72 & 2335 & O7.5Ib            &   50.20 & 49.13 & 39.68 & 0.70  \\
   & 26738 & 6.53 & 2.73 & -5.95 & 2016 & O9.2-9.7Iaf       &   50.07 & 48.53 & 38.03 & 2.09  \\
   & 23377 & 6.54 & 2.52 & -4.90 & 1781 & B0Ia+             &   49.71 & 47.03 & 39.40 & 3.35  \\

\hline
100& 55716 & 6.08 & 4.28 & -6.31 & 4860 & O2-3V((f))        &   49.93 & 49.52 & 46.04 & -0.02 \\
   & 48762 & 6.19 & 3.93 & -5.96 & 3840 & O3III(f)          &   50.03 & 49.53 & 45.89 & -0.46 \\
   & 42887 & 6.22 & 3.67 & -5.86 & 3266 & O3.5-4III(f)      &   50.03 & 49.42 & 45.34 & -0.25 \\
   & 37934 & 6.24 & 3.43 & -5.86 & 2809 & O5III(f)          &   50.00 & 49.24 & 40.75 & 0.21 \\
   & 32364 & 6.26 & 3.14 & -5.96 & 2371 & O7-7.5II          &   49.92 & 48.91 & 39.59 & 1.24 \\
   & 26545 & 6.28 & 2.77 & -6.26 & 1891 & O9.7Ia            &   49.71 & 47.68 & 36.77 & 2.76 \\
   & 30794 & 6.30 & 3.01 & -5.95 & 2155 & O7.5-8Ib          &   49.95 & 48.85 & 39.32 & 1.27 \\
   & 24478 & 6.32 & 2.60 & -6.06 & 1697 & B0Ia              &   49.67 & 46.97 & 35.67 & 2.77 \\

\hline                
80 & 53962 & 5.94 & 4.28 & -6.46 & 4704 & O2-3V((f))        &   49.79 & 49.36 & 45.68 & 0.10 \\
   & 47914 & 6.06 & 3.95 & -6.14 & 3759 & O3III(f)          &   49.88 & 49.37 & 45.57 & -0.21 \\
   & 42608 & 6.10 & 3.70 & -6.04 & 3204 & O3.5-4III(f)      &   49.89 & 49.27 & 45.08 & -0.02 \\
   & 37557 & 6.12 & 3.46 & -6.03 & 2782 & O5.5IV-III        &   49.86 & 49.08 & 40.48 & 0.60 \\
   & 33422 & 6.14 & 3.24 & -6.09 & 2437 & O7III             &   49.79 & 48.82 & 39.59 & 1.46 \\
   & 29185 & 6.16 & 2.98 & -6.24 & 2074 & O9Ib              &   49.68 & 48.28 & 38.14 & 2.57 \\
   & 33504 & 6.18 & 3.20 & -6.01 & 2337 & O6.5-7III         &   49.86 & 48.90 & 39.76 & 1.07 \\
   & 27792 & 6.21 & 2.85 & -6.17 & 1896 & O9-9.5Ia          &   49.71 & 48.15 & 37.72 & 2.50 \\
   & 23756 & 6.22 & 2.56 & -6.43 & 1592 & B0Ia              &   49.43 & 46.36 & 35.03 & 2.88 \\

\hline                                      
60 & 51724 & 5.70 & 4.32 & -6.80 & 4549 & O3-3.5V((f))      &   49.53 & 49.08 & 45.08 & 0.36 \\
   & 46496 & 5.84 & 3.99 & -6.44 & 3658 & O3.5III(f)        &   49.65 & 49.11 & 45.06 & 0.16 \\
   & 40591 & 5.90 & 3.69 & -6.32 & 3030 & O5V((f))          &   49.66 & 48.98 & 44.41 & 0.53 \\
   & 35715 & 5.93 & 3.44 & -6.32 & 2614 & O6.5III(f)        &   49.61 & 48.74 & 39.83 & 1.36 \\
   & 31615 & 5.96 & 3.20 & -6.40 & 2257 & O8II              &   49.52 & 48.37 & 38.72 & 2.40 \\
   & 36582 & 5.98 & 3.43 & -6.19 & 2559 & O6III(f)          &   49.70 & 48.88 & 40.16 & 0.85 \\
   & 31282 & 6.01 & 3.12 & -6.27 & 2104 & O8II-Ib           &   49.61 & 48.48 & 38.88 & 2.00 \\
   & 27165 & 6.02 & 2.87 & -6.48 & 1822 & O9.5Iab           &   49.40 & 47.28 & 36.46 & 3.03 \\
   & 23086 & 6.03 & 2.58 & -6.81 & 1542 & B0.5-0.7Ia        &   48.85 & 45.66 & 33.70 & 3.11 \\

\hline                                      
40 & 47041 & 5.34 & 4.34 & -7.31 & 4253 & O4V((f))          &   49.13 & 48.61 & 44.04 & 0.84 \\
   & 43614 & 5.50 & 4.04 & -6.95 & 3496 & O5V((f)           &   49.26 & 48.67 & 44.13 & 0.79 \\
   & 39374 & 5.57 & 3.79 & -6.82 & 2995 & O5.5V((f)         &   49.28 & 48.57 & 43.57 & 1.24 \\
   & 35589 & 5.61 & 3.58 & -6.79 & 2648 & O7V-IV            &   49.23 & 48.34 & 39.36 & 2.07 \\
   & 31810 & 5.65 & 3.34 & -6.84 & 2277 & O8.5III-II        &   49.12 & 47.81 & 37.99 & 3.07 \\
   & 35588 & 5.68 & 3.50 & -6.64 & 2475 & O6.5V((f))        &   49.33 & 48.45 & 39.52 & 1.73 \\
   & 29637 & 5.73 & 3.14 & -6.62 & 1995 & O9-9.5I           &   49.14 & 47.46 & 37.09 & 3.17 \\
   & 25426 & 5.75 & 2.86 & -6.96 & 1698 & B0.5Ia            &   48.66 & 45.66 & 34.81 & 3.25 \\
   & 22721 & 5.76 & 2.56 & -5.88 & 1312 & B0.7-1Ia          &   48.46 & 45.47 & 32.79 & 3.01 \\

\hline   
30 & 43303 & 5.06 & 4.35 & -7.75 & 4023 & O5.5V((f))          & 48.78 & 48.19 & 43.13 & 1.39 \\
   & 40652 & 5.23 & 4.07 & -7.39 & 3383 & O6V((f))            & 48.92 & 48.26 & 43.18 & 1.53 \\
   & 36413 & 5.33 & 3.78 & -7.24 & 2826 & O7V((f))            & 48.92 & 48.05 & 41.36 & 2.32 \\
   & 31334 & 5.40 & 3.45 & -7.25 & 2320 & O9III-II            & 48.75 & 47.11 & 37.07 & 3.39 \\
   & 34976 & 5.44 & 3.60 & -7.04 & 2508 & O7-7.5V((f))-III(f) & 49.01 & 48.06 & 38.99 & 2.44 \\
   & 29752 & 5.49 & 3.27 & -7.11 & 2060 & O9.5III             & 48.75 & 46.73 & 36.41 & 3.33 \\
   & 25565 & 5.49 & 3.00 & -7.36 & 1759 & B0.7I               & 48.10 & 44.95 & 34.15 & 3.27 \\
   & 23007 & 5.50 & 2.81 & -7.63 & 1579 & B1-1.5Ia            & 47.38 & 43.78 & 32.35 & 3.00 \\

\hline                     
\end{tabular}
\end{center}
\end{table*}

\setcounter{table}{0}

\begin{table*}[ht]
\begin{center}
\caption{Continued} \label{tab_smc}
\begin{tabular}{lccccclcccc}
\hline
$M$        &  $T_{\rm eff}$  &  $\log(L/L_{\odot})$ &  $\log g$ & $\log \dot{M}$ & $\varv_{\infty}$ & Spectral & $\log Q(\mathrm{H})$ & $\log Q(\ion{He}{i})$ & $\log Q(\ion{He}{ii})$ & EW(\ion{He}{ii}~1640) \\    
$M_{\odot}$  &  K          &                   &          &               & \kms       & Type  & [s$^{-1}$] & [s$^{-1}$] & [s$^{-1}$] & [\AA] \\  
\hline
20 & 37917 & 4.62 & 4.38 & -8.53 & 3738 & O7.5V((f))          & 48.12 & 47.25 & 41.34 & 2.79 \\
   & 36039 & 4.78 & 4.13 & -8.22 & 3210 & O8V((f))            & 48.22 & 47.19 & 41.29 & 3.04 \\
   & 32794 & 4.90 & 3.86 & -8.07 & 2743 & O9-9.5IV            & 48.14 & 46.48 & 37.77 & 3.38 \\
   & 29738 & 4.97 & 3.61 & -8.05 & 2352 & O9.7III-I           & 47.88 & 45.35 & 35.74 & 3.41 \\
   & 33001 & 5.01 & 3.75 & -7.82 & 2545 & O8.5-9IV            & 48.34 & 46.88 & 37.68 & 3.37 \\
   & 28987 & 5.04 & 3.50 & -7.90 & 2206 & B0III-Ib            & 47.89 & 45.23 & 35.28 & 3.41 \\
   & 25914 & 5.06 & 3.29 & -8.07 & 1951 & B0.7III-I           & 47.25 & 44.00 & 33.63 & 3.12 \\
   & 23164 & 5.07 & 3.08 & -8.30 & 1722 & B1Iab               & 46.52 & 42.90 & 32.35 & 2.75 \\
\hline
15 & 33908 & 4.29 & 4.40 & -9.20 & 3540 & O9.7V-IV            & 47.45 & 45.73 & 39.36 & 3.27 \\
   & 32436 & 4.44 & 4.17 & -8.92 & 3094 & O9.7V-IV            & 47.48 & 45.40 & 38.25 & 3.30 \\
   & 29956 & 4.55 & 3.92 & -8.79 & 2665 & B0-0.5V             & 47.24 & 44.59 & 36.34 & 3.22 \\
   & 27427 & 4.63 & 3.69 & -8.77 & 2331 & B0.7V               & 46.86 & 43.76 & 34.99 & 3.00 \\
   & 30154 & 4.68 & 3.81 & -8.51 & 2495 & B0V                 & 47.50 & 44.95 & 36.17 & 3.33 \\
   & 26269 & 4.71 & 3.53 & -8.64 & 2108 & B0.7V               & 46.74 & 43.44 & 34.18 & 2.94 \\
   & 23583 & 4.72 & 3.33 & -8.86 & 1877 & B1III               & 46.08 & 42.47 & 33.05 & 2.52 \\
\hline                     
\end{tabular}
\end{center}
\end{table*}

\begin{table*}[ht]
\begin{center}
\caption{Atmosphere model parameters (\teff, luminosity, surface gravity, mass-loss rate, wind terminal velocity), associated spectral types, ionizing fluxes and EW of \ion{He}{ii}~1640 for the Z=1/30 \zsun\ metallicity grid} \label{tab_1on30}
\begin{tabular}{lccccclcccc}
\hline
$M$        &  $T_{\rm eff}$  &  $\log(L/L_{\odot})$ &  $\log g$ & $\log \dot{M}$ & $\varv_{\infty}$ & Spectral & $\log Q(\mathrm{H})$ & $\log Q(\ion{He}{i})$ & $\log Q(\ion{He}{ii})$ & EW(\ion{He}{ii}~1640) \\    
$M_{\odot}$  &  K          &                   &          &               & \kms       & Type  & [s$^{-1}$] & [s$^{-1}$] & [s$^{-1}$] & [\AA] \\
\hline
150& 63880 & 6.38 & 4.41 & -6.71 & 5706 & O2V((f))        & 50.24 & 49.92 & 46.99 & 0.10 \\
   & 56234 & 6.47 & 4.09 & -6.36 & 4567 & O2V((f))        & 50.33 & 49.95 & 46.58 & -0.21 \\
   & 50008 & 6.50 & 3.85 & -6.20 & 3913 & O2V-III         & 50.36 & 49.90 & 46.13 & -0.51 \\
   & 43551 & 6.53 & 3.59 & -6.12 & 3343 & O3-3.5III(f)    & 50.36 & 49.80 & 45.56 & -0.81 \\
   & 38073 & 6.54 & 3.34 & -6.13 & 2868 & O4III(f)        & 50.33 & 49.66 & 45.04 & -0.97 \\
   & 32362 & 6.56 & 3.05 & -6.25 & 2428 & O7-7.5III(f)    & 50.27 & 49.37 & 44.06 & -0.84 \\
   & 37462 & 6.57 & 3.29 & -6.08 & 2765 & O4III-II        & 50.37 & 49.68 & 45.07 & -1.14 \\
   & 30808 & 6.59 & 2.93 & -6.22 & 2222 & O7.5II-Ib       & 50.30 & 49.35 & 43.71 & -1.16 \\
   & 26259 & 6.59 & 2.65 & -6.38 & 1890 & O9.5Ia          & 50.18 & 48.77 & 38.37 & 0.10 \\
   & 22831 & 6.60 & 2.40 & -6.75 & 1628 & O9.7-B0Ia+      & 50.05 & 47.63 & 36.57 & 0.84 \\

\hline
100& 60690 & 6.11 & 4.41 & -6.99 & 5264 & O2V((f))        & 49.96 & 49.62 & 46.11 & 0.17 \\
   & 55125 & 6.20 & 4.15 & -6.70 & 4420 & O2V((f))        & 50.05 & 49.66 & 45.90 & -0.05 \\
   & 49990 & 6.24 & 3.94 & -6.54 & 3875 & O2-3V((f))      & 50.09 & 49.63 & 45.57 & -0.25 \\
   & 43291 & 6.28 & 3.65 & -6.43 & 3223 & O3-4IV-III      & 50.09 & 49.53 & 45.08 & -0.51 \\
   & 38447 & 6.30 & 3.43 & -6.42 & 2832 & O5.5IV-III      & 50.07 & 49.39 & 44.61 & -0.57 \\
   & 35095 & 6.31 & 3.25 & -6.46 & 2522 & O6.5-7III(f)    & 50.04 & 49.25 & 44.24 & -0.56 \\
   & 40708 & 6.33 & 3.49 & -6.33 & 2872 & O4III(f)        & 50.14 & 49.52 & 45.02 & -0.78 \\
   & 32939 & 6.37 & 3.08 & -6.38 & 2224 & O7III           & 50.10 & 49.24 & 44.03 & -0.90 \\
   & 29505 & 6.38 & 2.89 & -6.50 & 1999 & O8.5II          & 50.02 & 48.93 & 39.80 & -0.36 \\
   & 26025 & 6.38 & 2.67 & -6.72 & 1760 & O9.7Ia          & 49.90 & 48.23 & 37.78 & 0.71 \\
   & 23362 & 6.38 & 2.48 & -6.95 & 1576 & B0-0.5Ia        & 49.71 & 46.64 & 35.57 & 1.21 \\

\hline                                                                          
80 & 58056 & 5.93 & 4.41 & -7.22 & 5050 & O2-3V((f))      & 49.78 & 49.42 & 45.56 & 0.23 \\
   & 53526 & 6.05 & 4.16 & -6.89 & 4284 & O2-3V((f))      & 49.89 & 49.48 & 45.46 & 0.02 \\
   & 47977 & 6.10 & 3.91 & -6.70 & 3640 & O3-3.5V((f))    & 49.93 & 49.44 & 45.15 & -0.20 \\
   & 42420 & 6.14 & 3.66 & -6.62 & 3112 & O4V-III         & 49.93 & 49.35 & 44.67 & -0.37 \\
   & 36002 & 6.17 & 3.34 & -6.63 & 2555 & O6.5V-III       & 49.89 & 49.13 & 44.12 & -0.38 \\
   & 38377 & 6.22 & 3.41 & -6.48 & 2626 & O5.5IV-III      & 49.99 & 49.31 & 44.55 & -0.66 \\
   & 34666 & 6.22 & 3.23 & -6.54 & 2363 & O7IV-III        & 49.94 & 49.13 & 44.07 & -0.53 \\
   & 30469 & 6.22 & 3.00 & -6.63 & 2066 & O8.5II          & 49.84 & 48.77 & 40.37 & 0.07 \\
   & 27043 & 6.23 & 2.79 & -6.84 & 1830 & O9.7Ia          & 49.71 & 48.02 & 37.71 & 0.54 \\
   & 22736 & 6.24 & 2.48 & -7.21 & 1525 & B0-0.5Ia        & 49.32 & 45.81 & 34.78 & 1.50 \\

\hline                                      
60 & 55241 & 5.70 & 4.44 & -7.52 & 4891 & O3-3.5V((f))    & 49.53 & 49.15 & 44.91 & 0.34 \\
   & 50768 & 5.84 & 4.15 & -7.14 & 4037 & O3-3.5V((f))    & 49.66 & 49.22 & 44.92 & 0.11 \\
   & 46172 & 5.90 & 3.93 & -6.98 & 3519 & O4V((f))        & 49.70 & 49.19 & 44.61 & -0.03 \\
   & 41427 & 5.94 & 3.70 & -6.90 & 3046 & O5.5V((f))      & 49.70 & 49.10 & 44.24 & -0.14 \\
   & 36834 & 5.98 & 3.45 & -6.86 & 2604 & O6.5IV-III      & 49.69 & 48.95 & 43.89 & -0.15 \\
   & 44191 & 5.99 & 3.75 & -6.76 & 3072 & O3.5III(f)      & 49.81 & 49.26 & 44.75 & -0.42 \\
   & 39182 & 6.02 & 3.51 & -6.73 & 2644 & O5.5IV          & 49.78 & 49.12 & 44.28 & -0.40 \\
   & 33815 & 6.03 & 3.25 & -6.80 & 2279 & O7.5IV-III      & 49.69 & 48.82 & 43.52 & -0.09 \\
   & 29662 & 6.05 & 3.00 & -6.92 & 1958 & O9.2II          & 49.60 & 48.34 & 38.79 & 0.78 \\
   & 27295 & 6.06 & 2.85 & -7.05 & 1797 & O9.7Iab/Ia      & 49.49 & 47.60 & 37.40 & 1.22 \\
   & 23110 & 6.08 & 2.54 & -7.44 & 1485 & B0.5Ia          & 49.09 & 45.58 & 34.84 & 1.59 \\

\hline  
40 & 50196 & 5.34 & 4.45 & -8.00 & 4521 & O4V((f))        & 49.14 & 48.70 & 43.97 & 0.51 \\
   & 47355 & 5.49 & 4.20 & -7.65 & 3856 & O4-5V((f))      & 49.28 & 48.80 & 44.01 & 0.33 \\
   & 43540 & 5.57 & 3.98 & -7.47 & 3364 & O5.5V((f))      & 49.32 & 48.77 & 43.69 & 0.23 \\
   & 39801 & 5.61 & 3.78 & -7.39 & 2979 & O6V((f))        & 49.32 & 48.68 & 43.43 & 0.21 \\
   & 36700 & 5.65 & 3.59 & -7.35 & 2635 & O7V((f))        & 49.30 & 48.55 & 43.21 & 0.26 \\
   & 40555 & 5.69 & 3.73 & -7.21 & 2839 & O5.5-6IV-V((f)) & 49.42 & 48.80 & 43.74 & 0.02 \\
   & 34683 & 5.73 & 3.42 & -7.21 & 2357 & O7.5IV-III      & 49.35 & 48.49 & 43.11 & 0.30 \\
   & 30737 & 5.74 & 3.20 & -7.34 & 2075 & O9.2II          & 49.21 & 47.91 & 39.99 & 1.22 \\
   & 27136 & 5.76 & 2.96 & -7.52 & 1795 & O9.7-B0Ia       & 48.97 & 46.36 & 36.37 & 1.66 \\
   & 23066 & 5.78 & 2.66 & -7.87 & 1498 & B0.7Ib          & 48.24 & 44.51 & 34.14 & 1.59 \\

\hline                     
\end{tabular}
\end{center}
\end{table*}

\setcounter{table}{1}

\begin{table*}[ht]
\begin{center}
\caption{Continued} \label{tab_1on30}
\begin{tabular}{lccccclcccc}
\hline
$M$        &  $T_{\rm eff}$  &  $\log(L/L_{\odot})$ &  $\log g$ & $\log \dot{M}$ & $\varv_{\infty}$ & Spectral & $\log Q(\mathrm{H})$ & $\log Q(\ion{He}{i})$ & $\log Q(\ion{He}{ii})$ & EW(\ion{He}{ii}~1640) \\    
$M_{\odot}$  &  K          &                   &          &               & \kms       & Type  & [s$^{-1}$] & [s$^{-1}$] & [s$^{-1}$] & [\AA] \\  
\hline

30 & 46315 & 5.06 & 4.47 & -8.42 & 4317 & O5.5V((f))      & 48.80 & 48.33 & 43.89 & 0.72 \\
   & 43782 & 5.23 & 4.20 & -8.04 & 3643 & O5.5V((f))      & 48.95 & 48.42 & 43.22 & 0.55 \\
   & 39538 & 5.33 & 3.93 & -7.86 & 3096 & O6.5V((f))      & 48.99 & 48.34 & 43.08 & 0.49 \\
   & 35296 & 5.39 & 3.66 & -7.79 & 2611 & O7-7.5V((f))    & 48.94 & 48.09 & 42.46 & 0.74 \\
   & 39570 & 5.45 & 3.80 & -7.59 & 2807 & O6-6.5V((f))    & 49.14 & 48.50 & 43.26 & 0.27 \\
   & 35792 & 5.48 & 3.60 & -7.57 & 2500 & O7.5V-IV((f))   & 49.08 & 48.27 & 42.84 & 0.48 \\
   & 30424 & 5.50 & 3.30 & -7.72 & 2093 & O9.5-9.7II      & 48.84 & 47.20 & 38.88 & 1.60 \\
   & 26689 & 5.53 & 3.04 & -7.90 & 1795 & B0Ib            & 48.52 & 45.55 & 35.86 & 1.76 \\
   & 22971 & 5.55 & 2.76 & -8.22 & 1519 & B1.5-2Ib/Ia     & 47.60 & 43.72 & 33.82 & 1.51 \\

\hline
20 & 40545 & 4.63 & 4.49 & -9.15 & 3972 & O7.5V((f))      & 48.23 & 47.61 & 42.30 & 1.02 \\
   & 38574 & 4.81 & 4.22 & -8.78 & 3375 & O8V((f))        & 48.38 & 47.67 & 42.09 & 0.94 \\
   & 35447 & 4.91 & 3.98 & -8.61 & 2938 & O8.5V((f))      & 48.36 & 47.40 & 41.24 & 1.22 \\
   & 32510 & 4.98 & 3.76 & -8.56 & 2574 & O9.5-9.7V-III   & 48.25 & 46.70 & 40.25 & 1.73 \\
   & 35580 & 5.02 & 3.87 & -8.37 & 2727 & O8.5V-III       & 48.52 & 47.61 & 41.86 & 1.03 \\
   & 31434 & 5.07 & 3.61 & -8.38 & 2338 & O9.7III-II      & 48.29 & 46.49 & 39.57 & 1.77 \\
   & 28161 & 5.08 & 3.41 & -8.52 & 2084 & B0.5III         & 47.78 & 44.76 & 36.23 & 1.77 \\
   & 25945 & 5.11 & 3.24 & -8.68 & 1886 & B1Ib            & 47.33 & 43.87 & 35.24 & 1.62 \\
   & 23210 & 5.16 & 2.99 & -8.82 & 1617 & B2Ib            & 46.85 & 42.89 & 33.86 & 1.35 \\

\hline
15 & 36381 & 4.30 & 4.51 & -9.77 & 3772 & O9V             & 47.68 & 46.68 & 40.18 & 1.51 \\
   & 34967 & 4.46 & 4.28 & -9.45 & 3290 & O9-9.5V-IV      & 47.78 & 46.57 & 40.36 & 1.64 \\
   & 32381 & 4.57 & 4.03 & -9.29 & 2831 & B0V             & 47.65 & 45.66 & 39.15 & 1.84 \\
   & 30070 & 4.64 & 3.83 & -9.25 & 2514 & B0.5V           & 47.38 & 44.64 & 37.46 & 1.77 \\
   & 32846 & 4.69 & 3.94 & -9.04 & 2674 & O9.5-9.7III     & 47.90 & 46.27 & 39.89 & 1.79 \\
   & 28380 & 4.73 & 3.65 & -9.12 & 2263 & B0.5-0.7V-III   & 47.16 & 44.04 & 36.42 & 1.69 \\
   & 25443 & 4.74 & 3.45 & -9.30 & 2020 & B1.5III-Ib      & 46.58 & 42.91 & 34.83 & 1.37 \\
   & 22183 & 4.75 & 3.20 & -9.61 & 1749 & B2II            & 45.79 & 41.58 & 32.38 & 0.96 \\

\hline                     
\end{tabular}
\end{center}
\end{table*}

\section{Additional spectroscopic sequences}
\label{ap_param}

\begin{figure*}[t]
\centering
\includegraphics[width=0.49\textwidth]{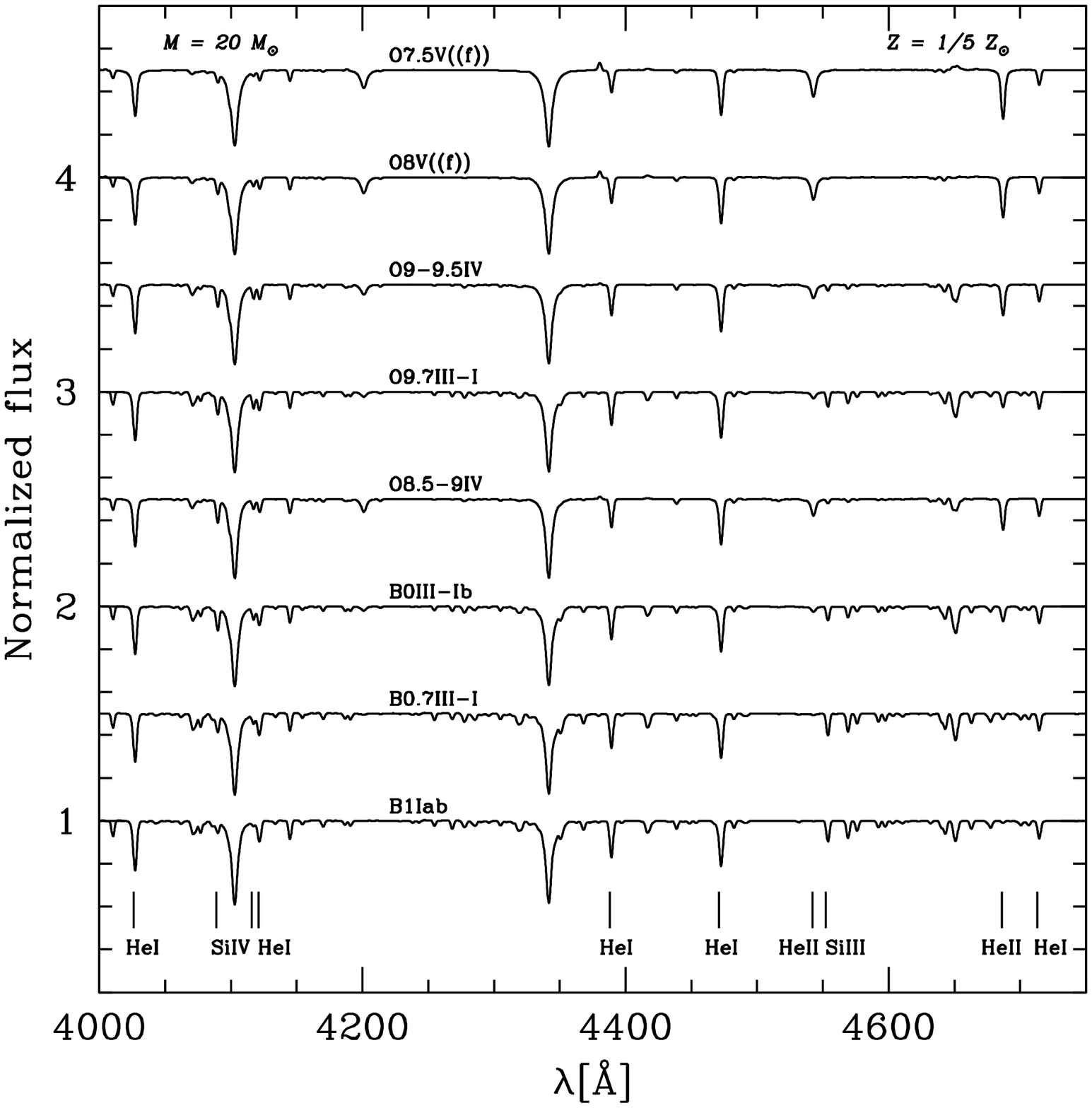}  
\includegraphics[width=0.49\textwidth]{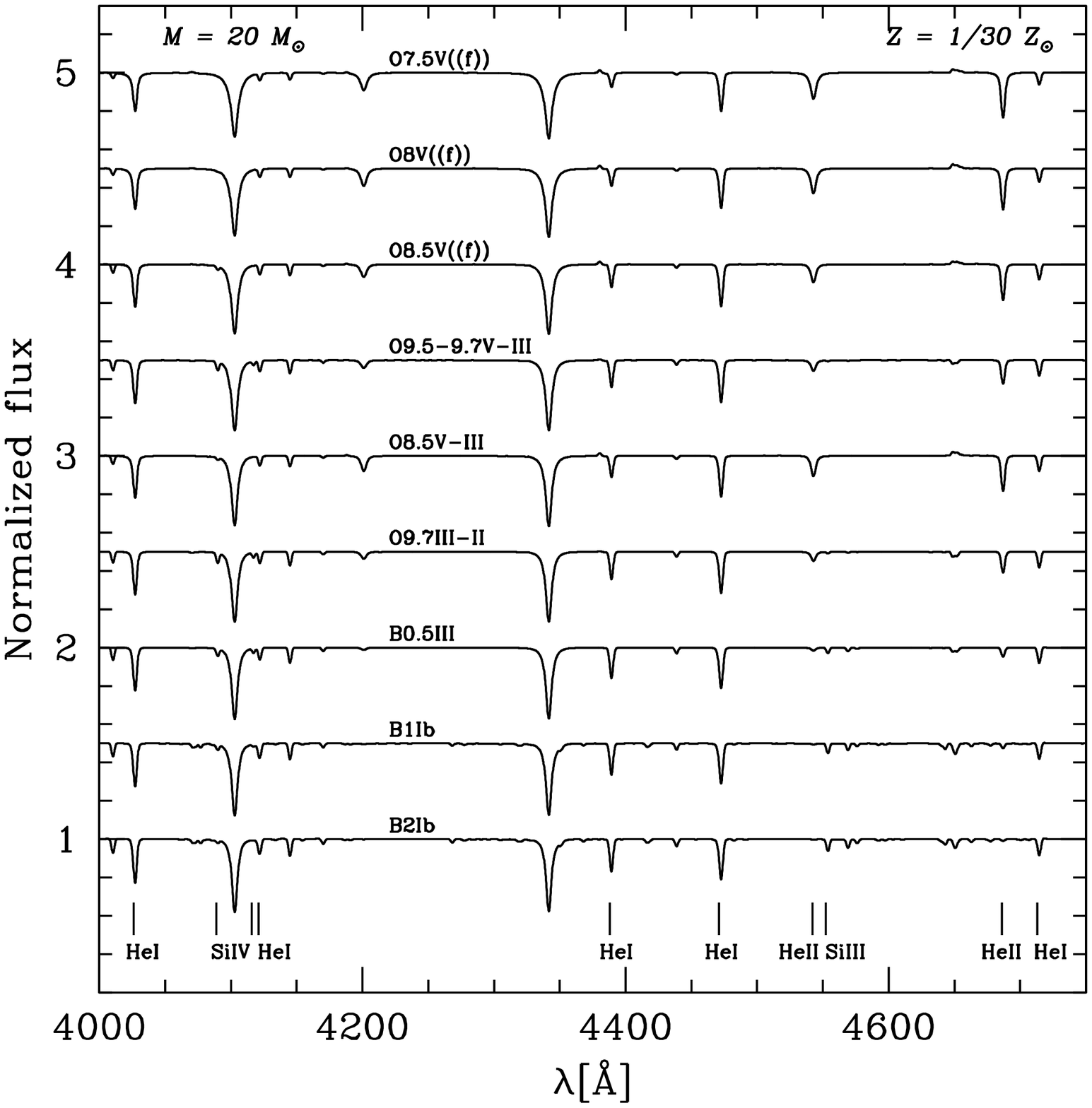}
\caption{Optical spectra of the sequence of models calculated along the 20 \msun\ track at SMC (left) and one-thirtieth solar (right) metallicity. The main diagnostic lines are indicated at the bottom of the figure.}
\label{fig_sv20opt}
\end{figure*}

\begin{figure*}[t]
\centering
\includegraphics[width=0.49\textwidth]{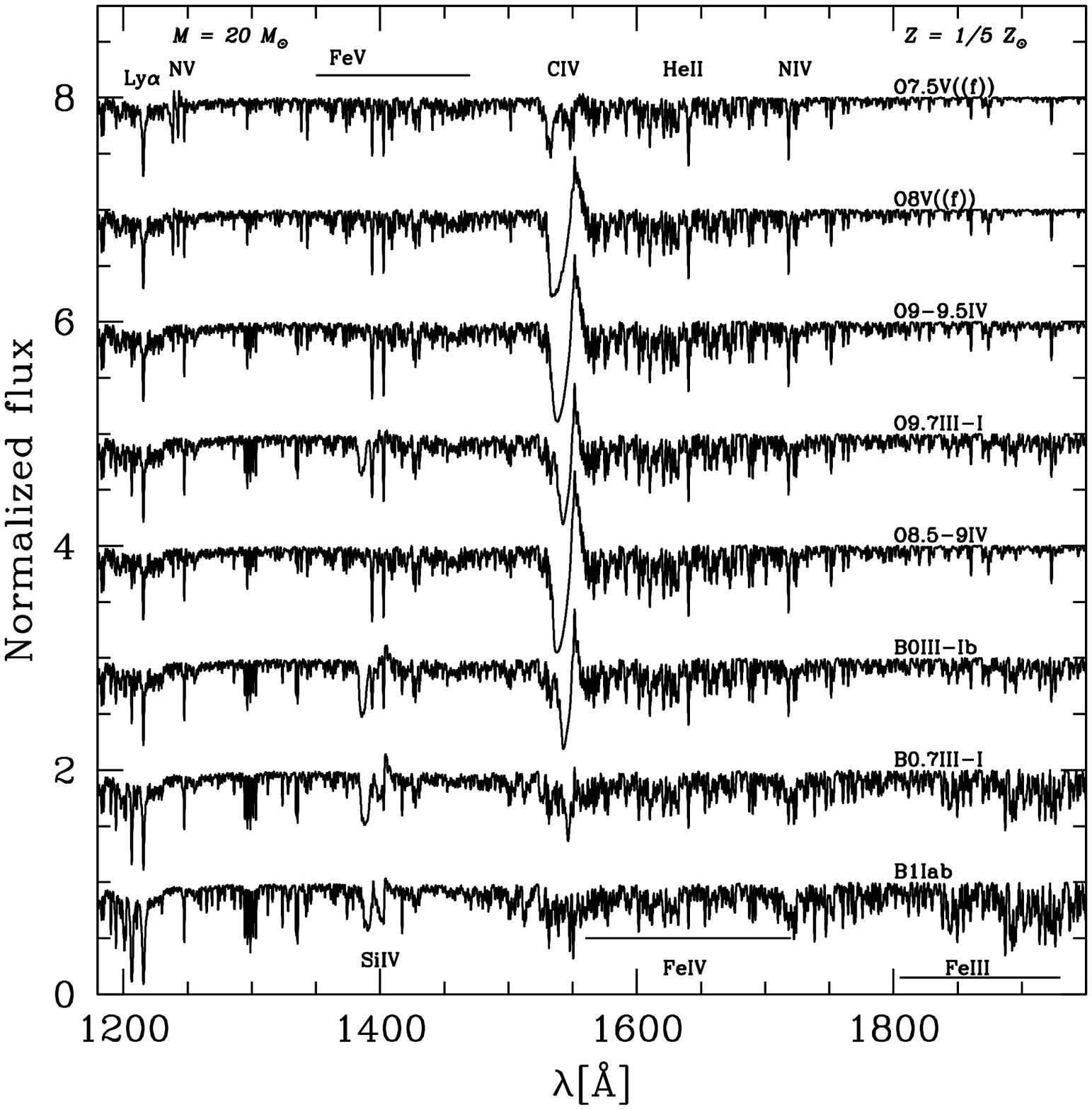}  
\includegraphics[width=0.49\textwidth]{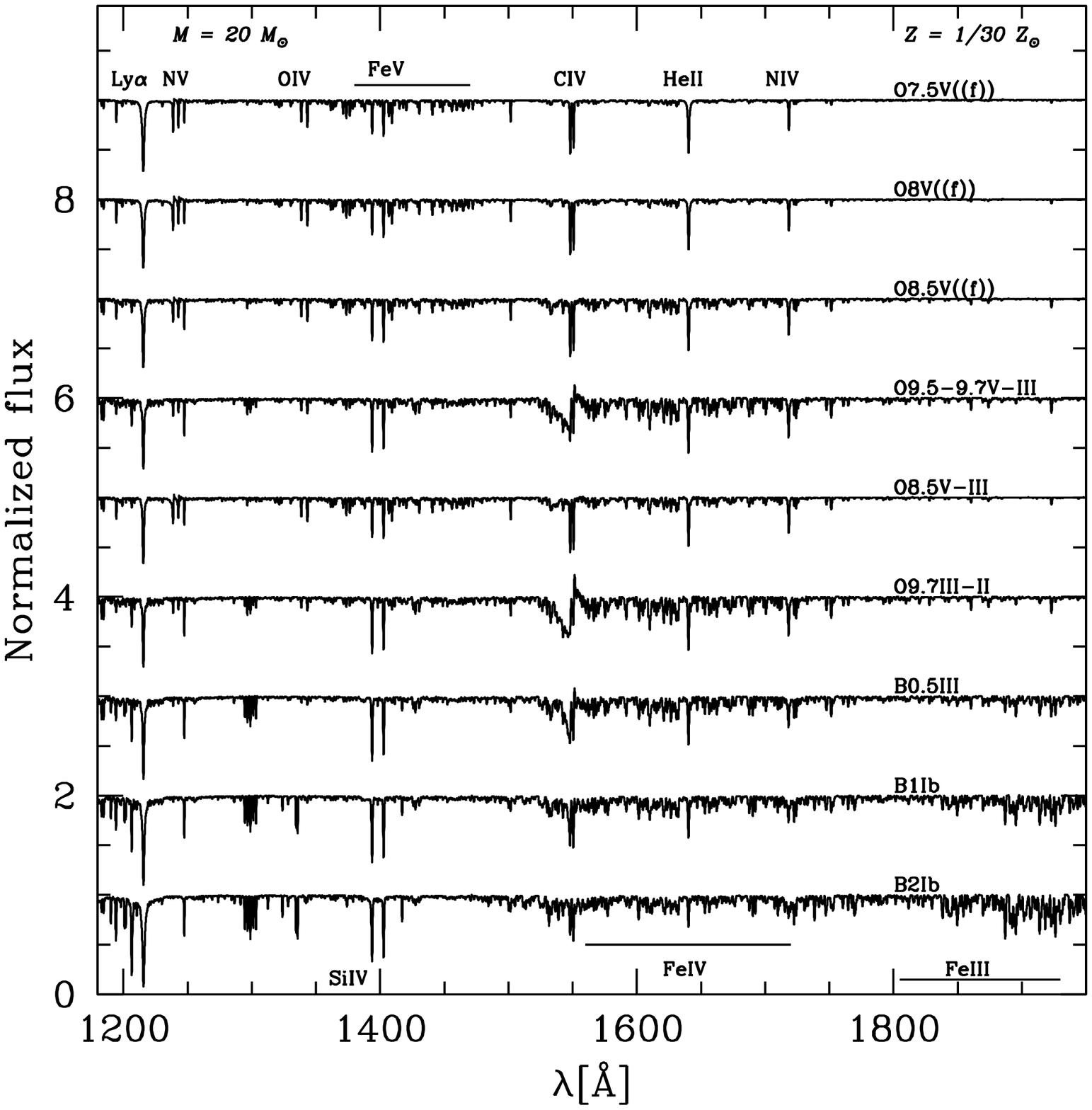}
\caption{Ultra-violet spectra of the sequence of models calculated along the 20 \msun\ track at SMC (left) and one-thirtieth solar (right) metallicity. The main diagnostic lines are indicated at the bottom of the figure.}
\label{fig_sv20uv}
\end{figure*}

\begin{figure*}[t]
  \centering
  \includegraphics[width=0.49\textwidth]{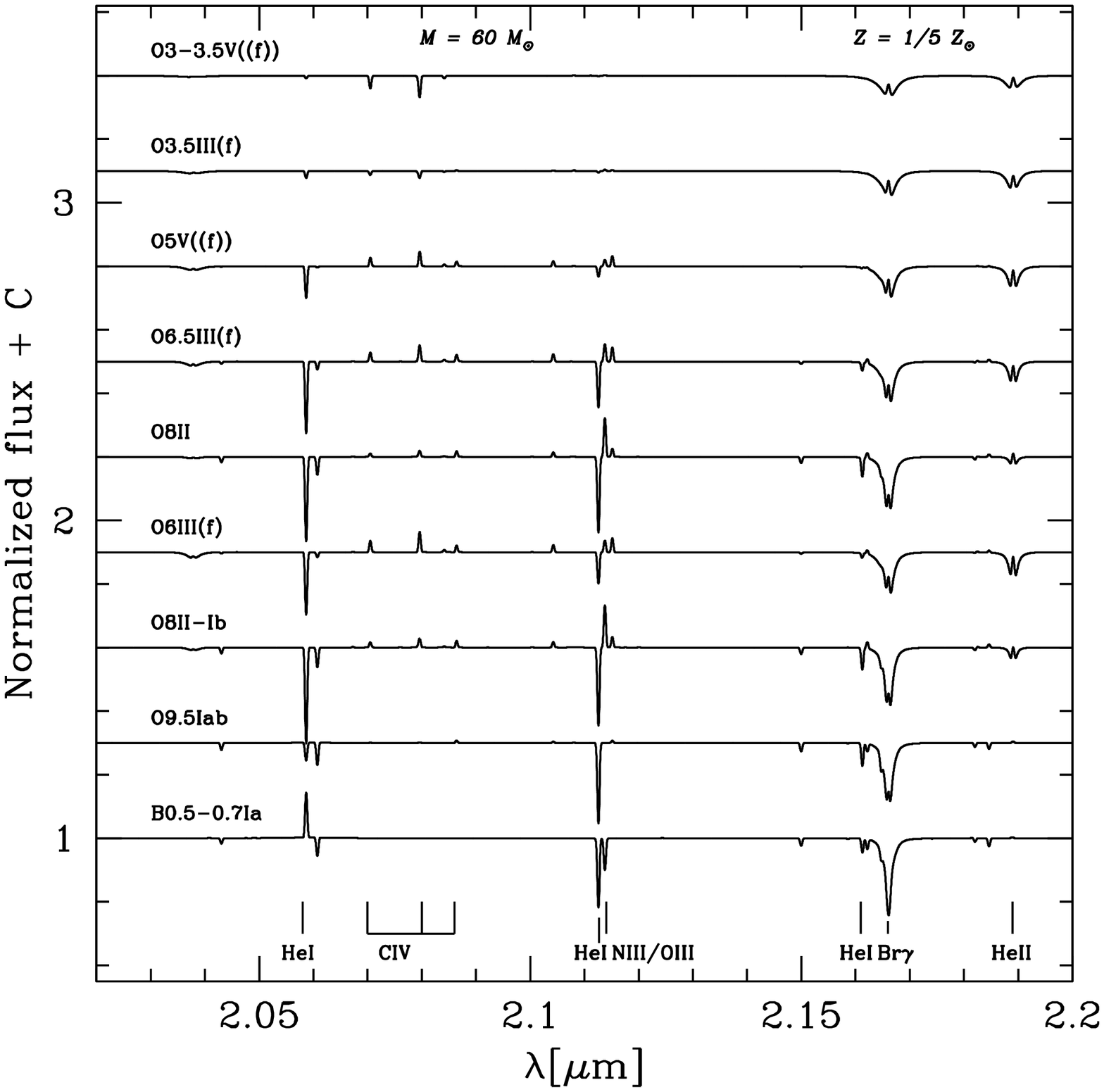}  
  \includegraphics[width=0.49\textwidth]{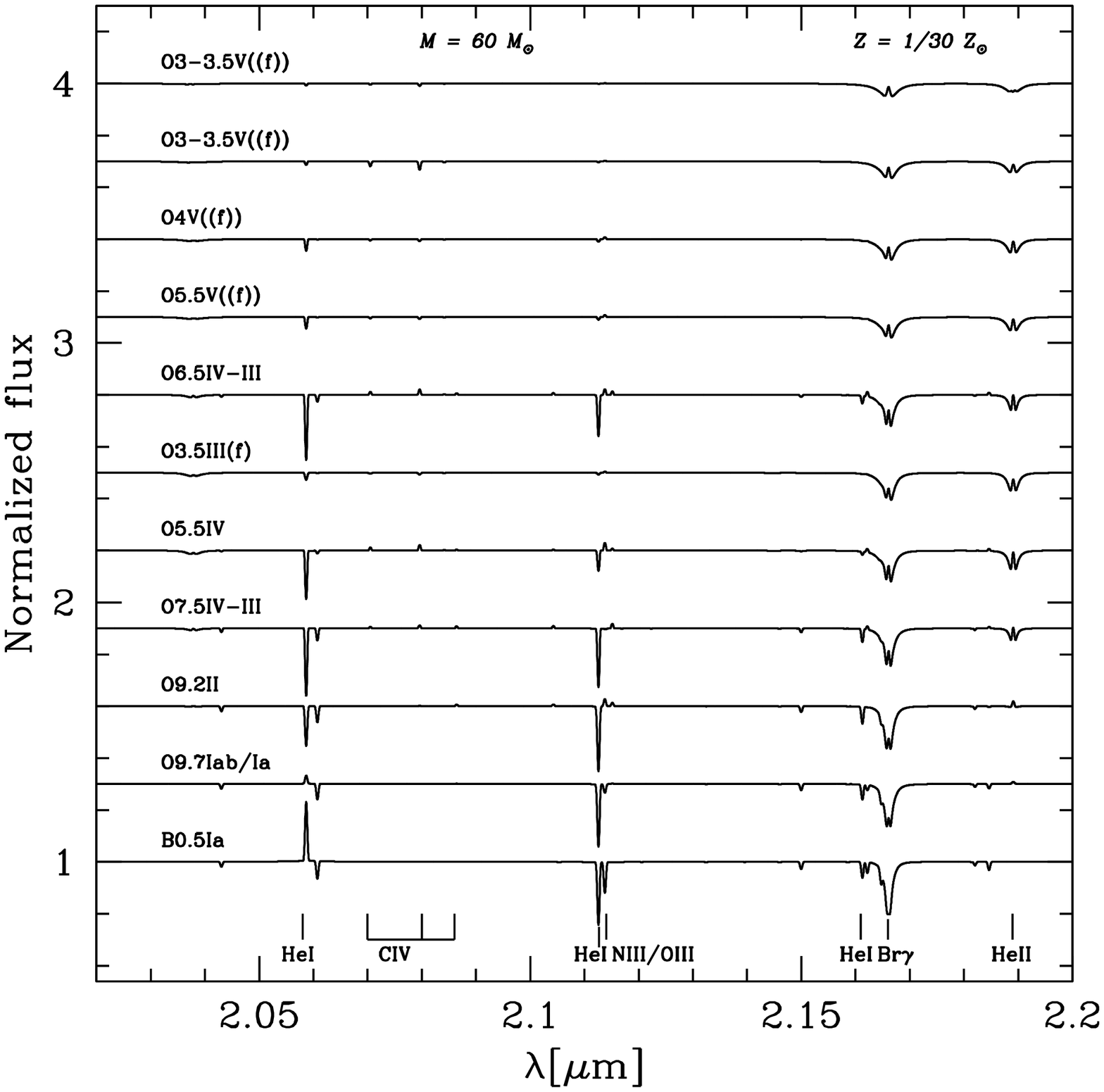}  
\caption{Infrared K-band spectra of the sequence of models calculated along the 60 \msun\ tracks at SMC (left panel) and one-thirtieth solar (right panel) metallicity. The main diagnostic lines are indicated.}
\label{fig_ir}
\end{figure*}

\begin{figure}[t]
\centering
\includegraphics[width=0.49\textwidth]{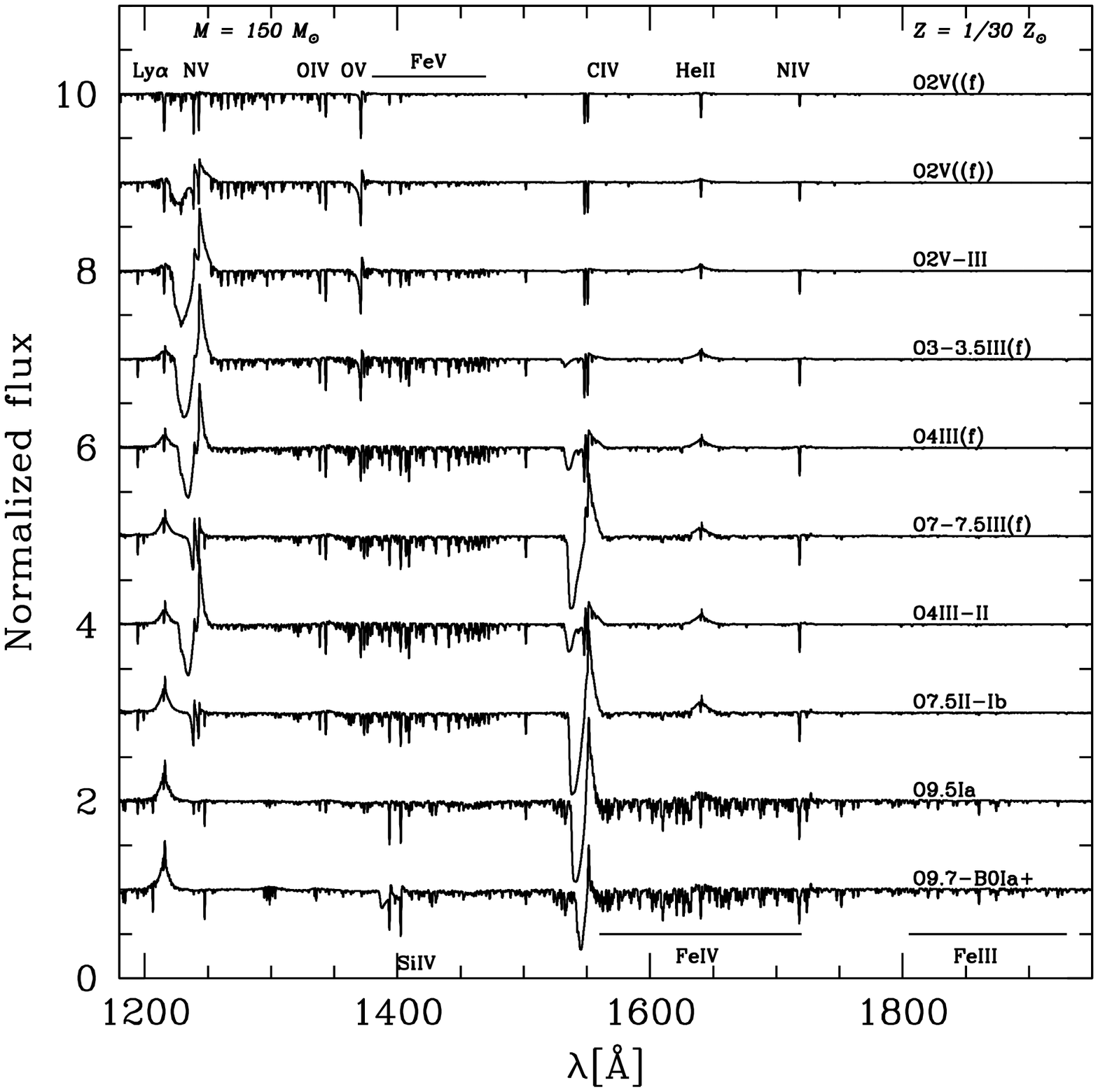}  
\caption{Ultraviolet spectra of the sequence of models calculated along the 150 \msun\ track at Z = 1/30 \zsun.}
\label{fig_sv150uv}
\end{figure}

We show in Fig.\ \ref{fig_sv20opt} and \ref{fig_sv20uv} the optical spectra of the 20~\msun\ sequence.
The UV spectra of the 150~\msun\ models are visible in Fig.\ \ref{fig_sv150uv}.
Finally, Fig.\ \ref{fig_ir} displays the K-band spectra of the 60~\msun\ models.

\section{Archival data}
\label{ap_archive}

In Table \ref{tab_archive} we give information on the archival spectra we used in Sect.~\ref{s_elt}. The data have been retrieved from the Polarbase database (\url{http://polarbase.irap.omp.eu/}, see \citealt{polarbase1,polarbase2}) and the ESO phase 3 archive facility (\url{http://archive.eso.org/wdb/wdb/adp/phase3_main/form}). Spectral types are given according to the GOSS catalog \citep{gosss,goss}.

\begin{table*}[ht]
\begin{center}
\caption{Galactic stars for which spectra were collected from archives to measure EWs. The following information is given: star's name, instrument, date of observation, spectral type, luminosity class, EW of \ion{He}{ii}~5412 and EW of \ion{He}{ii}~7065} \label{tab_archive}
\begin{tabular}{lcccccc}
\hline
Star            &   Instrument   &  Observation   &   ST           &  LC    &    EW(5412)  &  EW(7065) \\
                &                &   date          &           &        &  [\AA]     &  [\AA] \\
\hline
HD64568         & FEROS          & 13/01/2016   & O3     & V     & 1.03  & 0.06 \\
HD93205         & FEROS          & 03/05/2009   & O3.5   & V     & 1.00  & 0.09 \\
HD46223         & ESPaDOnS       & 22/02/2008   & O4     & V     & 1.06  & 0.15 \\
HD96715         & FEROS          & 25/12/2004   & O4     & V     & 1.09  & 0.16 \\
HD168076        & FEROS          & 09/05/2006   & O4     & III   & 0.88  & 0.09 \\
HD93250         & FEROS          & 20/09/2016   & O4     & III   & 1.07  & 0.08 \\
HD66811         & ESPaDOnS       & 14/02/2012   & O4     & I     & 0.98  & 0.13 \\
HD164794        & FEROS          & 22/09/2016   & O4     & V     & 1.05  & 0.13 \\
HD46150         & ESPaDOnS       & 31/01/2012   & O5     & V     & 1.00  & 0.23 \\
HD319699        & FEROS          & 22/03/2011   & O5     & V     & 1.04  & 0.28 \\
HD93843         & FEROS          & 21/04/2007   & O5     & III   & 0.91  & 0.30 \\
CPD474963       & FEROS          & 19/04/2007   & O5     & I     & 0.86  & 0.26 \\
HD93204         & FEROS          & 24/12/2004   & O5.5   & V     & 1.02  & 0.41 \\
HD305525        & ESPaDOnS       & 20/05/2012   & O5.5   & V     & 1.00  & 0.47 \\
BD164826        & FEROS          & 05/04/2015   & O5.5   & I     & 0.91  & 0.21 \\
HD42088         & ESPaDOnS       & 08/03/2010   & O6     & V     & 1.04  & 0.47 \\
HD303311        & FEROS          & 08/03/2017   & O6     & V     & 1.05  & 0.41 \\
HD101190        & ESPaDOnS       & 19/05/2005   & O6     & IV    & 0.72  & 0.57 \\
HD124314        & ESPaDOnS       & 20/04/2007   & O6     & IV    & 0.94  & 0.60 \\
HD152233        & ESPaDOnS       & 06/05/2004   & O6     & I     & 0.88  & 0.35 \\
HD199579        & ESPaDOnS       & 15/10/2008   & O6.5   & V     & 0.93  & 0.49 \\
HD344784        & NARVAL         & 07/11/2017   & O6.5   & V     & 0.89  & 0.41 \\
HD190864        & ESPaDOnS       & 05/06/2014   & O6.5   & III   & 1.02  & 0.61 \\
HD157857        & FEROS          & 20/04/2007   & O6.5   & II    & 0.99  & 0.65 \\
HD210839        & ESPaDOnS       & 22/08/2005   & O6.5   & I     & 0.81  & 0.55 \\
HD150958        & FEROS          & 21/04/2007   & O6.5   & I     & 0.75  & 0.34 \\
HD163758        & FEROS          & 12/05/2006   & O6.5   & I     & 0.90  & 0.61 \\
HD47839         & ESPaDOnS       & 03/02/2012   & O7     & V     & 0.77  & 0.46 \\
HD46485         & ESPaDOnS       & 14/02/2012   & O7     & V     & 0.87  & 0.69 \\
HD36879         & ESPaDOnS       & 06/09/2009   & O7     & V     & 0.93  & 0.65 \\
HD54662         & ESPaDOnS       & 19/10/2010   & O7     & V     & 0.95  & 0.53 \\
HD91824         & FEROS          & 24/12/2004   & O7     & V     & 0.87  & 0.50 \\
hd94963         & FEROS          & 21/04/2007   & O7     & II    & 0.85  & 0.68 \\
HD151515        & FEROS          & 26/06/2005   & O7     & II    & 0.80  & 0.61 \\
HD69464         & FEROS          & 16/05/2011   & O7     & I     & 0.76  & 0.70 \\
HD53975         & ESPaDOnS       & 09/01/2014   & O7.5   & V     & 0.81  & 0.65 \\
HD152590        & FEROS          & 26/06/2005   & O7.5   & V     & 0.82  & 0.58 \\
HD124979        & FEROS          & 20/05/2012   & O7.5   & IV    & 0.86  & 0.81 \\
HD34656         & ESPaDOnS       & 11/11/2011   & O7.5   & II    & 0.90  & 0.80 \\
HD120521        & FEROS          & 23/03/2011   & O7.5   & I     & 0.76  & 0.82 \\
HD192639        & NARVAL         & 09/08/2012   & O7.5   & I     & 0.85  & 0.82 \\
HD188001        & ESPaDOnS       & 23/07/2010   & O7.5   & I     & 0.81  & 0.78 \\
HD165246        & FEROS          & 20/08/2006   & O8     & V     & 0.71  & 0.82 \\
HD97848         & FEROS          & 02/05/2009   & O8     & V     & 0.66  & 0.70 \\
HD101191        & FEROS          & 19/05/2005   & O8     & V     & 0.72  & 0.68 \\
HD94024         & FEROS          & 15/05/2008   & O8     & IV    & 0.71  & 0.82 \\
HD36861         & ESPaDOnS       & 17/10/2010   & O8     & III   & 0.78  & 0.74 \\
HD115455        & FEROS          & 20/08/2006   & O8     & III   & 0.72  & 0.65 \\
HD319702        & FEROS          & 15/05/2008   & O8     & III   & 0.68  & 0.72 \\
HD162978        & ESPaDOnS       & 22/06/2012   & O8     & II    & 0.74  & 0.76 \\
HD96917         & FEROS          & 21/04/2007   & O8     & I     & 0.63  & 0.80 \\
HD151804        & FEROS          & 02/05/2004   & O8     & I     & 0.58  & 0.51 \\
HD14633         & ESPaDOnS       & 09/10/2009   & O8.5   & V     & 0.70  & 0.79 \\
HD46966         & ESPaDOnS       & 02/02/2012   & O8.5   & IV    & 0.67  & 0.61 \\
HD218195        & ESPaDOnS       & 05/07/2011   & O8.5   & III   & 0.66  & 0.83 \\
HD153426        & ESPaDOnS       & 04/07/2011   & O8.5   & III   & 0.57  & 0.66 \\
HD151003        & FEROS          & 14/05/2008   & O8.5   & III   & 0.45  & 0.69 \\
HD207198        & ESPaDOnS       & 26/07/2010   & O8.5   & II    & 0.60  & 0.88 \\
HD75211         & FEROS          & 05/05/2009   & O8.5   & II    & 0.65  & 0.89 \\
\hline                     
\end{tabular}
\end{center}
\end{table*}

\setcounter{table}{1}

\begin{table*}[ht]
\begin{center}
\caption{Continued} \label{tab_archive}
\begin{tabular}{lcccccc}
\hline
Star            &   Instrument   &  Observation   &   ST           &  LC    &    EW(5412)  &  EW(7065) \\
                &                &   date          &           &        &  [\AA]     &  [\AA] \\
\hline
HD125241        & FEROS          & 20/05/2012   & O8.5   & I     & 0.59  & 0.86 \\
HD149404        & FEROS          & 02/05/2005   & O8.5   & I     & 0.44  & 0.46 \\
HD112244        & FEROS          & 02/05/2004   & O8.5   & I     & 0.52  & 0.79 \\
HD303492        & FEROS          & 21/05/2012   & O8.5   & I     & 0.41  & 0.70 \\
HD214680        & ESPaDOnS       & 22/06/2005   & O9     & V     & 0.66  & 0.69 \\
HD149452        & FEROS          & 20/05/2012   & O9     & IV    & 0.70  & 0.91 \\
HD93028         & FEROS          & 24/12/2004   & O9     & IV    & 0.58  & 0.56 \\
HD24431         & ESPaDOnS       & 07/11/2011   & O9     & III   & 0.57  & 0.80 \\
HD93249         & FEROS          & 02/05/2009   & O9     & III   & 0.45  & 0.67 \\
HD105627        & FEROS          & 19/08/2013   & O9     & III   & 0.53  & 0.88 \\
HD207198        & ESPaDOnS       & 26/07/2011   & O9     & II    & 0.60  & 0.89 \\
HD209975        & ESPaDOnS       & 27/07/2011   & O9     & Ib    & 0.45  & 0.87 \\
HD30614         & ESPaDOnS       & 02/01/2013   & O9     & Ia    & 0.43  & 0.85 \\
HD152249        & FEROS          & 06/05/2004   & O9     & I     & 0.45  & 0.84 \\
HD173783        & FEROS          & 03/05/2009   & O9     & I     & 0.52  & 0.94 \\
HD148546        & FEROS          & 01/04/2008   & O9     & I     & 0.56  & 0.97 \\
BD661661        & NARVAL         & 06/09/2015   & O9.2   & V     & 0.42  & 0.66 \\
HD76341         & FEROS          & 24/12/2004   & O9.2   & IV    & 0.42  & 0.70 \\
HD96622         & FEROS          & 20/05/2012   & O9.2   & IV    & 0.40  & 0.73 \\
HD90087         & FEROS          & 04/05/2009   & O9.2   & III   & 0.47  & 0.95 \\
HD123008        & FEROS          & 14/05/2008   & O9.2   & I     & 0.51  & 0.97 \\
HD154368        & FEROS          & 02/05/2004   & O9.2   & I     & 0.42  & 0.84 \\
HD76968         & FEROS          & 24/12/2004   & O9.2   & I     & 0.37  & 0.76 \\
HD46202         & ESPaDOnS       & 02/02/2012   & O9.5   & V     & 0.49  & 0.63 \\
HD34078         & ESPaDOnS       & 27/02/2010   & O9.5   & V     & 0.50  & 0.59 \\
HD227757        & ESPaDOnS       & 26/07/2010   & O9.5   & V     & 0.52  & 0.59 \\
HD38666         & FEROS          & 14/11/2003   & O9.5   & V     & 0.48  & 0.59 \\
HD206183        & ESPaDOnS       & 23/07/2010   & O9.5   & IV/V  & 0.43  & 0.50 \\
HD166546        & FEROS          & 14/05/2008   & O9.5   & IV    & 0.36  & 0.67 \\
HD93027         & FEROS          & 02/05/2009   & O9.5   & IV    & 0.40  & 0.56 \\
HD163892        & ESPaDOnS       & 29/06/2018   & O9.5   & IV    & 0.42  & 0.77 \\
HD123056        & FEROS          & 03/05/2009   & O9.5   & IV    & 0.41  & 0.80 \\
HD167263        & FEROS          & 22/04/2005   & O9.5   & III   & 0.41  & 0.79 \\
HD52266         & FEROS          & 05/05/2009   & O9.5   & III   & 0.44  & 0.95 \\
HD152219        & FEROS          & 09/05/2004   & O9.5   & III   & 0.36  & 0.77 \\
HD36486         & ESPaDOnS       & 02/03/2016   & O9.5   & II    & 0.36  & 0.76 \\
HD188209        & ESPaDOnS       & 20/06/2016   & O9.5   & I     & 0.41  & 0.86 \\
HD54879         & ESPaDOnS       & 09/11/2014   & O9.7   & V     & 0.32  & 0.45 \\
HD207538        & ESPaDOnS       & 30/07/2008   & O9.7   & IV    & 0.29  & 0.65 \\
HD152200        & FEROS          & 09/05/2004   & O9.7   & IV    & 0.30  & 0.71 \\
HD189957        & ESPaDOnS       & 16/10/2008   & O9.7   & III   & 0.33  & 0.69 \\
HD55879         & ESPaDOnS       & 20/10/2008   & O9.7   & III   & 0.35  & 0.70 \\
HD154643        & ESPaDOnS       & 02/07/2011   & O9.7   & III   & 0.28  & 0.74 \\
HD68450         & FEROS          & 22/12/2004   & O9.7   & II    & 0.32  & 0.84 \\
HD154811        & FEROS          & 25/06/2005   & O9.7   & Ib    & 0.28  & 0.78 \\
HD167264        & FEROS          & 23/04/2005   & O9.7   & Iab   & 0.20  & 0.88 \\
HD75222         & FEROS          & 24/12/2004   & O9.7   & Iab   & 0.34  & 0.67 \\
HD152003        & FEROS          & 15/05/2008   & O9.7   & Iab   & 0.31  & 0.75 \\
HD195592        & NARVAL         & 14/09/2007   & O9.7   & Ia    & 0.30  & 0.76 \\
HD105056        & FEROS          & 02/05/2009   & O9.7   & Ia    & 0.28  & 0.38 \\
HD149438        & ESPaDOnS       & 08/02/2006   & B0    & V     & 0.22  & 0.50 \\
LS864           & FEROS          & 06/05/2012   & B0    & V     & 0.17  & 0.75 \\
HD190427        & NARVAL         & 27/11/2006   & B0    & III   & 0.39  & 0.66 \\
HD48434         & FEROS          & 07/01/2012   & B0    & III   & 0.16  & 0.88 \\
HD156134        & FEROS          & 11/05/2003   & B0    & Ib    & 0.20  & 0.85 \\
HD164402        & FEROS          & 22/04/2005   & B0    & Ib    & 0.16  & 0.76 \\
HD37128         & ESPaDOnS       & 14/10/2008   & B0    & Ia    & 0.17  & 0.77 \\
HD167756        & ESPaDOnS       & 18/06/2011   & B0    & Ia    & 0.17  & 0.78 \\
HD122879        & FEROS          & 23/04/2005   & B0    & Ia    & 0.19  & 0.92 \\
\hline                     
\end{tabular}
\end{center}
\end{table*}

\setcounter{table}{1}

\begin{table*}[ht]
\begin{center}
\caption{Continued} \label{tab_archive}
\begin{tabular}{lcccccc}
\hline
Star            &   Instrument   &  Observation   &   ST           &  LC    &    EW(5412)  &  EW(7065) \\
                &                &   date          &           &        &  [\AA]     &  [\AA] \\
\hline
HD36960         & ESPaDOnS       & 19/08/2008   & B0.5  & V     & 0.08  & 0.49 \\
HD211880        & NARVAL         & 10/11/2017   & B0.5  & V     & 0.14  & 0.43 \\
HD34816         & FEROS          & 10/01/2007   & B0.5  & IV    & 0.09  & 0.50 \\
HD194092        & NARVAL         & 13/10/2017   & B0.5  & III   & 0.06  & 0.44 \\
HD2619          & NARVAL         & 07/11/2017   & B0.5  & III   & 0.10  & 0.55 \\
HD191396        & NARVAL         & 11/10/2017   & B0.5  & II    & 0.08  & 0.66 \\
\hline                     
\end{tabular}
\end{center}
\end{table*}

\end{appendix}

\bibliographystyle{aa}
\bibliography{evol_lowZ}

\begin{thebibliography}{128}
\expandafter\ifx\csname natexlab\endcsname\relax\def\natexlab#1{#1}\fi

\bibitem[{{Abbott} \& {Hummer}(1985)}]{ah85}
{Abbott}, D.~C. \& {Hummer}, D.~G. 1985, \apj, 294, 286

\bibitem[{{Allen} {et~al.}(2008){Allen}, {Groves}, {Dopita}, {Sutherland}, \&
  {Kewley}}]{allen08}
{Allen}, M.~G., {Groves}, B.~A., {Dopita}, M.~A., {Sutherland}, R.~S., \&
  {Kewley}, L.~J. 2008, \apjs, 178, 20

\bibitem[{{Amard} {et~al.}(2016){Amard}, {Palacios}, {Charbonnel}, {Gallet}, \&
  {Bouvier}}]{amard16}
{Amard}, L., {Palacios}, A., {Charbonnel}, C., {Gallet}, F., \& {Bouvier}, J.
  2016, \aap, 587, A105

\bibitem[{{Angulo} {et~al.}(1999){Angulo}, {Arnould}, {Rayet}, {Descouvemont},
  {Baye}, {Leclercq-Willain}, {Coc}, {Barhoumi}, {Aguer}, {Rolfs}, {Kunz},
  {Hammer}, {Mayer}, {Paradellis}, {Kossionides}, {Chronidou}, {Spyrou},
  {degl'Innocenti}, {Fiorentini}, {Ricci}, {Zavatarelli}, {Providencia},
  {Wolters}, {Soares}, {Grama}, {Rahighi}, {Shotter}, \& {Lamehi
  Rachti}}]{nacre}
{Angulo}, C., {Arnould}, M., {Rayet}, M., {et~al.} 1999, \nphysa, 656, 3

\bibitem[{{Asplund} {et~al.}(2009){Asplund}, {Grevesse}, {Sauval}, \&
  {Scott}}]{asplund09}
{Asplund}, M., {Grevesse}, N., {Sauval}, A.~J., \& {Scott}, P. 2009, \araa, 47,
  481

\bibitem[{{Bj{\"o}rklund} {et~al.}(2020){Bj{\"o}rklund}, {Sundqvist}, {Puls},
  \& {Najarro}}]{bjorklund20}
{Bj{\"o}rklund}, R., {Sundqvist}, J.~O., {Puls}, J., \& {Najarro}, F. 2020,
  arXiv e-prints, arXiv:2008.06066

\bibitem[{{Bouret} {et~al.}(2003){Bouret}, {Lanz}, {Hillier}, {Heap}, {Hubeny},
  {Lennon}, {Smith}, \& {Evans}}]{jc03}
{Bouret}, J.~C., {Lanz}, T., {Hillier}, D.~J., {et~al.} 2003, \apj, 595, 1182

\bibitem[{{Bouret} {et~al.}(2015){Bouret}, {Lanz}, {Hillier}, {Martins},
  {Marcolino}, \& {Depagne}}]{jc15}
{Bouret}, J.~C., {Lanz}, T., {Hillier}, D.~J., {et~al.} 2015, \mnras, 449, 1545

\bibitem[{{Bouret} {et~al.}(2013){Bouret}, {Lanz}, {Martins}, {Marcolino},
  {Hillier}, {Depagne}, \& {Hubeny}}]{jc13}
{Bouret}, J.~C., {Lanz}, T., {Martins}, F., {et~al.} 2013, \aap, 555, A1

\bibitem[{{Bresolin} {et~al.}(2007){Bresolin}, {Urbaneja}, {Gieren},
  {Pietrzy{\'n}ski}, \& {Kudritzki}}]{bresolin07}
{Bresolin}, F., {Urbaneja}, M.~A., {Gieren}, W., {Pietrzy{\'n}ski}, G., \&
  {Kudritzki}, R.-P. 2007, \apj, 671, 2028

\bibitem[{{Brinchmann} {et~al.}(2008){Brinchmann}, {Kunth}, \&
  {Durret}}]{bri08}
{Brinchmann}, J., {Kunth}, D., \& {Durret}, F. 2008, \aap, 485, 657

\bibitem[{{Brott} {et~al.}(2011){Brott}, {de Mink}, {Cantiello}, {Langer}, {de
  Koter}, {Evans}, {Hunter}, {Trundle}, \& {Vink}}]{brott11}
{Brott}, I., {de Mink}, S.~E., {Cantiello}, M., {et~al.} 2011, \aap, 530, A115

\bibitem[{{Brown} {et~al.}(2002){Brown}, {Heap}, {Hubeny}, {Lanz}, \&
  {Lindler}}]{brown02}
{Brown}, T.~M., {Heap}, S.~R., {Hubeny}, I., {Lanz}, T., \& {Lindler}, D. 2002,
  \apjl, 579, L75

\bibitem[{{Camacho} {et~al.}(2016){Camacho}, {Garcia}, {Herrero}, \&
  {Sim{\'o}n-D{\'\i}az}}]{camacho16}
{Camacho}, I., {Garcia}, M., {Herrero}, A., \& {Sim{\'o}n-D{\'\i}az}, S. 2016,
  \aap, 585, A82

\bibitem[{{Cassata} {et~al.}(2013){Cassata}, {Le F{\`e}vre}, {Charlot},
  {Contini}, {Cucciati}, {Garilli}, {Zamorani}, {Adami}, {Bardelli}, {Le Brun},
  {Lemaux}, {Maccagni}, {Pollo}, {Pozzetti}, {Tresse}, {Vergani}, {Zanichelli},
  \& {Zucca}}]{cassata13}
{Cassata}, P., {Le F{\`e}vre}, O., {Charlot}, S., {et~al.} 2013, \aap, 556, A68

\bibitem[{{Castro} {et~al.}(2018){Castro}, {Oey}, {Fossati}, \&
  {Langer}}]{castro18}
{Castro}, N., {Oey}, M.~S., {Fossati}, L., \& {Langer}, N. 2018, \apj, 868, 57

\bibitem[{{Chandar} {et~al.}(2004){Chandar}, {Leitherer}, \&
  {Tremonti}}]{chandar04}
{Chandar}, R., {Leitherer}, C., \& {Tremonti}, C.~A. 2004, \apj, 604, 153

\bibitem[{{Chiosi} \& {Maeder}(1986)}]{cm86}
{Chiosi}, C. \& {Maeder}, A. 1986, \araa, 24, 329

\bibitem[{{Coc} {et~al.}(2004){Coc}, {Vangioni-Flam}, {Descouvemont},
  {Adahchour}, \& {Angulo}}]{Cocetal04}
{Coc}, A., {Vangioni-Flam}, E., {Descouvemont}, P., {Adahchour}, A., \&
  {Angulo}, C. 2004, \apj, 600, 544

\bibitem[{{Cohen} {et~al.}(2014){Cohen}, {Wollman}, {Leutenegger}, {Sundqvist},
  {Fullerton}, {Zsarg{\'o}}, \& {Owocki}}]{cohen14}
{Cohen}, D.~H., {Wollman}, E.~E., {Leutenegger}, M.~A., {et~al.} 2014, \mnras,
  439, 908

\bibitem[{{Conti} \& {Alschuler}(1971)}]{ca71}
{Conti}, P.~S. \& {Alschuler}, W.~R. 1971, \apj, 170, 325

\bibitem[{{Crowther}(2019)}]{paul19}
{Crowther}, P.~A. 2019, Galaxies, 7, 88

\bibitem[{{Crowther} {et~al.}(2016){Crowther}, {Caballero-Nieves}, {Bostroem},
  {Ma{\'\i}z Apell{\'a}niz}, {Schneider}, {Walborn}, {Angus}, {Brott},
  {Bonanos}, {de Koter}, {de Mink}, {Evans}, {Gr{\"a}fener}, {Herrero},
  {Howarth}, {Langer}, {Lennon}, {Puls}, {Sana}, \& {Vink}}]{paul16}
{Crowther}, P.~A., {Caballero-Nieves}, S.~M., {Bostroem}, K.~A., {et~al.} 2016,
  \mnras, 458, 624

\bibitem[{{de Mink} {et~al.}(2013){de Mink}, {Langer}, {Izzard}, {Sana}, \& {de
  Koter}}]{demink13}
{de Mink}, S.~E., {Langer}, N., {Izzard}, R.~G., {Sana}, H., \& {de Koter}, A.
  2013, \apj, 764, 166

\bibitem[{{Decressin} {et~al.}(2009){Decressin}, {Mathis}, {Palacios}, {Siess},
  {Talon}, {Charbonnel}, \& {Zahn}}]{dmp09}
{Decressin}, T., {Mathis}, S., {Palacios}, A., {et~al.} 2009, \aap, 495, 271

\bibitem[{{Donati} {et~al.}(1997){Donati}, {Semel}, {Carter}, {Rees}, \&
  {Collier Cameron}}]{polarbase1}
{Donati}, J.~F., {Semel}, M., {Carter}, B.~D., {Rees}, D.~E., \& {Collier
  Cameron}, A. 1997, \mnras, 291, 658

\bibitem[{{Dorn-Wallenstein} \& {Levesque}(2018)}]{dw18}
{Dorn-Wallenstein}, T.~Z. \& {Levesque}, E.~M. 2018, \apj, 867, 125

\bibitem[{{Dufton} {et~al.}(2019){Dufton}, {Evans}, {Hunter}, {Lennon}, \&
  {Schneider}}]{dufton19}
{Dufton}, P.~L., {Evans}, C.~J., {Hunter}, I., {Lennon}, D.~J., \& {Schneider},
  F.~R.~N. 2019, \aap, 626, A50

\bibitem[{{Evans} {et~al.}(2007){Evans}, {Bresolin}, {Urbaneja},
  {Pietrzy{\'n}ski}, {Gieren}, \& {Kudritzki}}]{evans07}
{Evans}, C.~J., {Bresolin}, F., {Urbaneja}, M.~A., {et~al.} 2007, \apj, 659,
  1198

\bibitem[{{Evans} {et~al.}(2019){Evans}, {Castro}, {Gonzalez}, {Garcia},
  {Bastian}, {Cioni}, {Clark}, {Davies}, {Ferguson}, {Kamann}, {Lennon},
  {Patrick}, {Vink}, \& {Weisz}}]{evans19}
{Evans}, C.~J., {Castro}, N., {Gonzalez}, O.~A., {et~al.} 2019, \aap, 622, A129

\bibitem[{{Fullerton}(2011)}]{fullerton11}
{Fullerton}, A.~W. 2011, in IAU Symposium, Vol. 272, Active OB Stars:
  Structure, Evolution, Mass Loss, and Critical Limits, ed. C.~{Neiner},
  G.~{Wade}, G.~{Meynet}, \& G.~{Peters}, 136--147

\bibitem[{{Gabler} {et~al.}(1992){Gabler}, {Gabler}, {Kudritzki}, \&
  {Mendez}}]{gabler92}
{Gabler}, R., {Gabler}, A., {Kudritzki}, R.~P., \& {Mendez}, R.~H. 1992, \aap,
  265, 656

\bibitem[{{Gabler} {et~al.}(1989){Gabler}, {Gabler}, {Kudritzki}, {Puls}, \&
  {Pauldrach}}]{gabler89}
{Gabler}, R., {Gabler}, A., {Kudritzki}, R.~P., {Puls}, J., \& {Pauldrach}, A.
  1989, \aap, 226, 162

\bibitem[{{Garcia}(2018)}]{garcia18}
{Garcia}, M. 2018, \mnras, 474, L66

\bibitem[{{Garcia} \& {Herrero}(2013)}]{garcia13}
{Garcia}, M. \& {Herrero}, A. 2013, \aap, 551, A74

\bibitem[{{Garcia} {et~al.}(2017){Garcia}, {Herrero}, {Najarro}, {Camacho},
  {Lennon}, {Urbaneja}, \& {Castro}}]{garcia17}
{Garcia}, M., {Herrero}, A., {Najarro}, F., {et~al.} 2017, in IAU Symposium,
  Vol. 329, The Lives and Death-Throes of Massive Stars, ed. J.~J. {Eldridge},
  J.~C. {Bray}, L.~A.~S. {McClelland}, \& L.~{Xiao}, 313--321

\bibitem[{{Garcia} {et~al.}(2019){Garcia}, {Herrero}, {Najarro}, {Camacho}, \&
  {Lorenzo}}]{garcia19}
{Garcia}, M., {Herrero}, A., {Najarro}, F., {Camacho}, I., \& {Lorenzo}, M.
  2019, \mnras, 484, 422

\bibitem[{{Garcia} {et~al.}(2014){Garcia}, {Herrero}, {Najarro}, {Lennon}, \&
  {Alejandro Urbaneja}}]{garcia14}
{Garcia}, M., {Herrero}, A., {Najarro}, F., {Lennon}, D.~J., \& {Alejandro
  Urbaneja}, M. 2014, \apj, 788, 64

\bibitem[{{G{\"o}tberg} {et~al.}(2017){G{\"o}tberg}, {de Mink}, \&
  {Groh}}]{got17}
{G{\"o}tberg}, Y., {de Mink}, S.~E., \& {Groh}, J.~H. 2017, \aap, 608, A11

\bibitem[{{G{\"o}tberg} {et~al.}(2018){G{\"o}tberg}, {de Mink}, {Groh},
  {Kupfer}, {Crowther}, {Zapartas}, \& {Renzo}}]{got18}
{G{\"o}tberg}, Y., {de Mink}, S.~E., {Groh}, J.~H., {et~al.} 2018, \aap, 615,
  A78

\bibitem[{{Gr{\"a}fener} \& {Vink}(2015)}]{gv15}
{Gr{\"a}fener}, G. \& {Vink}, J.~S. 2015, \aap, 578, L2

\bibitem[{{Groh} {et~al.}(2019){Groh}, {Ekstr{\"o}m}, {Georgy}, {Meynet},
  {Choplin}, {Eggenberger}, {Hirschi}, {Maeder}, {Murphy}, {Boian}, \&
  {Farrell}}]{groh19}
{Groh}, J.~H., {Ekstr{\"o}m}, S., {Georgy}, C., {et~al.} 2019, \aap, 627, A24

\bibitem[{{Groh} {et~al.}(2013){Groh}, {Meynet}, \& {Ekstr{\"o}m}}]{groh13}
{Groh}, J.~H., {Meynet}, G., \& {Ekstr{\"o}m}, S. 2013, \aap, 550, L7

\bibitem[{{Groh} {et~al.}(2014){Groh}, {Meynet}, {Ekstr{\"o}m}, \&
  {Georgy}}]{groh14}
{Groh}, J.~H., {Meynet}, G., {Ekstr{\"o}m}, S., \& {Georgy}, C. 2014, \aap,
  564, A30

\bibitem[{{Grunhut} {et~al.}(2017){Grunhut}, {Wade}, {Neiner}, {Oksala},
  {Petit}, {Alecian}, {Bohlender}, {Bouret}, {Henrichs}, {Hussain},
  {Kochukhov}, \& {MiMeS Collaboration}}]{grunhut17}
{Grunhut}, J.~H., {Wade}, G.~A., {Neiner}, C., {et~al.} 2017, \mnras, 465, 2432

\bibitem[{{Hainich} {et~al.}(2019){Hainich}, {Ramachandran}, {Shenar}, {Sand
  er}, {Todt}, {Gruner}, {Oskinova}, \& {Hamann}}]{hainich19}
{Hainich}, R., {Ramachandran}, V., {Shenar}, T., {et~al.} 2019, \aap, 621, A85

\bibitem[{{Herrero} {et~al.}(1992){Herrero}, {Kudritzki}, {Vilchez}, {Kunze},
  {Butler}, \& {Haser}}]{herrero92}
{Herrero}, A., {Kudritzki}, R.~P., {Vilchez}, J.~M., {et~al.} 1992, \aap, 261,
  209

\bibitem[{{Hillier} \& {Miller}(1998)}]{hm98}
{Hillier}, D.~J. \& {Miller}, D.~L. 1998, \apj, 496, 407

\bibitem[{{Hosek} {et~al.}(2014){Hosek}, {Kudritzki}, {Bresolin}, {Urbaneja},
  {Evans}, {Pietrzy{\'n}ski}, {Gieren}, {Przybilla}, \& {Carraro}}]{hosek14}
{Hosek}, Matthew~W., J., {Kudritzki}, R.-P., {Bresolin}, F., {et~al.} 2014,
  \apj, 785, 151

\bibitem[{{Iben}(1966)}]{iben1966}
{Iben}, Icko, J. 1966, \apj, 143, 516

\bibitem[{{Iliadis} {et~al.}(2001){Iliadis}, {D'Auria}, {Starrfield},
  {Thompson}, \& {Wiescher}}]{iliadis}
{Iliadis}, C., {D'Auria}, J.~M., {Starrfield}, S., {Thompson}, W.~J., \&
  {Wiescher}, M. 2001, \apjs, 134, 151

\bibitem[{{Izotov} {et~al.}(1999){Izotov}, {Chaffee}, {Foltz}, {Green},
  {Guseva}, \& {Thuan}}]{izotov99}
{Izotov}, Y.~I., {Chaffee}, F.~H., {Foltz}, C.~B., {et~al.} 1999, \apj, 527,
  757

\bibitem[{{Izotov} {et~al.}(1997){Izotov}, {Foltz}, {Green}, {Guseva}, \&
  {Thuan}}]{izotov97}
{Izotov}, Y.~I., {Foltz}, C.~B., {Green}, R.~F., {Guseva}, N.~G., \& {Thuan},
  T.~X. 1997, \apjl, 487, L37

\bibitem[{{Japelj} {et~al.}(2018){Japelj}, {Vergani}, {Salvaterra}, {Renzo},
  {Zapartas}, {de Mink}, {Kaper}, \& {Zibetti}}]{japelj18}
{Japelj}, J., {Vergani}, S.~D., {Salvaterra}, R., {et~al.} 2018, \aap, 617,
  A105

\bibitem[{{Kehrig} {et~al.}(2018){Kehrig}, {V{\'\i}lchez}, {Guerrero},
  {Iglesias-P{\'a}ramo}, {Hunt}, {Duarte-Puertas}, \&
  {Ramos-Larios}}]{kehrig18}
{Kehrig}, C., {V{\'\i}lchez}, J.~M., {Guerrero}, M.~A., {et~al.} 2018, \mnras,
  480, 1081

\bibitem[{{Kehrig} {et~al.}(2015){Kehrig}, {V{\'\i}lchez}, {P{\'e}rez-Montero},
  {Iglesias-P{\'a}ramo}, {Brinchmann}, {Kunth}, {Durret}, \& {Bayo}}]{kehrig15}
{Kehrig}, C., {V{\'\i}lchez}, J.~M., {P{\'e}rez-Montero}, E., {et~al.} 2015,
  \apjl, 801, L28

\bibitem[{{Keszthelyi} {et~al.}(2019){Keszthelyi}, {Meynet}, {Georgy}, {Wade},
  {Petit}, \& {David-Uraz}}]{zsolt19}
{Keszthelyi}, Z., {Meynet}, G., {Georgy}, C., {et~al.} 2019, \mnras, 485, 5843

\bibitem[{{Kippenhahn} {et~al.}(2012){Kippenhahn}, {Weigert}, \&
  {Weiss}}]{kww2012}
{Kippenhahn}, R., {Weigert}, A., \& {Weiss}, A. 2012, {Stellar Structure and
  Evolution}

\bibitem[{{Kobulnicky} {et~al.}(2014){Kobulnicky}, {Kiminki}, {Lundquist},
  {Burke}, {Chapman}, {Keller}, {Lester}, {Rolen}, {Topel}, {Bhattacharjee},
  {Smullen}, {Vargas {\'A}lvarez}, {Runnoe}, {Dale}, \& {Brotherton}}]{kobul14}
{Kobulnicky}, H.~A., {Kiminki}, D.~C., {Lundquist}, M.~J., {et~al.} 2014,
  \apjs, 213, 34

\bibitem[{{Krti{\v{c}}ka} \& {Kub{\'a}t}(2017)}]{kk17}
{Krti{\v{c}}ka}, J. \& {Kub{\'a}t}, J. 2017, \aap, 606, A31

\bibitem[{{Kub{\'a}tov{\'a}} {et~al.}(2019){Kub{\'a}tov{\'a}}, {Sz{\'e}csi},
  {Sander}, {Kub{\'a}t}, {Tramper}, {Krti{\v{c}}ka}, {Kehrig}, {Hamann},
  {Hainich}, \& {Shenar}}]{kub19}
{Kub{\'a}tov{\'a}}, B., {Sz{\'e}csi}, D., {Sander}, A.~A.~C., {et~al.} 2019,
  \aap, 623, A8

\bibitem[{{Langer} \& {Kudritzki}(2014)}]{lp14}
{Langer}, N. \& {Kudritzki}, R.~P. 2014, \aap, 564, A52

\bibitem[{{Lanz} \& {Hubeny}(2003)}]{lh03}
{Lanz}, T. \& {Hubeny}, I. 2003, \apjs, 146, 417

\bibitem[{{Leitherer} {et~al.}(2018){Leitherer}, {Byler}, {Lee}, \&
  {Levesque}}]{leitherer18}
{Leitherer}, C., {Byler}, N., {Lee}, J.~C., \& {Levesque}, E.~M. 2018, \apj,
  865, 55

\bibitem[{{Leitherer} {et~al.}(1992){Leitherer}, {Robert}, \&
  {Drissen}}]{leitherer92}
{Leitherer}, C., {Robert}, C., \& {Drissen}, L. 1992, \apj, 401, 596

\bibitem[{{Longland} {et~al.}(2010){Longland}, {Iliadis}, {Champagne},
  {Newton}, {Ugalde}, {Coc}, \& {Fitzgerald}}]{longland}
{Longland}, R., {Iliadis}, C., {Champagne}, A.~E., {et~al.} 2010, \nphysa, 841,
  1

\bibitem[{{Lucy}(2010)}]{lucy10}
{Lucy}, L.~B. 2010, \aap, 524, A41

\bibitem[{{Maeder}(1987)}]{maeder87}
{Maeder}, A. 1987, \aap, 178, 159

\bibitem[{{Maeder} \& {Meynet}(2000)}]{mm00}
{Maeder}, A. \& {Meynet}, G. 2000, \araa, 38, 143

\bibitem[{{Maeder} \& {Meynet}(2001)}]{mm01}
{Maeder}, A. \& {Meynet}, G. 2001, \aap, 373, 555

\bibitem[{{Mahy} {et~al.}(2020){Mahy}, {Sana}, {Abdul-Masih}, {Almeida},
  {Langer}, {Shenar}, {de Koter}, {de Mink}, {de Wit}, {Grin}, {Evans},
  {Moffat}, {Schneider}, {Barb{\'a}}, {Clark}, {Crowther}, {Gr{\"a}fener},
  {Lennon}, {Tramper}, \& {Vink}}]{mahy20}
{Mahy}, L., {Sana}, H., {Abdul-Masih}, M., {et~al.} 2020, \aap, 634, A118

\bibitem[{{Ma{\'\i}z Apell{\'a}niz} {et~al.}(2013){Ma{\'\i}z Apell{\'a}niz},
  {Sota}, {Morrell}, {Barb{\'a}}, {Walborn}, {Alfaro}, {Gamen}, {Arias}, \&
  {Gallego Calvente}}]{goss}
{Ma{\'\i}z Apell{\'a}niz}, J., {Sota}, A., {Morrell}, N.~I., {et~al.} 2013, in
  Massive Stars: From alpha to Omega, 198

\bibitem[{{Ma{\'\i}z Apell{\'a}niz} {et~al.}(2011){Ma{\'\i}z Apell{\'a}niz},
  {Sota}, {Walborn}, {Alfaro}, {Barb{\'a}}, {Morrell}, {Gamen}, \&
  {Arias}}]{gosss}
{Ma{\'\i}z Apell{\'a}niz}, J., {Sota}, A., {Walborn}, N.~R., {et~al.} 2011, in
  Highlights of Spanish Astrophysics VI, ed. M.~R. {Zapatero Osorio},
  J.~{Gorgas}, J.~{Ma{\'\i}z Apell{\'a}niz}, J.~R. {Pardo}, \& A.~{Gil de Paz},
  467--472

\bibitem[{{Marchenko} {et~al.}(2007){Marchenko}, {Foellmi}, {Moffat},
  {Martins}, {Bouret}, \& {Depagne}}]{march07}
{Marchenko}, S.~V., {Foellmi}, C., {Moffat}, A.~F.~J., {et~al.} 2007, \apjl,
  656, L77

\bibitem[{{Markova} {et~al.}(2018){Markova}, {Puls}, \& {Langer}}]{markova18}
{Markova}, N., {Puls}, J., \& {Langer}, N. 2018, \aap, 613, A12

\bibitem[{{Martins}(2011)}]{martins11}
{Martins}, F. 2011, Bulletin de la Societe Royale des Sciences de Liege, 80, 29

\bibitem[{{Martins}(2018)}]{classif}
{Martins}, F. 2018, \aap, 616, A135

\bibitem[{{Martins} {et~al.}(2015){Martins}, {Herv{\'e}}, {Bouret},
  {Marcolino}, {Wade}, {Neiner}, {Alecian}, {Grunhut}, \& {Petit}}]{mimesO}
{Martins}, F., {Herv{\'e}}, A., {Bouret}, J.-C., {et~al.} 2015, \aap, 575, A34

\bibitem[{{Martins} {et~al.}(2009){Martins}, {Hillier}, {Bouret}, {Depagne},
  {Foellmi}, {Marchenko}, \& {Moffat}}]{martins09}
{Martins}, F., {Hillier}, D.~J., {Bouret}, J.~C., {et~al.} 2009, \aap, 495, 257

\bibitem[{{Martins} \& {Palacios}(2013)}]{mp13}
{Martins}, F. \& {Palacios}, A. 2013, \aap, 560, A16

\bibitem[{{Martins} \& {Palacios}(2017)}]{mp17}
{Martins}, F. \& {Palacios}, A. 2017, \aap, 598, A56

\bibitem[{{Martins} {et~al.}(2002){Martins}, {Schaerer}, \&
  {Hillier}}]{martins02}
{Martins}, F., {Schaerer}, D., \& {Hillier}, D.~J. 2002, \aap, 382, 999

\bibitem[{{Mathys}(1988)}]{mathys88}
{Mathys}, G. 1988, \aaps, 76, 427

\bibitem[{{McConnachie} {et~al.}(2005){McConnachie}, {Irwin}, {Ferguson},
  {Ibata}, {Lewis}, \& {Tanvir}}]{mcco05}
{McConnachie}, A.~W., {Irwin}, M.~J., {Ferguson}, A.~M.~N., {et~al.} 2005,
  \mnras, 356, 979

\bibitem[{{Melena} {et~al.}(2008){Melena}, {Massey}, {Morrell}, \&
  {Zangari}}]{melena08}
{Melena}, N.~W., {Massey}, P., {Morrell}, N.~I., \& {Zangari}, A.~M. 2008, \aj,
  135, 878

\bibitem[{{Moe} \& {Di Stefano}(2013)}]{moedistef13}
{Moe}, M. \& {Di Stefano}, R. 2013, \apj, 778, 95

\bibitem[{{Mokiem} {et~al.}(2007{\natexlab{a}}){Mokiem}, {de Koter}, {Evans},
  {Puls}, {Smartt}, {Crowther}, {Herrero}, {Langer}, {Lennon}, {Najarro},
  {Villamariz}, \& {Vink}}]{mokiem07a}
{Mokiem}, M.~R., {de Koter}, A., {Evans}, C.~J., {et~al.} 2007{\natexlab{a}},
  \aap, 465, 1003

\bibitem[{{Mokiem} {et~al.}(2006){Mokiem}, {de Koter}, {Evans}, {Puls},
  {Smartt}, {Crowther}, {Herrero}, {Langer}, {Lennon}, {Najarro}, {Villamariz},
  \& {Yoon}}]{mokiem06}
{Mokiem}, M.~R., {de Koter}, A., {Evans}, C.~J., {et~al.} 2006, \aap, 456, 1131

\bibitem[{{Mokiem} {et~al.}(2007{\natexlab{b}}){Mokiem}, {de Koter}, {Vink},
  {Puls}, {Evans}, {Smartt}, {Crowther}, {Herrero}, {Langer}, {Lennon},
  {Najarro}, \& {Villamariz}}]{mokiem07}
{Mokiem}, M.~R., {de Koter}, A., {Vink}, J.~S., {et~al.} 2007{\natexlab{b}},
  \aap, 473, 603

\bibitem[{{Nanayakkara} {et~al.}(2019){Nanayakkara}, {Brinchmann}, {Boogaard},
  {Bouwens}, {Cantalupo}, {Feltre}, {Kollatschny}, {Marino}, {Maseda},
  {Matthee}, {Paalvast}, {Richard}, \& {Verhamme}}]{nanaya19}
{Nanayakkara}, T., {Brinchmann}, J., {Boogaard}, L., {et~al.} 2019, \aap, 624,
  A89

\bibitem[{{Negueruela} {et~al.}(2010){Negueruela}, {Clark}, \&
  {Ritchie}}]{negue10}
{Negueruela}, I., {Clark}, J.~S., \& {Ritchie}, B.~W. 2010, \aap, 516, A78

\bibitem[{{Palacios} {et~al.}(2010){Palacios}, {Gebran}, {Josselin}, {Martins},
  {Plez}, {Belmas}, \& {L{\`e}bre}}]{pollux10}
{Palacios}, A., {Gebran}, M., {Josselin}, E., {et~al.} 2010, \aap, 516, A13

\bibitem[{{Palmerio} {et~al.}(2019){Palmerio}, {Vergani}, {Salvaterra}, {Sand
  ers}, {Japelj}, {Vidal-Garc{\'\i}a}, {D'Avanzo}, {Corre}, {Perley},
  {Shapley}, {Boissier}, {Greiner}, {Le Floc'h}, \& {Wiseman}}]{palmeiro19}
{Palmerio}, J.~T., {Vergani}, S.~D., {Salvaterra}, R., {et~al.} 2019, \aap,
  623, A26

\bibitem[{{Patr{\'\i}cio} {et~al.}(2016){Patr{\'\i}cio}, {Richard}, {Verhamme},
  {Wisotzki}, {Brinchmann}, {Turner}, {Christensen}, {Weilbacher}, {Blaizot},
  {Bacon}, {Contini}, {Lagattuta}, {Cantalupo}, {Cl{\'e}ment}, \&
  {Soucail}}]{patricio16}
{Patr{\'\i}cio}, V., {Richard}, J., {Verhamme}, A., {et~al.} 2016, \mnras, 456,
  4191

\bibitem[{{Perley} {et~al.}(2016){Perley}, {Tanvir}, {Hjorth}, {Laskar},
  {Berger}, {Chary}, {de Ugarte Postigo}, {Fynbo}, {Kr{\"u}hler}, {Levan},
  {Micha{\l}owski}, \& {Schulze}}]{perley16}
{Perley}, D.~A., {Tanvir}, N.~R., {Hjorth}, J., {et~al.} 2016, \apj, 817, 8

\bibitem[{{Petit} {et~al.}(2014){Petit}, {Louge}, {Th{\'e}ado}, {Paletou},
  {Manset}, {Morin}, {Marsden}, \& {Jeffers}}]{polarbase2}
{Petit}, P., {Louge}, T., {Th{\'e}ado}, S., {et~al.} 2014, \pasp, 126, 469

\bibitem[{{Raghavan} {et~al.}(2010){Raghavan}, {McAlister}, {Henry}, {Latham},
  {Marcy}, {Mason}, {Gies}, {White}, \& {ten Brummelaar}}]{rag10}
{Raghavan}, D., {McAlister}, H.~A., {Henry}, T.~J., {et~al.} 2010, \apjs, 190,
  1

\bibitem[{{Ramachandran} {et~al.}(2019){Ramachandran}, {Hamann}, {Oskinova},
  {Gallagher}, {Hainich}, {Shenar}, {Sand er}, {Todt}, \& {Fulmer}}]{rama19}
{Ramachandran}, V., {Hamann}, W.~R., {Oskinova}, L.~M., {et~al.} 2019, \aap,
  625, A104

\bibitem[{{Repolust} {et~al.}(2004){Repolust}, {Puls}, \&
  {Herrero}}]{repolust04}
{Repolust}, T., {Puls}, J., \& {Herrero}, A. 2004, \aap, 415, 349

\bibitem[{{Rivero Gonz{\'a}lez} {et~al.}(2012){Rivero Gonz{\'a}lez}, {Puls},
  {Massey}, \& {Najarro}}]{rg12}
{Rivero Gonz{\'a}lez}, J.~G., {Puls}, J., {Massey}, P., \& {Najarro}, F. 2012,
  \aap, 543, A95

\bibitem[{{Ross} {et~al.}(2015){Ross}, {Holtzman}, {Saha}, \&
  {Anthony-Twarog}}]{ross15}
{Ross}, T.~L., {Holtzman}, J., {Saha}, A., \& {Anthony-Twarog}, B.~J. 2015,
  \aj, 149, 198

\bibitem[{{Sakashita} {et~al.}(1959){Sakashita}, {{\^O}no}, \&
  {Hayashi}}]{soh1959}
{Sakashita}, S., {{\^O}no}, Y., \& {Hayashi}, C. 1959, Progress of Theoretical
  Physics, 21, 315

\bibitem[{{Sander} {et~al.}(2015){Sander}, {Shenar}, {Hainich},
  {G{\'\i}menez-Garc{\'\i}a}, {Todt}, \& {Hamann}}]{sander15}
{Sander}, A., {Shenar}, T., {Hainich}, R., {et~al.} 2015, \aap, 577, A13

\bibitem[{{Sander} {et~al.}(2017){Sander}, {Hamann}, {Todt}, {Hainich}, \&
  {Shenar}}]{sander17}
{Sander}, A.~A.~C., {Hamann}, W.~R., {Todt}, H., {Hainich}, R., \& {Shenar}, T.
  2017, \aap, 603, A86

\bibitem[{{Sander} {et~al.}(2020){Sander}, {Vink}, \& {Hamann}}]{sander20}
{Sander}, A. A.~C., {Vink}, J.~S., \& {Hamann}, W.~R. 2020, \mnras, 491, 4406

\bibitem[{{Saxena} {et~al.}(2019){Saxena}, {Pentericci}, {Mirabelli},
  {Schaerer}, {Schneider}, {Cullen}, {Amorin}, {Bolzonella}, {Bongiorno},
  {Carnall}, {Castellano}, {Cucciati}, {Fontana}, {Fynbo}, {Garilli},
  {Gargiulo}, {Guaita}, {Hathi}, {Hutchison}, {Marchi}, {McLeod}, {McLure},
  {Papovich}, {Pozzetti}, {Talia}, \& {Zamorani}}]{saxena19}
{Saxena}, A., {Pentericci}, L., {Mirabelli}, M., {et~al.} 2019, arXiv e-prints,
  arXiv:1911.09999

\bibitem[{{Schaerer}(2003)}]{schaerer03}
{Schaerer}, D. 2003, \aap, 397, 527

\bibitem[{{Schaerer} \& {de Koter}(1997)}]{costar3}
{Schaerer}, D. \& {de Koter}, A. 1997, \aap, 322, 598

\bibitem[{{Schaerer} {et~al.}(1996){Schaerer}, {de Koter}, {Schmutz}, \&
  {Maeder}}]{costar1}
{Schaerer}, D., {de Koter}, A., {Schmutz}, W., \& {Maeder}, A. 1996, \aap, 310,
  837

\bibitem[{{Schaerer} {et~al.}(2019){Schaerer}, {Fragos}, \&
  {Izotov}}]{schaerer19}
{Schaerer}, D., {Fragos}, T., \& {Izotov}, Y.~I. 2019, \aap, 622, L10

\bibitem[{{Schmutz} \& {Hamann}(1986)}]{schmutzhamann86}
{Schmutz}, W. \& {Hamann}, W.~R. 1986, \aap, 166, L11

\bibitem[{{Senchyna} {et~al.}(2017){Senchyna}, {Stark}, {Vidal-Garc{\'\i}a},
  {Chevallard}, {Charlot}, {Mainali}, {Jones}, {Wofford}, {Feltre}, \&
  {Gutkin}}]{senchyna17}
{Senchyna}, P., {Stark}, D.~P., {Vidal-Garc{\'\i}a}, A., {et~al.} 2017, \mnras,
  472, 2608

\bibitem[{{Sim{\'o}n-D{\'\i}az}(2020)}]{sergio20}
{Sim{\'o}n-D{\'\i}az}, S. 2020, arXiv e-prints, arXiv:2001.04853

\bibitem[{{Sota} {et~al.}(2014){Sota}, {Ma{\'{\i}}z Apell{\'a}niz}, {Morrell},
  {Barb{\'a}}, {Walborn}, {Gamen}, {Arias}, \& {Alfaro}}]{sota14}
{Sota}, A., {Ma{\'{\i}}z Apell{\'a}niz}, J., {Morrell}, N.~I., {et~al.} 2014,
  \apjs, 211, 10

\bibitem[{{Sota} {et~al.}(2011){Sota}, {Ma{\'{\i}}z Apell{\'a}niz}, {Walborn},
  {Alfaro}, {Barb{\'a}}, {Morrell}, {Gamen}, \& {Arias}}]{sota11}
{Sota}, A., {Ma{\'{\i}}z Apell{\'a}niz}, J., {Walborn}, N.~R., {et~al.} 2011,
  \apjs, 193, 24

\bibitem[{{Stanway} {et~al.}(2020){Stanway}, {Chrimes}, {Eldridge}, \&
  {Stevance}}]{stanway20}
{Stanway}, E.~R., {Chrimes}, A.~A., {Eldridge}, J.~J., \& {Stevance}, H.~F.
  2020, \mnras [\eprint[arXiv]{2004.11913}]

\bibitem[{{Steidel} {et~al.}(2016){Steidel}, {Strom}, {Pettini}, {Rudie},
  {Reddy}, \& {Trainor}}]{steidel16}
{Steidel}, C.~C., {Strom}, A.~L., {Pettini}, M., {et~al.} 2016, \apj, 826, 159

\bibitem[{{Sz{\'e}csi} {et~al.}(2015){Sz{\'e}csi}, {Langer}, {Yoon}, {Sanyal},
  {de Mink}, {Evans}, \& {Dermine}}]{szecsi15}
{Sz{\'e}csi}, D., {Langer}, N., {Yoon}, S.-C., {et~al.} 2015, \aap, 581, A15

\bibitem[{{Tramper} {et~al.}(2014){Tramper}, {Sana}, {de Koter}, {Kaper}, \&
  {Ram{\'\i}rez-Agudelo}}]{tramper14}
{Tramper}, F., {Sana}, H., {de Koter}, A., {Kaper}, L., \&
  {Ram{\'\i}rez-Agudelo}, O.~H. 2014, \aap, 572, A36

\bibitem[{{Vergani} {et~al.}(2015){Vergani}, {Salvaterra}, {Japelj}, {Le
  Floc'h}, {D'Avanzo}, {Fernandez-Soto}, {Kr{\"u}hler}, {Melandri}, {Boissier},
  {Covino}, {Puech}, {Greiner}, {Hunt}, {Perley}, {Petitjean}, {Vinci},
  {Hammer}, {Levan}, {Mannucci}, {Campana}, {Flores}, {Gomboc}, \&
  {Tagliaferri}}]{vergani15}
{Vergani}, S.~D., {Salvaterra}, R., {Japelj}, J., {et~al.} 2015, \aap, 581,
  A102

\bibitem[{{Vink} {et~al.}(2001){Vink}, {de Koter}, \& {Lamers}}]{vink01}
{Vink}, J.~S., {de Koter}, A., \& {Lamers}, H.~J.~G.~L.~M. 2001, \aap, 369, 574

\bibitem[{{Walborn}(1972)}]{walborn72}
{Walborn}, N.~R. 1972, \aj, 77, 312

\bibitem[{{Walborn} \& {Fitzpatrick}(1990)}]{wf90}
{Walborn}, N.~R. \& {Fitzpatrick}, E.~L. 1990, \pasp, 102, 379

\bibitem[{{Walborn} {et~al.}(2002){Walborn}, {Howarth}, {Lennon}, {Massey},
  {Oey}, {Moffat}, {Skalkowski}, {Morrell}, {Drissen}, \& {Parker}}]{walborn02}
{Walborn}, N.~R., {Howarth}, I.~D., {Lennon}, D.~J., {et~al.} 2002, \aj, 123,
  2754

\bibitem[{{Wofford} {et~al.}(2014){Wofford}, {Leitherer}, {Chandar}, \&
  {Bouret}}]{wofford14}
{Wofford}, A., {Leitherer}, C., {Chandar}, R., \& {Bouret}, J.-C. 2014, \apj,
  781, 122

\bibitem[{{Xu} {et~al.}(2013{\natexlab{a}}){Xu}, {Goriely}, {Jorissen}, {Chen},
  \& {Arnould}}]{Xu2013a}
{Xu}, Y., {Goriely}, S., {Jorissen}, A., {Chen}, G.~L., \& {Arnould}, M.
  2013{\natexlab{a}}, \aap, 549, A106

\bibitem[{{Xu} {et~al.}(2013{\natexlab{b}}){Xu}, {Takahashi}, {Goriely},
  {Arnould}, {Ohta}, \& {Utsunomiya}}]{Xu2013b}
{Xu}, Y., {Takahashi}, K., {Goriely}, S., {et~al.} 2013{\natexlab{b}}, \nphysa,
  918, 61

\bibitem[{{Yoon} {et~al.}(2006){Yoon}, {Langer}, \& {Norman}}]{yoon06}
{Yoon}, S.~C., {Langer}, N., \& {Norman}, C. 2006, \aap, 460, 199

\end{thebibliography}

\end{document}